\documentclass[aip,amsmath,amssymb,preprint]{revtex4-1}

\usepackage{graphicx}
\usepackage{dcolumn}
\usepackage{bm}

\usepackage[utf8]{inputenc}
\usepackage[T1]{fontenc}
\usepackage{mathptmx}
\usepackage{etoolbox}
\usepackage[dvipsnames]{xcolor}
\usepackage{subcaption}

\makeatletter
\def\@email#1#2{%
 \endgroup
 \patchcmd{\titleblock@produce}
  {\frontmatter@RRAPformat}
  {\frontmatter@RRAPformat{\produce@RRAP{*#1\href{mailto:#2}{#2}}}\frontmatter@RRAPformat}
  {}{}
}%
\makeatother
\begin{document}

\preprint{AIP/123-QED}

\title[Treatment of Thermal Non-Equilibrium Dissociation Rates: Application to $\text H_2$]{Treatment of Thermal Non-Equilibrium Dissociation Rates: Application to $\text H_2$}
\author{Alex T. Carroll} \email{amoricar@caltech.edu}
\author{Jacob Wolmer}
\author{Guillaume Blanquart}
\affiliation{Department of Mechanical and Civil Engineering, California Institute of Technology, Pasadena, California 91125}
\author{Aaron M. Brandis}
\affiliation{NASA Ames Research Center, Mountain View, California 94035}
\author{Brett A. Cruden}
\affiliation{AMA, Inc., NASA Ames Research Center, Mountain View, California 94035}
\date{\today}

\begin{abstract}
This work presents a detailed description of the thermochemical non-equilibrium dissociation of diatomic molecules, and applies this theory to the case of $\rm H_2$ dissociation. The master equations are used to derive corresponding aggregate rate constant expressions that hold for any degree of thermochemical non-equilibrium. These general expressions are analyzed in three key limits/ regimes: the thermal equilibrium limit, the quasi-steady-state (QSS) regime, and the pre-QSS regime. Under several simplifying assumptions, an analytical source term expression that holds in all of these regimes, and is only a function of the translational temperature, $T_{\rm t}$, and the fraction of dissociation, $\phi_{\rm A}$, is proposed. This expression has two input parameters: the QSS dissociation rate constant in the absence of recombination, $k_{\rm d,nr}(T_{\rm t})$, and a pre-QSS correction factor, $\eta(T_{\rm t})$. The value of $\eta(T_{\rm t})$ is evaluated by comparing the predictions of the proposed expression against existing master equation simulations of a 0-D isothermal and isochoric reactor for the case of $\rm H_2$ dissociation with the third-bodies $\rm H_2$, H, and He. Despite its simple functional form, the proposed expression is able to reproduce the master equation results for the majority of the tested conditions. The best fit of $k_{\rm d,nr}(T_{\rm t})$ is then evaluated by conducting a detailed literature review. Data from a wide range of experimental and computational studies are considered for the third-bodies $\rm H_2$, H, and inert gases, and fits that are valid from 200 to 20,000 K are proposed. From this review, the uncertainty of the proposed fits are estimated to be less than a factor of two.
\end{abstract}

\maketitle

\section{Introduction}
\label{sec:intro}

The dissociation of diatomic species is a relevant physical process for many fields of research, including astrophysics~\cite{Draine1983,Hollenbach1989,Chang1991,Bell2018,Kristensen2023}, plasma physics~\cite{Matveyev1995,VPSilakov1996,Capitelli2002,Shakhatov2018,Smith2024}, detonations~\cite{Taylor2013,Voelkel2016,Shi2016,Vargas2022,Baumgart2025}, and hypersonic entry flows~\cite{Park2011,Park2012,Palmer2014,Erb2019,Carroll2023_conv,Scoggins2024,Steer2024,Steuer2024,Rataczak2024}. In each of these applications, large temperature gradients lead to internal energy excitation and dissociation of molecular species.
Understanding and accurately modeling these non-equilibrium phenomena is complicated by the fact that the processes of internal energy relaxation and dissociation are inherently coupled. For example, high internal energy states are preferentially dissociated, which lead to non-Boltzmann state distributions~\cite{Marrone1963,Macheret1994}.

To obtain a detailed description of these non-equilibrium phenomena, ab-initio computational methods can be used. The most popular of these methods involves first constructing accurate potential energy surfaces (PESs) that are then used in quasi-classical trajectory (QCT) calculations to obtain state-specific cross-sections/ rate constants that explicitly describe the internal energy excitation and dissociation processes. An overview on the earliest uses of the QCT method is provided by Porter~\cite{Porter1974}. These state-specific rates can then be used in master equation simulations, where the transitions/ reactions between internal energy states are explicitly computed. A related method is the direct molecular simulation (DMS) approach proposed by Koura~\cite{Koura1997}, where trajectory calculations on a given PES are performed during a flow simulation.

Recently, several studies in the hypersonic flow literature have used either the master equation approach~\cite{Kim2009,Kim2012,Panesi2013,Andrienko2015} or the DMS approach~\cite{Norman2013,Valentini2014,Schwartzentruber2018} to study non-equilibrium dissociation at high temperatures (> 10,000 K). These studies have highlighted several key physical features in a dissociating post-shock flow. Namely, the internal state distribution of the dissociating molecule initially begins at a Boltzmann distribution defined at the pre-shock temperature. Then, excited internal energy states become populated, eventually leading to a local balance between excitation and dissociation known as the quasi-steady-state (QSS) phenomena. Finally, the distribution evolves towards thermochemical equilibrium, defined by a Boltzmann distribution at the post-shock temperature.

While the physical insights gained from these ab-initio studies are valuable, they are unfortunately too computationally expensive to use in most practical computational fluid dynamics (CFD) calculations. The best strategy for incorporating the necessary physics from these detailed approaches into computationally tractable kinetic models for CFD calculations is still an open question. Some models, such as the Park two-temperature~\cite{Park1988,Park1989,Park1989NonequilibriumHA}, the modified Marrone-Treanor (MMT)~\cite{Chaudhry2019,Chaudhry2020} and the Singh-Schwartzentruber~\cite{Singh2020_1,Singh2020_2} models, rely on a multi-temperature framework where rates are described as a function of the translational temperature and one or more internal energy temperatures. Others~\cite{Magin2012,Sahai2017,Venturi2020} rely on the more general multi-group/ coarse-graining approach described by Liu et al.~\cite{Liu2015}, where individual rovibrational states of the dissociating molecules are binned together into a number of groups. While the multi-temperature and coarse-graining approaches have been used successfully in the past to reproduce the dissociation dynamics from detailed ab-initio calculations, there are still a few limitations. Namely, they rely on multiple coupled input parameters that must be evaluated consistently, and they also require the solution of additional species and/ or internal energy conservation equations.

Separately, the task of verifying dissociation rate constants across multiple different sources from the literature can be difficult. For hypersonic entry flows in particular, the dissociation rates for $\rm N_2$/ $\rm O_2$ and $\rm CO_2$ chemistry have been studied and reviewed in the past due to their relevance for Earth and Mars entry flows~\cite{Park1993,Park1994,Park2001,Sarma2000,Pietanza2021}. Comparatively less work has been done to review and validate relevant data for $\rm H_2$/ He systems. $\rm H_2$/ He chemistry is of particular interest in the planetary entry community, as the recently published 2023-2032 decadal survey for planetary exploration identified probe missions to the giant plants Saturn and Uranus (whose atmospheres are composed primarily of $\rm H_2$/ He) as priorities~\cite{NAP26522}. As highlighted in the recent aerothermal heating uncertainty studies by Palmer et al.~\cite{Palmer2014} and Rataczak et al.~\cite{Rataczak2024}, uncertainties in $\rm H_2$ dissociation rate constants can have a significant impact on predicted convective and radiative heating rates. This presents a problem, as the accurate prediction of heating rates is crucial for the design of mass efficient thermal protection systems for entry probes. Many computational studies in the ice and gas giant entry flow literature~\cite{Palmer2014,Higdon2018,Liu2021,Hansson2021,Carroll2023_conv,Carroll2023,Coelho2023,Rataczak2024} have used the dissociation rate constants proposed by Leibowitz~\cite{Leibowitz1973_1}, despite the fact that the Leibowitz rates are only based on a single set of shock tube experiments (namely those of Jacobs et al.~\cite{Jacobs1967}). Reviews of experimentally-reported $\rm H_2$ dissociation rate constants have been performed by Cohen and Westburg~\cite{Cohen1983} and Baulch et al.~\cite{Baulch1972,Baulch2005}. However, these reviews were primarily concerned with temperature ranges relevant for combustion applications (1,000 - 3,000 K), and did not consider the results of any computational studies. As will be shown later, computational studies are essential for constraining the gaps in the experimental data sets, especially at the high temperatures most relevant for hypersonic entry flow applications.

In light of these issues, the objective of the present work is twofold. First, to develop and validate a computationally inexpensive dissociation model that still captures the necessary physics from detailed ab-initio approaches. Second, to perform an extensive review of the corresponding model inputs for the case of $\rm H_2$ dissociation with the third-bodies $\rm H_2$, H, and He/ inert gases. The paper is organized as follows. First, section~\ref{sec:noneq} discusses the theory of non-equilibrium dissociation and the mathematical formulation of the master equations. From the master equation formulation, a simplified non-equilibrium rate constant expression (that is only a function of the translational temperature and the fraction of dissociation) is proposed. Next, in section~\ref{sec:applicationH2}, the proposed expression is validated for the case of $\rm H_2$ dissociation against existing master equation results. Then, in section~\ref{sec:review}, a review of the available rate constant data for $\rm H_2$ dissociation from both experimental and computational sources is presented. Following this review, new rate constant fits that are valid across the temperature range from 200 - 20,000 K are proposed.

\section{Theory of Non-Equilibrium Dissociation}
\label{sec:noneq}

Non-equilibrium effects on dissociation are first described in terms of state-specific rates and the corresponding master equations in section \ref{sec:StS}. Then, three important limits/ regimes of dissociation rates are discussed: the thermal equilibrium limit (\ref{sec:thermal}), the quasi-steady-state (QSS) regime (\ref{sec:QSS}), and finally the pre-QSS regime (\ref{sec:preQSS}). Throughout this section, only the case of dissociation in a single-component mixture/ an infinitely dilute bath of a single third-body is considered. The case of dissociation in a multi-component mixture is discussed separately in Appendix~\ref{sec:mixtures}.

\subsection{State-Specific Rates and Master Equations}
\label{sec:StS}

For an arbitrary homonuclear diatomic molecule $\rm A_2$, the dissociation/ recombination reaction through collisions with a third-body, M, can be written as follows
\begin{equation}
    \rm A_2 + M \leftrightarrow 2A + M.
    \label{eqn:A2bulk}
\end{equation}
The chemical source terms for this reaction are given by
\begin{equation}
    \frac{dn_{\rm A_2}}{dt} = -n_{\rm A_2}n_{\rm M}k_{\rm d} + n_{\rm A}^2n_{\rm M}k_{\rm r}
    \label{eqn:dnA2}
\end{equation}
and
\begin{equation}
    \frac{dn_{\rm A}}{dt} = -2\frac{dn_{\rm A_2}}{dt} = 2n_{\rm A_2}n_{\rm M}k_{\rm d} - 2n_{\rm A}^2n_{\rm M}k_{\rm r},
    \label{eqn:dnA}
\end{equation}
where $t$ is time, $n_{\rm i}$ is the number density for species i, and $k_{\rm d}$ and $k_{\rm r}$ are the dissociation and recombination rate constants, respectively.

While reaction~\eqref{eqn:A2bulk} and Eq.~\eqref{eqn:dnA2} and \eqref{eqn:dnA} are adequate for describing the dissociation/ recombination of $\rm A_2$ in an aggregate sense, they do not consider the detailed internal energy distributions of any of the reactants or products. Therefore, to capture the evolution of the rovibrational distribution of $\rm A_2$, the same dissociation process of reaction~\eqref{eqn:A2bulk} is written in terms of the reactions of the individual internal energy states of $\rm A_2$,
\begin{equation}
    \rm A_2 {\it(J,\nu)} + M \leftrightarrow 2A + M,
    \label{eqn:A2jvdiss}
\end{equation}
along with a separate reaction to describe the relaxation of the internal energy modes of $\rm A_2$,
\begin{equation}
    \rm A_2 {\it(J,\nu)} + M \leftrightarrow A_2 {\it(J',\nu')} + M.
    \label{eqn:A2jvrelax}
\end{equation}
Here, $J$ and $\nu$ represent the rotational and vibrational quantum numbers, respectively. Throughout this work, only the electronic ground states of $\rm A_2$ and A are considered. The corresponding chemical source terms/ master equations for these state-specific reactions are given by
\begin{eqnarray}
    \frac{d n_{\rm A_2 \it(J,\nu)}}{dt} =
    && n_{\rm M} \sum_{\nu'=0}^{\nu_{\rm max}} \sum_{J'=0}^{J_{\rm max}(\nu')} n_{\rm A_2 \it (J',\nu')} k(J',\nu' \rightarrow J,\nu) 
    + n_{\rm M} n_{\rm A}^2 k(c \rightarrow J,\nu) \nonumber\\ &&
    - n_{\rm M} \sum_{\nu'=0}^{\nu_{\rm max}} \sum_{J'=0}^{J_{\rm max}(\nu')} n_{\rm A_2 \it (J,\nu)} k(J,\nu \rightarrow J',\nu') 
    - n_{\rm M} n_{\rm A_2 \it(J,\nu)} k(J,\nu \rightarrow c)
\label{eqn:masterA2jv}
\end{eqnarray}
and
\begin{equation}
    \frac{d n_{\rm A}}{dt} = 2n_{\rm M} \sum_{\nu=0}^{\nu_{\rm max}} \sum_{J=0}^{J_{\rm max}(\nu)} n_{\rm A_2 \it (J,\nu)} k(J,\nu \rightarrow c) - 2 n_{\rm M} n_{\rm A}^2\sum_{\nu=0}^{\nu_{\rm max}} \sum_{J=0}^{J_{\rm max}(\nu)} k(c \rightarrow J,\nu).
    \label{eqn:masterA}
\end{equation}
In these equations, the $(J,\nu)$ state-specific relaxation rate constants are denoted as $k(J,\nu \rightarrow J',\nu')$ and $k(J',\nu' \rightarrow J,\nu)$, and the state-specific dissociation and recombination rate constants are denoted as $k(J,\nu \rightarrow c)$ and $k(c \rightarrow J,\nu)$ respectively, where $c$ indicates the dissociated state of $\rm A_2$.

The forward and backward rate constants for the state-specific processes are related via micro-reversibility as
\begin{equation}
    \frac{k(J,\nu \rightarrow c)}{k(c \rightarrow J,\nu)} = \frac{n_{\rm A, eq}^2}{n_{\rm A_2 \it(J,\nu), \rm eq}} = \frac{Q_{\rm t,A}^2Q_{\rm A}^2}{Q_{\rm t,A_2}Q_{\rm A_2 \it(J,\nu)}}\exp \left(-\frac{\theta_{\rm d,A_2}}{T_{\rm t}} \right)
    \label{eqn:A2microdiss}
\end{equation}
and
\begin{equation}
    \frac{k(J,\nu \rightarrow J',\nu')}{k(J',\nu' \rightarrow J,\nu)} = \frac{n_{\rm A_2 \it(J',\nu'), \rm eq}}{n_{\rm A_2 \it(J,\nu), \rm eq}} = \frac{Q_{\rm A_2 \it(J',\nu')}}{Q_{\rm A_2 \it(J,\nu)}}.
    \label{eqn:A2microrelax}
\end{equation}
Here, the subscript eq refers to the locally defined equilibrium state evaluated at the translational temperature, $T_{\rm t}$, and $\theta_{\rm d,A_2}$ is the characteristic dissociation temperature of $\rm A_2$ measured from the ground state. $Q_{\rm t,i} = (2\pi m_{\rm i} k_{\rm B} T_{\rm t}/h^2)^{3/2}$ is the translational partition function with species' mass $m_i$ and the Boltzmann and Planck constants, $k_{\rm B}$ and $h$, respectively, while $Q_{\rm i}$ is the internal partition function for species i evaluated at $T_{\rm t}$. The state-specific internal partition function, $Q_{\rm A_2 \it(J,\nu)}$, is related to the bulk $\rm A_2$ internal partition function, $Q_{\rm A_2}$, as
\begin{equation}
    Q_{\rm A_2} = \sum_{\nu=0}^{\nu_{\rm max}} \sum_{J=0}^{J_{\rm max}(\nu)} Q_{\rm A_2 \it(J,\nu)}.
    \label{eqn:QA2}
\end{equation}
For the case of H and $\rm H_2$ (considering only electronic ground states), $Q_{\rm H}=2$ and $Q_{{\rm H_2}(J,\nu)} = g_{\rm s} (2J+1) \exp( -\theta_{\rm H_2 \it(J,\nu)} / T_{\rm t} )$, where $\theta_{{\rm H_2}(J,\nu)}$ is the characteristic rovibrational temperature of ${\rm H_2}(J,\nu)$, and the statistical weight due to nuclear spin-splitting, $g_{\rm s}$, is 1/4 for all even $J$ (para-hydrogen) and 3/4 for all odd $J$ (ortho-hydrogen)~\cite{Colonna2012,Popovas2016}.

The state-specific rate constants, $k(J,\nu \rightarrow c)$, $k(c \rightarrow J,\nu)$, $k(J,\nu \rightarrow J',\nu')$, and $k(J',\nu' \rightarrow J,\nu)$, are all only functions of the translational temperature, $T_{\rm t}$. This is a direct consequence of the fact that the reaction of a given $\rm A_2 \it(J,\nu)$ state is only a function of its impact with the third-body, M. Therefore, if the internal energy of M does not change during a collision, then the state-specific rate constants can only be a function of $T_{\rm t}$. In the event that M is another molecule that could simultaneously undergo a relaxation or reaction, the state-specific rate constants can still be written as just functions of $T_{\rm t}$, provided that the rate constants are written to index the internal energy states of both $\rm A_2 \it(J,\nu)$ and M$(J,\nu)$~\cite{Kim2012}. For the sake of simplicity, the less general case of M with a fixed internal energy will be considered for the analysis presented here.

Comparing the atomic master equation of Eq.~\eqref{eqn:masterA} to Eq.~\eqref{eqn:dnA}, it is evident that the aggregate rate constants can be expressed in terms of the state-specific rate constants as
\begin{equation}
    k_{\rm d} = \sum_{\nu=0}^{\nu_{\rm max}}\sum_{J=0}^{J_{\rm max}(\nu)} \frac{n_{\rm A_2 \it(J,\nu)}}{n_{\rm A_2}} k(J,\nu \rightarrow c)
    \label{eqn:A2kd_sum}
\end{equation}
and
\begin{equation}
    k_{\rm r} = \sum_{\nu=0}^{\nu_{\rm max}}\sum_{J=0}^{J_{\rm max}(\nu)} k(c \rightarrow J,\nu),
    \label{eqn:A2kr_sum}
\end{equation}
where 
\begin{equation}
    n_{\rm A_2} = \sum_{\nu=0}^{\nu_{\rm max}}\sum_{J=0}^{J_{\rm max}(\nu)} n_{\rm A_2 \it(J,\nu)}.
    \label{eqn:A2Jv_sum}
\end{equation}
Using the micro-reversibility relation of Eq.~\eqref{eqn:A2microdiss}, Eq.~\eqref{eqn:A2kr_sum} can equivalently be written as
\begin{equation}
    k_{\rm r} = K_{\rm eq}^{-1} \sum_{\nu=0}^{\nu_{\rm max}}\sum_{J=0}^{J_{\rm max}(\nu)} \frac{Q_{\rm A_2 \it(J,\nu)}}{Q_{\rm A_2}} k(J,\nu \rightarrow c).
    \label{eqn:A2kr_sum2}
\end{equation}
Here, we have used the fact that $n_{\rm A_2,eq}$ and $n_{\rm A_2{\it(J,\nu)},eq}$ are related via their internal partition functions as
\begin{equation}
    \frac{n_{\rm A_2 \it(J,\nu) \rm ,eq}}{n_{\rm A_2,eq}} = \frac{Q_{\rm A_2 \it(J,\nu)}}{Q_{\rm A_2}},
    \label{eqn:A2Boltz}
\end{equation}
and that $n_{\rm A_2,eq}$ and $n_{\rm A,eq}$ are related via the equilibrium constant, $K_{\rm eq}$, as
\begin{equation}
    K_{\rm eq} = \frac{n_{\rm A,eq}^2}{n_{\rm A_2,eq}} = \frac{Q_{\rm t,A}^2Q_{\rm A}^2}{Q_{\rm t,A_2}Q_{\rm A_2}}\exp \left(-\frac{\theta_{\rm d,A_2}}{T_{\rm t}} \right).
    \label{eqn:A2Keq}
\end{equation}
Because Eq.~\eqref{eqn:A2kd_sum} and \eqref{eqn:A2kr_sum2} come from the master equations of Eq.~\eqref{eqn:masterA2jv} and \eqref{eqn:masterA}, these results are general, and hold for any degree of thermochemical non-equilibrium. 

In a compression/ shocked flow environment, the rovibrational distribution of $\rm A_2$ initially begins as a Boltzmann distribution at the pre-shock temperature, which then evolves towards a Boltzmann distribution at the post-shock equilibrium temperature. In the general case, this evolution can occur through a series of non-Boltzmann distributions, in contrast to the assumption of a typical multi-temperature model, which assumes that it must occur through a series of Boltzmann distributions. Because the distribution of $\rm A_2 \it(J,\nu)$ can continually change in this post-shock environment, Eq.~\eqref{eqn:A2kd_sum} gives that the effective aggregate $k_{\rm d}$ can continually change as well. Therefore, unlike the state-specific rate constants, $k_{\rm d}$ is a function of the instantaneous distribution of $\rm A_2 \it(J,\nu)$.

In contrast, Eq.~\eqref{eqn:A2kr_sum} gives that $k_{\rm r}$ must always be a function of the state-specific rates alone, and therefore, can only ever be a function of $T_{\rm t}$. This will be of particular importance when interpreting the results of the QSS analysis presented in section~\ref{sec:QSS}.

\subsection{Thermal Rates}
\label{sec:thermal}

In the limit of thermal equilibrium, the rovibrational distribution of $\rm A_2$ can be described simply by Eq.~\eqref{eqn:A2Boltz}. Therefore, substituting Eq.~\eqref{eqn:A2Boltz} into Eq.~\eqref{eqn:A2kd_sum} gives the thermal dissociation rate constant,
\begin{equation}
    k_{\rm d, th} = \sum_{\nu=0}^{\nu_{\rm max}}\sum_{J=0}^{J_{\rm max}(\nu)} \frac{Q_{\rm A_2 \it(J,\nu)}}{Q_{\rm A_2}} k(J,\nu \rightarrow c).
    \label{eqn:A2kd_thermal}
\end{equation}
This thermal rate is referred to as the ``one-way'' rate by Park~\cite{Park1989NonequilibriumHA} and Kim and Boyd~\cite{Kim2012,Kim2013}. Taking the ratio of Eq.~\eqref{eqn:A2kd_thermal} over the general expression for $k_{\rm r}$ given by Eq.~\eqref{eqn:A2kr_sum2}, the expected result that macroscopic detailed balance must be satisfied at thermal equilibrium is obtained,
\begin{equation}
    \frac{k_{\rm d, th}}{k_{\rm r}} = K_{\rm eq}.
    \label{eqn:A2DEB}
\end{equation}
Since Eq.~\eqref{eqn:A2kr_sum2} is a general relation for $k_{\rm r}$, this result implies that regardless of the degree of non-equilibrium, $k_{\rm r}$ must always satisfy Eq.~\eqref{eqn:A2DEB}.

\subsection{QSS Rates}
\label{sec:QSS}
\subsubsection{Formulation and Solution for $\bm{A_2(J,\nu)}$}
A second regime of importance for the distribution of $\rm A_2 \it(J,\nu)$ and therefore $k_{\rm d}$ is the quasi-steady-state or QSS condition. The QSS condition arises when the rate of change of $n_{\rm A_2{\it(J,\nu)}}$ is significantly smaller than the total production term in the right-hand side of Eq.~\eqref{eqn:masterA2jv},
\begin{equation}
    \left| \frac{d n_{\rm A_2 \it(J,\nu)}}{dt} \right| \ll n_{\rm M} \sum_{\nu'=0}^{\nu_{\rm max}} \sum_{J'=0}^{J_{\rm max}(\nu')} n_{\rm A_2 \it (J',\nu')} k(J',\nu' \rightarrow J,\nu) + n_{\rm M} n_{\rm A}^2 k(c \rightarrow J,\nu),
\label{eqn:masterA2jv_QSS1}
\end{equation}
or equivalently the total consumption term,
\begin{equation}
    \left| \frac{d n_{\rm A_2 \it(J,\nu)}}{dt} \right| \ll n_{\rm M} \sum_{\nu'=0}^{\nu_{\rm max}} \sum_{J'=0}^{J_{\rm max}(\nu')} n_{\rm A_2 \it (J,\nu)} k(J,\nu \rightarrow J',\nu') + n_{\rm M} n_{\rm A_2 \it(J,\nu)} k(J,\nu \rightarrow c).
\label{eqn:masterA2jv_QSS2}
\end{equation}
When this condition is satisfied, the total consumption term is approximately equal to the total production term such that
\begin{eqnarray}
    && \sum_{\nu'=0}^{\nu_{\rm max}} \sum_{J'=0}^{J_{\rm max}(\nu')} n_{\rm A_2 \it (J,\nu)} k(J,\nu \rightarrow J',\nu') + n_{\rm A_2 \it(J,\nu)} k(J,\nu \rightarrow c) \nonumber \\
    && \approx \sum_{\nu'=0}^{\nu_{\rm max}} \sum_{J'=0}^{J_{\rm max}(\nu')} n_{\rm A_2 \it (J',\nu')} k(J',\nu' \rightarrow J,\nu) + n_{\rm A}^2 k(c \rightarrow J,\nu).
\label{eqn:masterA2jv_QSS3}
\end{eqnarray}

To obtain the corresponding dissociation rate constant at QSS, a similar setup to the one proposed by Park~\cite{Park1989NonequilibriumHA} is used here. In particular, the local equilibrium normalized number densities, 
\begin{equation}
    \phi_{\rm A_2 \it(J,\nu)} \equiv  \frac{n_{\rm A_2 \it (J,\nu)}} {n_{\rm A_2 \it (J,\nu), \rm eq}},
    \label{eqn:phiA2}
\end{equation}
\begin{equation}
    \phi_{\rm A} \equiv \frac{n_{\rm A}}{n_{\rm A, eq}},
    \label{eqn:phiA}
\end{equation}
and
\begin{equation}
    \chi \equiv \frac{n_{\rm A_2}}{n_{\rm A_2,eq}}
    \label{eqn:chi}
\end{equation}
are defined. For convenience, the equilibrium normalized fractional density for ${\rm A_2}(J,\nu)$ is also defined as
\begin{equation}
    \psi_{\rm A_2 \it(J,\nu)} \equiv  \frac{\phi_{\rm A_2 \it (J,\nu)}} {\chi}.
    \label{eqn:psiA2}
\end{equation}
Then, Eq.~\eqref{eqn:masterA2jv_QSS3} is rewritten using these definitions and Eq.~\eqref{eqn:A2microdiss}, \eqref{eqn:A2microrelax}, and \eqref{eqn:A2Boltz} to get
\begin{eqnarray}
    && \phi_{\rm A_2 \it(J,\nu)} \left[ \sum_{\nu'=0}^{\nu_{\rm max}} \sum_{J'=0}^{J_{\rm max}(\nu')} k(J,\nu \rightarrow J',\nu') + k(J,\nu \rightarrow c) \right]
    - \sum_{\nu'=0}^{\nu_{\rm max}} \sum_{J'=0}^{J_{\rm max}(\nu')} \phi_{\rm A_2 \it(J',\nu')} k(J,\nu \rightarrow J',\nu') \nonumber \\
    && = \phi_{\rm A}^2 k(J,\nu \rightarrow c).
\label{eqn:masterA2jv_QSS4}
\end{eqnarray}
For $N$ total rovibrational states of $\rm A_2$, Eq.~\eqref{eqn:masterA2jv_QSS4} gives $N$ constraint equations. However, Eq.~\eqref{eqn:A2Jv_sum} gives an additional constraint for the total number density of $\rm A_2$ that must be satisfied as well. Written in terms of $\phi_{\rm A_2 \it(J,\nu)}$, this additional constraint is expressed as
\begin{equation}
    \sum_{\nu=0}^{\nu_{\rm max}} \sum_{J=0}^{J_{\rm max}(\nu)} \phi_{\rm A_2 \it(J,\nu)} n_{\rm A_2 \it(J,\nu) \rm ,eq} = \chi n_{\rm A_2,eq},
    \label{eqn:A2chi_n}
\end{equation}
or equivalently
\begin{equation}
    \sum_{\nu=0}^{\nu_{\rm max}} \sum_{J=0}^{J_{\rm max}(\nu)} \phi_{\rm A_2 \it(J,\nu)} \frac{Q_{\rm A_2 \it(J,\nu)}}{Q_{\rm A_2}} = \chi.
    \label{eqn:A2chi_Q}
\end{equation}
This gives an over-constrained system with $N+1$ equations for $N$ unknowns. To resolve this over-constraint, the QSS condition given by Eq.~\eqref{eqn:masterA2jv_QSS4} for the ground state, $\rm A_2$($J$=0,$\nu$=0), is dropped. This follows from the understanding that the QSS condition arises as a competition between the production process of excitation and the consumption process of dissociation, as the magnitudes of the excitation and dissociation terms in Eq.~\eqref{eqn:masterA2jv_QSS4} are initially much larger than the magnitudes of the de-excitation and recombination terms in compression/ shocked flows. As a consequence, the QSS condition is least likely to be satisfied by the ground state, as unlike all other excited rovibrational states of $\rm A_2 \it(J,\nu)$, there is no excitation term that can repopulate the ground state and compete with the dissociation process.

Equations~\eqref{eqn:masterA2jv_QSS4} and \eqref{eqn:A2chi_Q} can then be written in the matrix form
\begin{equation}
       \bm{M} \vec{\phi}_{\rm A_2} = \vec{C}\phi_{\rm A}^2 + \vec{D} \chi.
       \label{eqn:A2mat}
\end{equation}
The entries in the first row of matrix $\bm{M}$ are given by $Q_{\rm A_2 \it(J,\nu)}/Q_{\rm A_2}$, and the first entries of the column vectors $\vec{C}$ and $\vec{D}$ are 0 and 1, respectively. For all other rows, the entries of $\bm{M}$ are given by the coefficients in the left-hand side of Eq.~\eqref{eqn:masterA2jv_QSS4}, the entries of $\vec{C}$ are given by the coefficients in the right-hand side of Eq.~\eqref{eqn:masterA2jv_QSS4}, and $\vec{D}=0$. The solution to Eq.~\eqref{eqn:A2mat} can be written as the sum of two terms,
\begin{equation}
    \vec{\psi}_{\rm A_2,nr} \equiv \bm{M}^{-1}\vec{D}
    \label{eqn:A2hom}
\end{equation}
and
\begin{equation}
    \vec{\psi}_{\rm A_2,r} \equiv \bm{M}^{-1}\vec{C},
    \label{eqn:A2par}
\end{equation}
such that
\begin{equation}
    \vec{\phi}_{\rm A_2} = \vec{\psi}_{\rm A_2,nr}\chi + \vec{\psi}_{\rm A_2,r} \phi_{\rm A}^2.
    \label{eqn:phi}
\end{equation}
$\vec{\psi}_{\rm A_2,nr}$ represents the QSS solution in the absence of recombination/ in the non-recombining (nr) limit, and $\vec{\psi}_{\rm A_2,r}$ represents the contribution of recombination (r) to the QSS solution. In Park~\cite{Park1989NonequilibriumHA} as well as the subsequent master equation studies by Kim et al.~\cite{Kim2009,Kim2010} and Kim and Boyd~\cite{Kim2012,Kim2013}, the solution to Eq.~\eqref{eqn:A2mat} was instead described as the sum of a homogeneous component, $\vec{\psi}_{\rm A_2,h}\equiv {\bm M}^{-1} \vec{D} \chi$, and a particular component, $\vec{\psi}_{\rm A_2,p} \equiv {\bm M}^{-1} \vec{C}$. The formulation of the present work differs in that $\chi$ is pulled out of the definition of the homogeneous solution, such that $\vec{\psi}_{\rm A_2,nr}$ and $\vec{\psi}_{\rm A_2,r}$ both represent particular solutions to Eq.~\eqref{eqn:A2mat} that are only functions of ${\bm M}$, $\vec{D}$, and $\vec{C}$.

Two key observations are made here regarding the QSS formulation described by Eq.~\eqref{eqn:A2hom}-\eqref{eqn:phi}. Firstly, $\bm M$, $\vec{C}$, and $\vec{D}$ are only functions of the state-specific rates, which in turn are only functions of $T_{\rm t}$. As a result, $\vec{\psi}_{\rm A_2,nr}$ and $\vec{\psi}_{\rm A_2,r}$ are also only functions of $T_{\rm t}$. Therefore, even though the QSS distribution may correspond to internal rovibrational temperatures, $T_{\rm r}$ and $T_{\rm v}$, that are not equal to $T_{\rm t}$, the QSS solution is none-the-less functionalized by $T_{\rm t}$ alone. Secondly, since the QSS assumption of approximately equal production and consumption terms as given by Eq.~\eqref{eqn:masterA2jv_QSS3} is true at equilibrium, Eq.~\eqref{eqn:phi} must also be valid at equilibrium. By construction, $\phi_{\rm A}$ and $\chi$ are equal to 1 and $\vec{\phi}_{\rm A_2}$ is the unity vector at equilibrium. Hence, Eq.~\eqref{eqn:phi} implies that
\begin{equation}
    1 = \vec{\psi}_{\rm A_2,nr} + \vec{\psi}_{\rm A_2,r}.
    \label{eqn:phi2}
\end{equation}

\subsubsection{Solution for $\bm{k_d}$}
\label{sec:kdQSS}

To compute the dissociation rate constant at QSS, the general expression for $k_{\rm d}$ given by Eq.~\eqref{eqn:A2kd_sum} is first rewritten in terms of $\psi_{{\rm A_2}(J,\nu)}$ using Eq.~\eqref{eqn:A2Boltz}, \eqref{eqn:phiA2}, \eqref{eqn:chi}, and \eqref{eqn:psiA2} as
\begin{equation}
    k_{\rm d}
    = \sum_{\nu=0}^{\nu_{\rm max}}\sum_{J=0}^{J_{\rm max}(\nu)} \psi_{\rm A_2 \it (J,\nu)} \frac{Q_{\rm A_2 {\it(J,\nu)}}}{Q_{\rm A_2}} k(J,\nu \rightarrow c).
    \label{eqn:A2kd_sum2}
\end{equation}
Separately, Eq.~\eqref{eqn:psiA2}, \eqref{eqn:phi}, and \eqref{eqn:phi2} are combined to give 
\begin{equation}
    \vec{\psi}_{\rm A_2} = \vec{\psi}_{\rm A_2,nr} \left( 1 - \frac{\phi_{\rm A}^2}{\chi} \right ) + \frac{\phi_{\rm A}^2}{\chi}.
    \label{eqn:phi3}
\end{equation}
Equation~\eqref{eqn:phi3} is then substituted into Eq.~\eqref{eqn:A2kd_sum2} to give the final expression for $k_{\rm d}$ at QSS,
\begin{equation}
    k_{\rm d} = \sum_{\nu=0}^{\nu_{\rm max}} \sum_{J=0}^{J_{\rm max}(\nu)} \left( \psi_{\rm A_2{\it (J,\nu)},nr} \left( 1 - \frac{\phi_{\rm A}^2}{\chi} \right ) + \frac{\phi_{\rm A}^2}{\chi} \right) \frac{Q_{\rm A_2 \it(J,\nu)}}{Q_{\rm A_2}} k(J,\nu \rightarrow c).
    \label{eqn:kd}
\end{equation}
Since $\vec{\psi}_{\rm A_2,nr}$ represents the QSS solution in the absence of recombination, the corresponding dissociation rate constant in the non-recombining limit, $k_{\rm d,nr}$, can be computed simply from Eq.~\eqref{eqn:A2kd_sum2} as
\begin{equation}
    k_{\rm d, nr} = \sum_{\nu=0}^{\nu_{\rm max}} \sum_{J=0}^{J_{\rm max}(\nu)} \psi_{\rm A_2 \it (J,\nu), \rm nr} \frac{Q_{\rm A_2 \it(J,\nu)}}{Q_{\rm A_2}} k(J,\nu \rightarrow c).
    \label{eqn:kdnr}
\end{equation}
Using Eq.~\eqref{eqn:kdnr} in addition to the expression for the thermal rate constant, Eq.~\eqref{eqn:A2kd_thermal}, Eq.~\eqref{eqn:kd} can be written more compactly as
\begin{equation}
    k_{\rm d} = k_{\rm d,nr} \left( 1 - \frac{\phi_{\rm A}^2}{\chi} \right ) + k_{\rm d,th} \frac{\phi_{\rm A}^2}{\chi}.
    \label{eqn:kdnrth}
\end{equation}
Here, the coefficient
\begin{equation}
    \frac{\phi_{\rm A}^2}{\chi} = \frac{n_{\rm A}^2}{n_{\rm A_2}}\frac{n_{\rm A_2,eq}}{n_{\rm A,eq}^2} = \frac{n_{\rm A}^2}{n_{\rm A_2}}\frac{1}{K_{\rm eq}}
    \label{eqn:prog}
\end{equation}
is effectively a progress variable that varies from 0 in the limit of no dissociation to 1 in the limit of equilibrium. As such, Eq.~\eqref{eqn:kdnrth} gives a description of $k_{\rm d}$ that varies between the two limiting rates of $k_{\rm d,nr}$ and $k_{\rm d,th}$. Since $k_{\rm d, nr}$ and $k_{\rm d, th}$ are only functions of $T_{\rm t}$, the description of $k_{\rm d}$ given by Eq.~\eqref{eqn:kdnrth} is only a function of $T_{\rm t}$ and $\phi_{\rm A}^2/\chi$. Despite this lack of dependence on any other internal temperatures (e.g., $T_{\rm r}$ or $T_{\rm v}$), the $k_{\rm d}$ given by Eq.~\eqref{eqn:kdnrth} only satisfies macroscopic detailed balance with $k_{\rm r}$ exactly in the limit where $\phi_{\rm A}^2/\chi=1$ and hence $k_{\rm d} = k_{\rm d,th}$. This implies that at this limit of the QSS condition, thermal and chemical equilibria are reached {\it simultaneously}.

Finally, it is worth noting that the description of $k_{\rm d}$ given by Eq.~\eqref{eqn:kdnrth} is valid everywhere where the QSS assumption of Eq.~\eqref{eqn:masterA2jv_QSS3} is valid, namely everywhere outside of the pre-QSS region. Some master equation studies in the literature, such as the study by Venturi et al.~\cite{Venturi2020} for the dissociation of $\rm O_2$ with M = O, distinguish between a ``QSS'' and ``post-QSS'' region. The discussion above suggests that it would be more appropriate to classify these regions as the non-recombining (nr) and recombining (r) regions, where $k_{\rm d}\approx k_{\rm d,nr}$ or $k_{\rm d,nr} < k_{\rm d} \leq k_{\rm d,th}$, respectively.

\subsubsection{Chemical Source Term}
\label{sec:QSSsrc}

While Eq.~\eqref{eqn:kdnrth} provides an expression for $k_{\rm d}$ that is valid for all regimes where the QSS assumption of Eq.~\eqref{eqn:masterA2jv_QSS3} is valid, it implies that separate fits for both $k_{\rm d,nr}$ and $k_{\rm d,th}$ in terms of $T_{\rm t}$ are required. However, if we are only concerned with reproducing the total chemical source term for each species, i.e., $dn_{\rm A_2}/dt$ and $dn_{\rm A}/dt$, as is typically the case in CFD calculations, a further simplification can be made that eliminates the dependence on $k_{\rm d,th}$. In particular, if Eq.~\eqref{eqn:kdnrth} is substituted into the general aggregate source term equation, Eq.~\eqref{eqn:dnA}, $dn_{\rm A}/dt$ can be written as
\begin{equation}
    \frac{dn_{\rm A}}{dt} = 2n_{\rm A_2}n_{\rm M}\left(k_{\rm d,nr} \left( 1 - \frac{\phi_{\rm A}^2}{\chi} \right ) + k_{\rm d,th} \frac{\phi_{\rm A}^2}{\chi} \right) - 2n_{\rm A}^2n_{\rm M}k_{\rm r}.
    \label{eqn:dnA_2}
\end{equation}
If Eq.~\eqref{eqn:prog} is then used to substitute for $\phi_{\rm A}^2/\chi$, Eq.~\eqref{eqn:dnA_2} can be written as
\begin{equation}
\begin{split}
    \frac{dn_{\rm A}}{dt}
    & = 2n_{\rm A_2}n_{\rm M}k_{\rm d,nr} - 2 n_{\rm A}^2n_{\rm M}\frac{k_{\rm d,nr}}{K_{\rm eq}}.
\end{split}
\label{eqn:dnA_3}
\end{equation}
Therefore, with no additional simplifying assumptions, the chemical source term from the full formulation of Eq.~\eqref{eqn:dnA_2} can be computed equivalently as solely a function of $k_{\rm d,nr}$. What's more, the dependence on $k_{\rm d,nr}$ in Eq.~\eqref{eqn:dnA_3} is exactly the same as it would be if one had ``incorrectly'' assumed that $k_{\rm d}=k_{\rm d,nr}$ and applied macroscopic detailed balance to compute the consumption term. However, the above analysis clearly shows that the consumption term with $k_{\rm d,nr}/K_{\rm eq}$ does not correspond to the physical recombination rate constant, $k_{\rm r}$, but instead comes from the negative term of the dissociation rate constant given by Eq.~\eqref{eqn:kdnrth}.

While not explicitly stated or justified, the alternate formulation given by Eq.~\eqref{eqn:dnA_3} is inherently what Park~\cite{Park1989NonequilibriumHA} and Kim~\cite{Kim2009,Kim2010,Kim2012,Kim2013} have used. In their formulation, the recombination rate constant was simply defined as
\begin{equation}
    k_{\rm r, Park} = \sum_{\nu=0}^{\nu_{\rm max}} \sum_{J=0}^{J_{\rm max}(\nu)} (1-\psi_{\rm A_2 \it (J,\nu), \rm r}) \frac{n_{\rm A_2,eq}}{n_{\rm A,eq}^2} \frac{Q_{\rm A_2 \it(J,\nu)}}{Q_{\rm A_2}} k(J,\nu \rightarrow c).
    \label{eqn:krpark}
\end{equation}
Using Eq.~\eqref{eqn:A2Keq}, \eqref{eqn:phi2}, and \eqref{eqn:kdnr}, one can show that the above expression is equivalent to
\begin{equation}
    k_{\rm r, Park} = \frac{k_{\rm d,nr}}{K_{\rm eq}}.
    \label{eqn:krparkDEB}
\end{equation}
Therefore, $k_{\rm r,Park}$ is defined such that it satisfies macroscopic detailed balance with $k_{\rm d,nr}$ by construction.

There are two key implications of these results. First, the same $k_{\rm d,nr}$ expression should be used to compute both the production and consumption terms of Eq.~\eqref{eqn:dnA_3}. Second, since $k_{\rm d,nr}$ is only a function of $T_{\rm t}$ in the QSS regime, one significant reason to use a fit of $k_{\rm d,nr}$ that depends on other variables is to capture dissociation in regions where the QSS assumption is not valid, i.e., in the pre-QSS regime.

\subsubsection{Transition Between $\bm k_{\bm d,nr}$ and $\bm k_{\bm d,th}$}
\label{sec:kdnrth_trans}

Leveraging the above analysis, it is possible to determine explicitly the degree of dissociation that is reached when $k_{\rm d}$ exceeds the $k_{\rm d,nr}$ limit. Mathematically, this limit is defined as the degree of dissociation reached when $k_{\rm d}$ exceeds some $\delta$ fraction of $k_{\rm d,nr}$, i.e., $k_{\rm d}=k_{\rm d,nr}(1+\delta)$. Substituting this limit for $k_{\rm d}$ into Eq.~\eqref{eqn:kdnrth} gives the limiting value for $\phi_{\rm A}^2/\chi$ as
\begin{equation}
    \frac{\phi_{\rm A}^2}{\chi} \biggr|_{\rm nr} = \frac{\delta k_{\rm d,nr}}{k_{\rm d,th}-k_{\rm d,nr}}.
    \label{eqn:phiA2chi_nr}
\end{equation}
Next, the degree of dissociation, $\alpha$, is defined as
\begin{equation}
    \alpha \equiv \frac{n_{\rm A}}{\tilde{n}_{\rm A}},
    \label{eqn:alpha}
\end{equation}
where $\tilde{n}_{\rm A} \equiv 2n_{\rm A_2}+n_{\rm A}$ is the total number density of A atoms. From this, the number densities of $\rm A_2$ and A can be written as
\begin{equation}
    n_{\rm A_2} = \frac{1-\alpha}{2}\tilde{n}_{\rm A}, \quad n_{\rm A} = \alpha\tilde{n}_{\rm A}\,.
    \label{eqn:alpha_n}
\end{equation}
By construction, $\alpha$ is related to $\phi_{\rm A}$ as
\begin{equation}
    \frac{\alpha}{\alpha_{\rm eq}} = \frac{n_{\rm A}}{\tilde{n}_{\rm A}} \frac{\tilde{n}_{\rm A}}{n_{\rm A,eq}} = \phi_{\rm A},
    \label{eqn:alphanorm2}
\end{equation}
such that $\phi_{\rm A}$ is a direct measure of the fraction of dissociation relative to the locally defined equilibrium at $T_{\rm t}$.

Using Eq.~\eqref{eqn:alpha_n} along with Eq.~\eqref{eqn:prog}, $\phi_{\rm A}^2/\chi$ takes the form
\begin{equation}
    \frac{\phi_{\rm A}^2}{\chi} = \frac{2\alpha^2\tilde{n}_{\rm A}}{(1-\alpha)K_{\rm eq}},
\end{equation}
with the corresponding solution for $\alpha$ given by
\begin{equation}
    \alpha = \frac{1}{2} \left[-\frac{K_{\rm eq}}{2\tilde{n}_{\rm A}}\frac{\phi_{\rm A}^2}{\chi} + \sqrt{\frac{K_{\rm eq}}{2\tilde{n}_{\rm A}}\frac{\phi_{\rm A}^2}{\chi} \left(\frac{K_{\rm eq}}{2\tilde{n}_{\rm A}}\frac{\phi_{\rm A}^2}{\chi}+4 \right)} \right].
    \label{eqn:alphasol}
\end{equation}
To account for the fact that the equilibrium degree of dissociation can vary as a function of $T_{\rm t}$ and $\tilde{n}_{\rm A}$, $\alpha$ is normalized by $\alpha_{\rm eq}$ (computed using Eq.~\eqref{eqn:alphasol} with $\phi_{\rm A}^2/\chi$ = 1) and the value of $\phi_A^2/\chi$ in the non-recombining limit from Eq.~\eqref{eqn:phiA2chi_nr} is substituted to give
\begin{equation}
    \phi_{\rm A}\big|_{\rm nr} = \frac{\alpha}{\alpha_{\rm eq}}\biggr|_{\rm nr} = \frac{\phi_{\rm A}^2}{\chi}\biggr|_{\rm nr} \frac{\sqrt{1+8\tilde{n}_{\rm A}/K_{\rm eq}/(\phi_{\rm A}^2/\chi)|_{\rm nr}}-1}{\sqrt{1+8\tilde{n}_{\rm A}/K_{\rm eq}}-1}.
    \label{eqn:alphanorm}
\end{equation}

Practically, to use this analysis, one first selects a threshold value of $\delta$ to define the $k_{\rm d, nr}$ limit. Then, the corresponding ratio $(\phi_A^2/\chi)|_{\rm nr}$ is evaluated using Eq.~\eqref{eqn:phiA2chi_nr}. Finally, Eq.~\eqref{eqn:alphanorm} is used to give the fraction of dissociation that occurs in the non-recombining limit. Importantly, these results are only a function of $T_{\rm t}$ and $\tilde{n}_{\rm A}$, and hence it is not necessary to solve the full system of master equations to know how these variables will impact the transition of $k_{\rm d}$ from the non-recombining to the thermal limit.

\subsubsection{Solution for $\bm e_{\bm rv}$}
\label{sec:QSServ}

Since Eq.~\eqref{eqn:phi3} gives a complete description of the ${\rm A_2}(J,\nu)$ distribution in the QSS regime, the corresponding evolution of the average rovibrational energy of $\rm A_2$ can be computed explicitly as well. In general, the average rovibrational energy of $\rm A_2$ is given by
\begin{equation}
    e_{\rm rv} = k_{\rm B} \sum_{\nu=0}^{\nu_{\rm max}} \sum_{J=0}^{J_{\rm max}(\nu)} \frac{n_{\rm A_2 \it (J,\nu)}}{n_{\rm A_2}} \theta_{\rm A_2 \it(J,\nu)}
    = k_{\rm B} \sum_{\nu=0}^{\nu_{\rm max}} \sum_{J=0}^{J_{\rm max}(\nu)} \psi_{\rm A_2 {\it (J,\nu)}} \frac{Q_{\rm A_2 \it (J,\nu)}}{Q_{\rm A_2}} \theta_{\rm A_2 \it(J,\nu)},
    \label{eqn:erv}
\end{equation}
where $\theta_{{\rm A_2}(J,\nu)}$ is the characteristic rovibrational temperature of ${\rm A_2}(J,\nu)$. Using Eq.~\eqref{eqn:phi3}, $e_{\rm rv}$ in the QSS regime can then be expressed as
\begin{equation}
    e_{\rm rv} = e_{\rm rv,nr} \left(1-\frac{\phi_{\rm A}^2}{\chi}\right) + e_{\rm rv,th} \frac{\phi_{\rm A}^2}{\chi},
    \label{eqn:erv_QSS}
\end{equation}
where $e_{\rm rv}$ in the non-recombining and thermal limits has been defined as
\begin{equation}
    e_{\rm rv,nr} \equiv k_{\rm B} \sum_{\nu=0}^{\nu_{\rm max}} \sum_{J=0}^{J_{\rm max}(\nu)} \psi_{\rm A_2 {\it (J,\nu)},nr} \frac{Q_{\rm A_2 \it (J,\nu)}}{Q_{\rm A_2}} \theta_{\rm A_2 \it(J,\nu)}
    \label{eqn:erv_nr}
\end{equation}
and
\begin{equation}
    e_{\rm rv,th} \equiv k_{\rm B} \sum_{\nu=0}^{\nu_{\rm max}} \sum_{J=0}^{J_{\rm max}(\nu)} \frac{Q_{\rm A_2 \it (J,\nu)}}{Q_{\rm A_2}} \theta_{\rm A_2 \it(J,\nu)},
    \label{eqn:erv_th}
\end{equation}
respectively. As with $k_{\rm d,nr}$ and $k_{\rm d,th}$, Eq.~\eqref{eqn:erv_nr} and \eqref{eqn:erv_th} give that $e_{\rm rv,nr}$ and $e_{\rm rv,th}$ are only functions of $T_{\rm t}$, and hence $e_{\rm rv}$ in the QSS regime is functionalized by $T_{\rm t}$ and $\phi_{\rm A}^2/\chi$ alone.

Unlike $k_{\rm d,nr}$, $e_{\rm rv,nr}$ is not a commonly reported quantity in the literature. Fortunately, it is still possible to obtain estimates of $e_{\rm rv,nr}$ if energy-equivalent rotational and vibrational temperatures, $T_{\rm r}$ and $T_{\rm v}$, are reported. In a vibration-preferred framework where $\theta_{\rm r, A_2 \it(J,\nu)}\equiv \theta_{\rm A_2 \it(J,\nu)}-\theta_{\rm A_2 \it(J={\rm 0},\nu)}$ and $\theta_{\rm v, A_2 \it(J,\nu)}\equiv \theta_{\rm A_2 \it(J={\rm 0},\nu)}$, $T_{\rm r}$ and $T_{\rm v}$ are defined such that the average rotational and vibrational energies, $e_{\rm r}$ and $e_{\rm v}$, are given by
\begin{equation}
    e_{\rm r} \equiv k_{\rm B} \sum_{\nu=0}^{\nu_{\rm max}} \sum_{J=0}^{J_{\rm max}(\nu)} \frac{n_{\rm A_2 \it (J,\nu)}}{n_{\rm A_2}} \theta_{\rm r, A_2 \it(J,\nu)}
    \equiv k_{\rm B} \sum_{\nu=0}^{\nu_{\rm max}} \sum_{J=0}^{J_{\rm max}(\nu)} \frac{Q_{\rm A_2 \it (J,\nu)}(T_{\rm r},T_{\rm v})}{Q_{\rm A_2}(T_{\rm r},T_{\rm v})} \theta_{\rm r, A_2 \it(J,\nu)} 
    \label{eqn:er}
\end{equation}
and
\begin{equation}
    e_{\rm v} \equiv k_{\rm B} \sum_{\nu=0}^{\nu_{\rm max}} \sum_{J=0}^{J_{\rm max}(\nu)} \frac{n_{\rm A_2 \it (J,\nu)}}{n_{\rm A_2}} \theta_{\rm v, A_2 \it(J,\nu)}
    \equiv k_{\rm B} \sum_{\nu=0}^{\nu_{\rm max}} \sum_{J=0}^{J_{\rm max}(\nu)} \frac{Q_{\rm A_2 \it (J,\nu)}(T_{\rm r},T_{\rm v})}{Q_{\rm A_2}(T_{\rm r},T_{\rm v})} \theta_{\rm v, A_2 \it(J,\nu)},
    \label{eqn:ev}
\end{equation}
where
\begin{equation}
    e_{\rm rv} = e_{\rm r} + e_{\rm v}
    \label{eqn:erv_sum}
\end{equation}
by construction. Therefore, as long as estimates of $T_{\rm r,nr}(T_{\rm t})$ and $T_{\rm v,nr}(T_{\rm t})$ can be obtained, $e_{\rm rv,nr}$ can be computed equivalently as
\begin{equation}
    e_{\rm rv,nr} = k_{\rm B} \sum_{\nu=0}^{\nu_{\rm max}} \sum_{J=0}^{J_{\rm max}(\nu)} \frac{Q_{\rm A_2 \it (J,\nu)}(T_{\rm r,nr},T_{\rm v,nr})}{Q_{\rm A_2}(T_{\rm r,nr},T_{\rm v,nr})} \theta_{\rm rv, A_2 \it(J,\nu)}.
    \label{eqn:erv_nr2}
\end{equation}

\subsection{Pre-QSS Rates}
\label{sec:preQSS}

\subsubsection{Formulation}
The theory of the previous sections provides a complete description of $k_{\rm d}$ and $e_{\rm rv}$ under the QSS assumption up to and including thermal equilibrium. However, this description is of course only valid once the QSS condition of Eq.~\eqref{eqn:masterA2jv_QSS3} is met. The region before QSS is established is referred to as the ``pre-QSS'' region. If it is assumed that recombination is negligible in the pre-QSS region as it occurs before the non-recombining QSS limit, the master equations for ${\rm A_2}(J,\nu)$ from Eq.~\eqref{eqn:masterA2jv} can be simplified to
\begin{eqnarray}
    \frac{d n_{\rm A_2 \it(J,\nu)}}{dt} = && n_{\rm M}\sum_{\nu'=0}^{\nu_{\rm max}} \sum_{J'=0}^{J_{\rm max}(\nu')} n_{\rm A_2 \it (J',\nu')} k(J',\nu' \rightarrow J,\nu) - n_{\rm M}\sum_{\nu'=0}^{\nu_{\rm max}} \sum_{J'=0}^{J_{\rm max}(\nu')} n_{\rm A_2 \it (J,\nu)} k(J,\nu \rightarrow J',\nu') \nonumber \\
    && - n_{\rm M}n_{\rm A_2 \it(J,\nu)} k(J,\nu \rightarrow c).
\label{eqn:masterA2jv_preQSS1}
\end{eqnarray}
The inherent time-dependence and coupled nature of Eq.~\eqref{eqn:masterA2jv_preQSS1} implies that it is only possible to obtain an exact solution in the pre-QSS by solving Eq.~\eqref{eqn:masterA2jv_preQSS1} for all states of ${\rm A_2}(J,\nu)$ simultaneously. However, under several simplifying assumptions, useful insights into the pre-QSS behavior can still be obtained.

The isothermal and isochoric assumptions are first made such that all equilibrium number densities can be treated as constants. Under this assumption, Eq.~\eqref{eqn:masterA2jv_preQSS1} can be rewritten in terms of $\phi_{\rm A_2(J,\nu)}$ as
\begin{eqnarray}
    \frac{d \phi_{\rm A_2 \it(J,\nu)}}{dt} = &&n_{\rm M} \sum_{\nu'=0}^{\nu_{\rm max}} \sum_{J'=0}^{J_{\rm max}(\nu')} \phi_{\rm A_2 \it(J',\nu')} k(J,\nu \rightarrow J',\nu') \nonumber \\
    && - n_{\rm M} \phi_{\rm A_2 \it(J,\nu)} \left[ \sum_{\nu'=0}^{\nu_{\rm max}} \sum_{J'=0}^{J_{\rm max}(\nu')} k(J,\nu \rightarrow J',\nu') + k(J,\nu \rightarrow c) \right],
\label{eqn:masterA2jv_preQSS2}
\end{eqnarray}
or equivalently in matrix form as
\begin{equation}
    \frac{d \vec{\phi}_{\rm A_2}}{dt} = -n_{\rm M}\bm{\tilde{M}}\vec{\phi}_{\rm A_2},
    \label{eqn:A2mat_preQSS}
\end{equation}
where the entries of $\bm{\tilde{M}}$ correspond to the state-specific rate constant terms in the right-hand side of Eq.~\eqref{eqn:masterA2jv_preQSS2}. This definition of $\bm{\tilde{M}}$ is nearly identical to the definition of $\bm{M}$ used previously in the QSS formulation of Eq.~\eqref{eqn:A2mat}, except that the first row of $\bm{\tilde{M}}$ does not correspond to the constraint equation of Eq.~\eqref{eqn:A2chi_Q} as it does for $\bm{M}$. Next, the rate of change of $\chi$ is expressed in terms of $\vec{\phi}_{\rm A_2}$ by taking the time derivative of Eq.~\eqref{eqn:A2chi_Q} and applying Eq.~\eqref{eqn:A2mat_preQSS} to get
\begin{equation}
    \frac{d \chi}{dt} = \sum_{\nu=0}^{\nu_{\rm max}} \sum_{J=0}^{J_{\rm max}(\nu)} \frac{d \phi_{\rm A_2 \it(J,\nu)}}{dt} \frac{Q_{\rm A_2 \it(J,\nu)}}{Q_{\rm A_2}} 
    = -{n_{\rm M}}^t\vec{Q}_{\rm A_2}\bm{\tilde{M}}\vec{\phi}_{\rm A_2}.
    \label{eqn:dchidt}
\end{equation}
Here, $^t\vec{Q}_{\rm A_2}$ corresponds to the transpose of the vector with entries given by $Q_{\rm A_2 \it(J,\nu)}/Q_{\rm A_2}$, and Eq.~\eqref{eqn:A2chi_Q} gives that $^t\vec{Q}_{\rm A_2}\vec{\psi}_{\rm A_2}=1$. These results can be combined to express the time rate of change of $\vec{\psi}_{\rm A_2}$ as
\begin{equation}
    \frac{d \vec{\psi}_{\rm A_2}}{dt} = \chi^{-1}\frac{d \vec{\phi}_{\rm A_2}}{dt} + \vec{\phi}_{\rm A_2}\frac{d \chi^{-1}}{dt} = -n_{\rm M} \left[ \tilde{M} \vec{\psi}_{\rm A_2} - \left( ^t\vec{Q}_{\rm A_2} \bm{\tilde{M}}\vec{\psi}_{\rm A_2} \right) \vec{\psi}_{\rm A_2} \right].
    \label{eqn:dpsidt}
\end{equation}
It can be shown that all eigenvalues of $\bm{\tilde{M}}$ are negative, and hence as $t\rightarrow\infty$, the solution to Eq.~\eqref{eqn:dpsidt} converges to a steady state given by
\begin{equation}
    \tilde{M} \vec{\psi}_{\rm A_2, \infty} = \left( ^t\vec{Q}_{\rm A_2} \bm{\tilde{M}}\vec{\psi}_{\rm A_2,\infty} \right) \vec{\psi}_{\rm A_2,\infty}.
    \label{eqn:preQSSsteady}
\end{equation}
In other words, $\vec{\psi}_{\rm A_2,\infty}$ is an eigenvector of $\bm{\tilde{M}}$ with the corresponding eigenvalue
\begin{equation}
    \lambda_0 =\ ^t\vec{Q}_{\rm A_2} \bm{\tilde{M}}\vec{\psi}_{\rm A_2,\infty}.
    \label{eqn:lambda0}
\end{equation}
A linear stability analysis of Eq.~\eqref{eqn:A2mat_preQSS} would show that $\lambda_0$ is the smallest eigenvalue and hence corresponds to the slowest decaying mode of $\bm{\tilde{M}}$. 

\subsubsection{Approximate Solution for $\bm A_2(J,\nu)$ and $\bm k_{\bm d}$}
Until this point in the analysis of the pre-QSS, only the isothermal and isochoric assumptions have been made. To compute a solution for the evolution of $A_2(J,\nu)$ without numerically integrating the master equations, three additional approximations are made here. Firstly, the nonlinear term of Eq.~\eqref{eqn:dpsidt} is approximated using Eq.~\eqref{eqn:lambda0} as 
\begin{equation}
    \left( ^t\vec{Q}_{\rm A_2} \bm{\tilde{M}} \vec{\psi}_{\rm A_2} \right) \vec{\psi}_{\rm A_2} \approx
    \lambda_0 \vec{\psi}_{\rm A_2}.
    \label{eqn:lambdalinear}
\end{equation}
This is equivalent to assuming that all but the slowest modes of $\bm{\tilde{M}}$ have decayed. Secondly, it is assumed that $\vec{\psi}_{\rm A_2,\infty}\approx \vec{\psi}_{\rm A_2,nr}$. $\vec{\psi}_{\rm A_2,\infty}$ represents the steady solution to the pre-QSS master equations (Eq.~\eqref{eqn:masterA2jv_preQSS2}), whereas $\vec{\psi}_{\rm A_2,nr}$ is the solution to the QSS master equations (Eq.~\eqref{eqn:masterA2jv_QSS4}) in the absence of recombination. Due to the fact that $\bm{M}$ and $\bm{\tilde{M}}$ are not exactly equal, $\vec{\psi}_{\rm A_2,\infty}$ and $ \vec{\psi}_{\rm A_2,nr}$ are not formally equivalent. For practical applications however, they are the same. Thirdly, the third-body M is treated as inert such that $n_{\rm M}$ is constant. 

By multiplying Eq.~\eqref{eqn:preQSSsteady} by $n_{\rm M}$ and adding it to Eq.~\eqref{eqn:dpsidt}, then applying Eq.~\eqref{eqn:lambdalinear}, the following expression is obtained
\begin{equation}
    \frac{d \vec{\psi}_{\rm A_2}}{dt} = -n_{\rm M} \left(\bm{\tilde{M}} -\lambda_0 \bm{I} \right)\left( \vec{\psi}_{\rm A_2} - \vec{\psi}_{\rm A_2,\infty} \right).
\end{equation}
Using the fact that $d\vec{\psi}_{\rm A_2,nr}/dt = 0$ and applying the approximation that $\vec{\psi}_{\rm A_2,\infty}\approx \vec{\psi}_{\rm A_2,nr}$ then gives
\begin{equation}
    \frac{d}{dt}\left( \vec{\psi}_{\rm A_2} - \vec{\psi}_{\rm A_2,nr} \right) =-n_{\rm M} \left(\bm{\tilde{M}} -\lambda_0 \bm{I} \right)\left( \vec{\psi}_{\rm A_2} - \vec{\psi}_{\rm A_2,nr} \right),
    \label{eqn:dpsidt_2}
\end{equation}
where the time-dependent solution is given by
\begin{equation}
    \vec{\psi}_{\rm A_2} = \vec{\psi}_{\rm A_2,nr} - \exp\left[-n_{\rm M}(\bm{\tilde{M}} -\lambda_0 \bm{I})t\right] \left( \vec{\psi}_{\rm A_2,nr} - \vec{\psi}_{\rm A_2,0} \right).
    \label{eqn:psisol}
\end{equation}
Under the discussed assumptions, the pre-QSS internal state distribution of $\rm A_2$ is only a function of $n_{\rm M}$, $T_{\rm t}$, and $t$. In particular, the pre-QSS solution has a simple exponential dependence on both $n_{\rm M}$ and $t$. Mathematically, the convergence rate of this solution is controlled by the smallest eigenvalue of $(\bm{\tilde{M}} -\lambda_0 \bm{I})$~\cite{Haug1987}. Hence, this matrix can be approximated by $(\lambda_1-\lambda_0){\bm I}$, where $\lambda_1$ is the second smallest eigenvalue of $\bm{\tilde{M}}$. As will be shown in section~\ref{sec:kdextract}, the resulting functional form from this approximation is sufficient for capturing the majority of the pre-QSS behavior for $\rm H_2$ dissociation. Then, Eq.~\eqref{eqn:psisol} is substituted into the general expression for $k_{\rm d}$ given by Eq.~\eqref{eqn:A2kd_sum2} to yield
\begin{eqnarray}
    k_{\rm d,pre-QSS} &&= k_{\rm d,nr} - \exp(-n_{\rm M}(\lambda_1-\lambda_0) t) (k_{\rm d,nr} - k_{\rm d,0}) \nonumber \\
    && = k_{\rm d,nr}[1 - (1-\epsilon) \exp(-n_{\rm M}(\lambda_1-\lambda_0) t)].
    \label{eqn:kdpreQSS}
\end{eqnarray}
Here, $k_{\rm d,0}$ corresponds to the dissociation rate constant evaluated at the post-shock $T_{\rm t}$ but weighted by the pre-shock Boltzmann distribution, and $\epsilon \equiv k_{\rm d,0}/k_{\rm d,nr}$. For any practical pre-shock temperatures (e.g., 100 - 300 K), $k_{\rm d,0}$ will be magnitudes smaller than $k_{\rm d,nr}$, such that $\epsilon \ll 1$.

While Eq.~\eqref{eqn:kdpreQSS} gives a compact expression for $k_{\rm d,pre-QSS}$, its dependence on $t$ makes it inconvenient for practical use. To remedy this, an alternate formulation is proposed where $t$ is replaced with the fraction of dissociation, $\phi_{\rm A}$. This is done by considering the chemical source term for A in the pre-QSS,
\begin{equation}
    \frac{d n_{\rm A}}{dt} = 2 n_{\rm A_2} n_{\rm M} k_{\rm d,pre-QSS}.
    \label{eqn:dnAdt_preQSS}
\end{equation}
Equation~\eqref{eqn:dnAdt_preQSS} is then rewritten using Eq.~\eqref{eqn:alpha_n} and Eq.~\eqref{eqn:kdpreQSS} as
\begin{equation}
    \frac{d \alpha}{dt} = (1-\alpha) n_{\rm M} k_{\rm d,nr} [1 - (1-\epsilon) \exp(-n_{\rm M}(\lambda_1-\lambda_0) t)],
    \label{eqn:dalphadt_preQSS}
\end{equation}
where $\alpha$ is the degree of dissociation as defined previously in Eq.~\eqref{eqn:alpha}. Next, using a first-order Taylor expansion to approximate $\exp(-n_{\rm M}(\lambda_1-\lambda_0) t) \approx 1 - n_{\rm M}(\lambda_1-\lambda_0) t$, the solution to Eq.~\eqref{eqn:dalphadt_preQSS} (in the limit of a negligible $\epsilon$) is given by
\begin{equation}
    t = \frac{1}{n_{\rm M}}\sqrt{-\frac{2 \ln(1-\alpha)}{k_{\rm d,nr} (\lambda_1-\lambda_0)}}.
    \label{eqn:talpha}
\end{equation}
Substituting this result back into Eq.~\eqref{eqn:kdpreQSS} gives
\begin{equation}
    k_{\rm d,pre-QSS} = k_{\rm d,nr}\left[1 - (1-\epsilon) \exp\left( -\sqrt{-\frac{2 (\lambda_1-\lambda_0) \ln(1-\alpha)}{k_{\rm d,nr}}} \right) \right].
    \label{eqn:kdpreQSS3}
\end{equation}
Finally, because the equilibrium degree of dissociation, $\alpha_{\rm eq}$, is close to 1 (i.e., complete dissociation) for the majority of the conditions where pre-QSS effects would be relevant, $\alpha/\alpha_{\rm eq}=\phi_{\rm A}$ is approximated as $\alpha \approx \phi_{\rm A}$. Performing this substitution in Eq.~\eqref{eqn:kdpreQSS3} yields the final expression for the pre-QSS dissociation rate constant,
\begin{equation}
    k_{\rm d,pre-QSS} = k_{\rm d,nr}\left[1 - (1-\epsilon) \exp\left( -\sqrt{-\frac{\ln(1-\phi_{\rm A})}{\eta}} \right) \right],
    \label{eqn:kdpreQSS4}
\end{equation}
where $\eta (T_{\rm t}) \equiv k_{\rm d,nr}/(2(\lambda_1-\lambda_0))$. For most uses, the $\epsilon$ in Eq.~\eqref{eqn:kdpreQSS4} can be neglected. However, when Eq.~\eqref{eqn:kdpreQSS4} is used in a numerical integration (as in a CFD calculation), a non-zero value of $\epsilon$ is required such that $k_{\rm d,pre-QSS}$ is not 0 at $\phi_{\rm A} = 0$. Practically, for the case of $\rm H_2$ dissociation discussed later in section~\ref{sec:prednum}, it was found that there is no sensitivity to the value of $\epsilon$ below the threshold of $\epsilon \leq 10^{-3}$.

A key implication of these results is that under the discussed assumptions, $k_{\rm d,pre-QSS}$ can be expressed simply as $k_{\rm d,nr}$ modified by a multiplicative non-equilibrium factor that is only a function of $T_{\rm t}$ and $\phi_{\rm A}$. Importantly, the dependence of $k_{\rm d,pre-QSS}$ on $n_{\rm M}$ as seen in Eq.~\eqref{eqn:kdpreQSS} drops out when $t$ is substituted by $\phi_{\rm A}$. This means that as long as an accurate fit of $\eta (T_{\rm t})$ can be obtained, the fraction of dissociation that occurs in the pre-QSS can be estimated easily as just a function of $T_{\rm t}$. Analogously to the QSS analysis presented in section~\ref{sec:kdnrth_trans}, the fraction of dissociation that occurs in the pre-QSS region can be defined as the values of $\phi_{\rm A}=\alpha/\alpha_{\rm eq}$ for which $k_{\rm d,pre-QSS}$ is within some $\delta$ fraction of $k_{\rm d,nr}$, i.e., $k_{\rm d,pre-QSS}=k_{\rm d,nr}(1-\delta)$. Substituting this constraint into Eq.~\eqref{eqn:kdpreQSS4} with $\epsilon=0$ gives the desired result,
\begin{equation}
    \phi_{\rm A} \big|_{\rm pre-QSS} = \frac{\alpha}{\alpha_{\rm eq}} \biggr|_{\rm pre-QSS} = 1 - \exp ( -\eta (\ln(\delta))^2 ).
    \label{eqn:fracpreQSS}
\end{equation}

\subsubsection{Chemical Source Term}
\label{sec:preQSSsrc}

There are two key implications of the analysis presented above for the calculation of the chemical source term. First, since Eq.~\eqref{eqn:kdpreQSS4} gives that $k_{\rm d,pre-QSS}$ is simply $k_{\rm d,nr}$ multiplied by a correction factor, it is easy to incorporate the expression for $k_{\rm d,pre-QSS}$ into the previous QSS rate constant expression. Namely, by replacing $k_{\rm d,nr}$ in Eq.~\eqref{eqn:kdnrth} with $k_{\rm d,pre-QSS}$, a complete description of $k_{\rm d}$ that is valid from the pre-QSS to the thermal limit can be obtained. This result is given by
\begin{equation}
    k_{\rm d} = k_{\rm d,nr} \left[1 - (1-\epsilon)\exp\left( -\sqrt{-\frac{\ln(1-\phi_{\rm A})}{\eta}} \right) \right] \left( 1 - \frac{\phi_{\rm A}^2}{\chi} \right) + k_{\rm d,th} \frac{\phi_{\rm A}^2}{\chi}.
    \label{eqn:kdprenrth}
\end{equation}
As expected, in the limit of no dissociation when $\phi_{\rm A}=\chi=0$, this expression gives that $k_{\rm d}=\epsilon k_{\rm d,nr} = k_{\rm d,0}$, and in the limit of equilibrium when $\phi_{\rm A}=\chi=1$, this expression gives that $k_{\rm d} = k_{\rm d,th}$.

Second, by combining this description of $k_{\rm d}$ with the formula for $k_{\rm r}$ from Eq.~\eqref{eqn:A2DEB}, the general source terms for $\rm A_2$ and A can be computed directly from Eq.~\eqref{eqn:dnA2} and \eqref{eqn:dnA}. Equivalently, by taking advantage of the same source term equivalence as discussed for Eq.~\eqref{eqn:dnA_2} and \eqref{eqn:dnA_3}, the chemical source term can be computed practically as
\begin{equation}
    \frac{dn_{\rm A}}{dt} = 2n_{\rm A_2}n_{\rm M}k_{\rm d,pre-QSS} - 2 n_{\rm A}^2n_{\rm M}\frac{k_{\rm d,pre-QSS}}{K_{\rm eq}}.
    \label{eqn:dnA_preQSS}
\end{equation}
This gives a complete description of the non-equilibrium chemical source term that is only a function of the translational temperature, $T_{\rm t}$, and the fraction of dissociation, $\phi_{\rm A}$, via the expression for $k_{\rm d,pre-QSS}$ from Eq.~\eqref{eqn:kdpreQSS4}.

\subsubsection{Approximate Solution for $\bm e_{\bm rv}$}
\label{sec:preQSServ}

Analogously to the QSS case, the evolution of the average rovibrational energy, $e_{\rm rv}$, can also be computed from the approximate solution of the $\rm A_2 \it (J,\nu)$ distribution in the pre-QSS regime. Using the same substitutions as Eq.~\eqref{eqn:kdpreQSS4} for $\bm{\tilde{M}}$ and $t$, the rovibrational distribution of $\rm A_2$ can be computed from Eq.~\eqref{eqn:psisol} as
\begin{equation}
    \vec{\psi}_{\rm A_2} = \vec{\psi}_{\rm A_2,nr} - \exp\left(- \sqrt{-\frac{\ln(1-\phi_{\rm A})}{\eta}} \right) \left( \vec{\psi}_{\rm A_2,nr} - \vec{\psi}_{\rm A_2,0} \right).
    \label{eqn:psisol2}
\end{equation}
Substituting this result into Eq.~\eqref{eqn:erv} and using the definition of $e_{\rm rv,nr}$ from Eq.~\eqref{eqn:erv_nr} then gives
\begin{equation}
    e_{\rm rv,pre-QSS} = e_{\rm rv,nr}\left[1 - \left(1-\frac{e_{\rm rv,th,0}}{e_{\rm rv,nr}}\right) \exp\left( -\sqrt{-\frac{\ln(1-\phi_{\rm A})}{\eta}} \right) \right],
    \label{eqn:erv_preQSS}
\end{equation}
where $e_{\rm rv,th,0}$ corresponds to $e_{\rm rv,th}$ evaluated via Eq.~\eqref{eqn:erv_th} at the pre-shock temperature. Similarly to the discussion for $k_{\rm d,pre-QSS}$, this result gives that $e_{\rm rv,pre-QSS}$ is equal to $e_{\rm rv,nr}$ multiplied by a pre-QSS correction factor. Therefore, by applying this correction factor in the previous QSS description of $e_{\rm rv}$ from Eq.~\eqref{eqn:erv_QSS}, the corresponding complete description of $e_{\rm rv}$ from the pre-QSS to the thermal limit is given by
\begin{equation}
    e_{\rm rv} = e_{\rm rv,nr} \left[1 - \left(1-\frac{e_{\rm rv,th,0}}{e_{\rm rv,nr}}\right) \exp\left( -\sqrt{-\frac{\ln(1-\phi_{\rm A})}{\eta}} \right) \right]  \left(1-\frac{\phi_{\rm A}^2}{\chi}\right) + e_{\rm rv,th} \frac{\phi_{\rm A}^2}{\chi}.
    \label{eqn:erv_preQSS2}
\end{equation}

\section{Application to the Dissociation of $\rm \bf H_2$}
\label{sec:applicationH2}

In this section, the QSS and pre-QSS theories described in section~\ref{sec:noneq} are applied to the case of $\rm H_2$ dissociation. In section~\ref{sec:kdextract}, the rate expressions are first validated against master equation results from the literature. Then, in section~\ref{sec:fracdiss}, the fraction of dissociation that occurs in the QSS and pre-QSS regimes are estimated as a function of temperature and number density. Finally in section~\ref{sec:prednum}, the QSS and pre-QSS source term expressions are used to compute number density and average rovibrational energy profiles which are compared directly to the master equation results.

\subsection{Extracted Rate Constants}
\label{sec:kdextract}

\begin{figure*}[hbt!]
    \centering
    \begin{subfigure}[b]{0.345\textwidth}
        \centering
        \includegraphics[width=\textwidth,trim={0cm 0cm 1.3cm 0cm},clip]{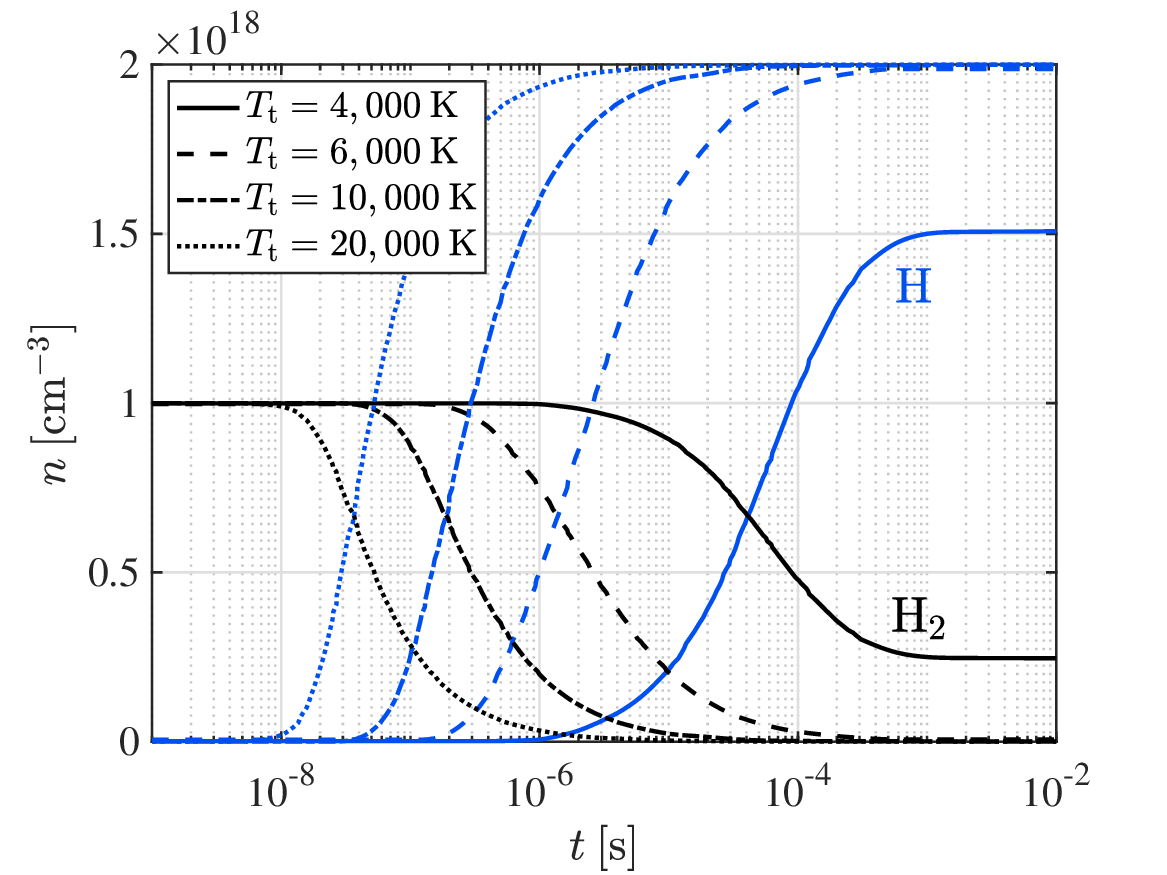}
        \caption{M = $\rm H_2$}
    \end{subfigure}
    \hfill
    \begin{subfigure}[b]{0.32\textwidth}
        \centering
        \includegraphics[width=\textwidth,trim={1.3cm 0cm 1.3cm 0cm},clip]{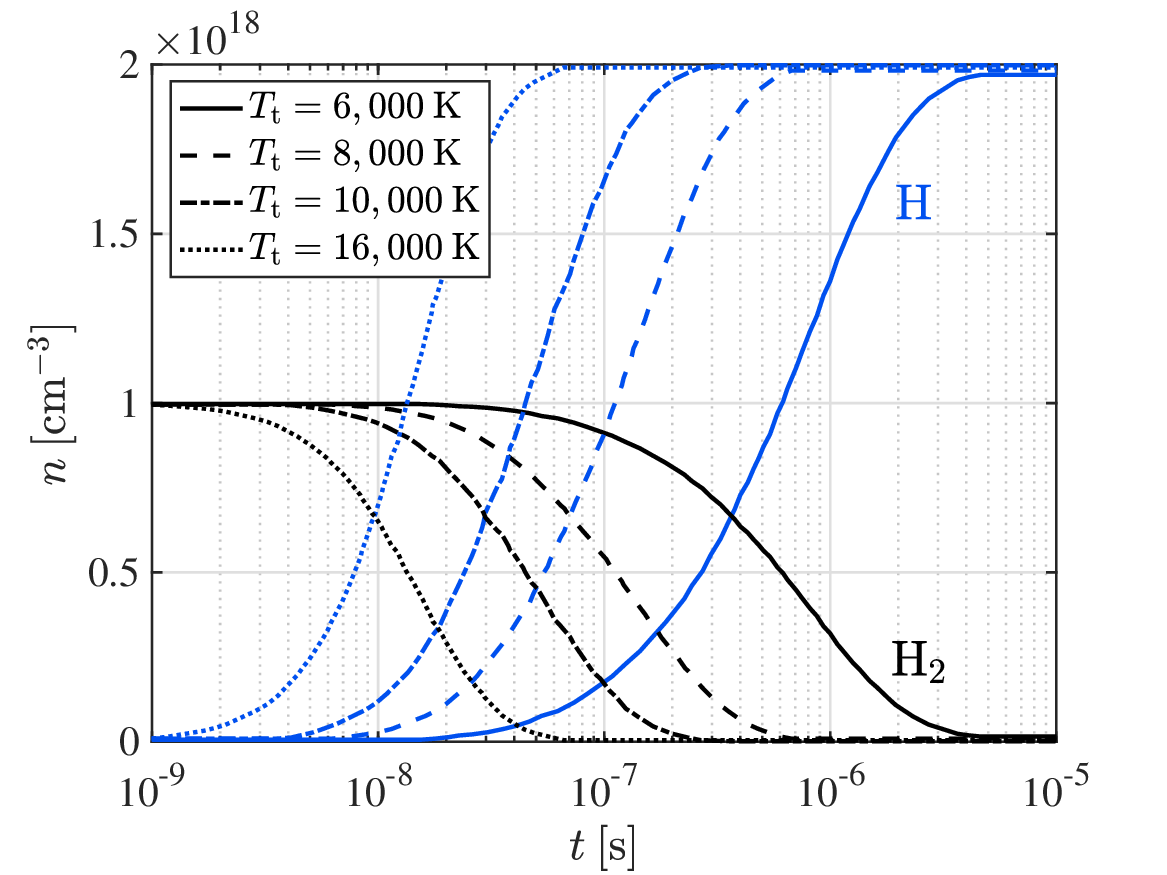}
        \caption{M = H}
    \end{subfigure}
    \hfill
    \begin{subfigure}[b]{0.32\textwidth}
        \centering
        \includegraphics[width=\textwidth,trim={1.3cm 0cm 1.3cm 0cm},clip]{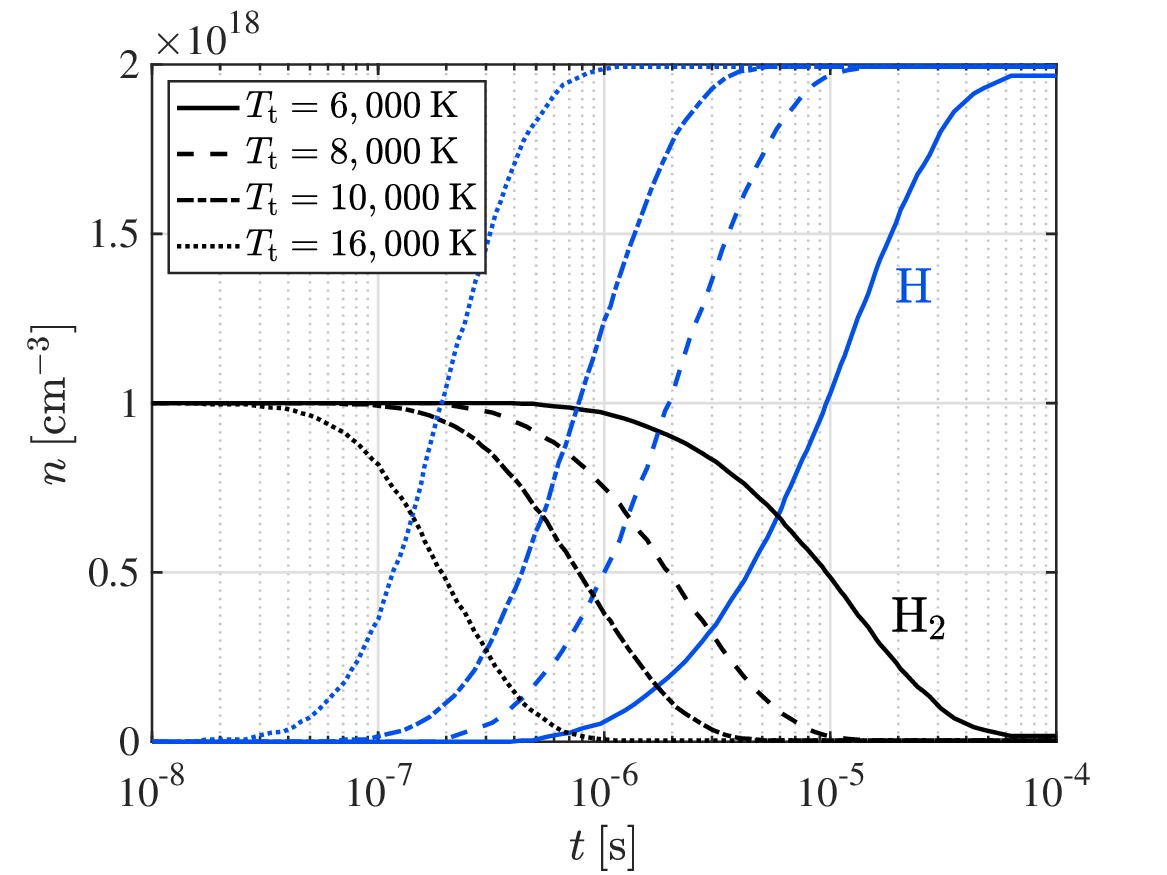}
        \caption{M = He}
    \end{subfigure}
    \caption{Number density profiles for the master equation calculations of Kim and Boyd~\cite{Kim2012} (M = $\rm H_2$) and Kim~\cite{Kim2015} (M = H and He).}
    \label{fig:n_master}
\end{figure*}

\begin{figure*}[hbt!]
    \centering
    \begin{subfigure}[b]{0.345\textwidth}
        \centering
        \includegraphics[width=\textwidth,trim={0cm 0cm 1.3cm 0cm},clip]{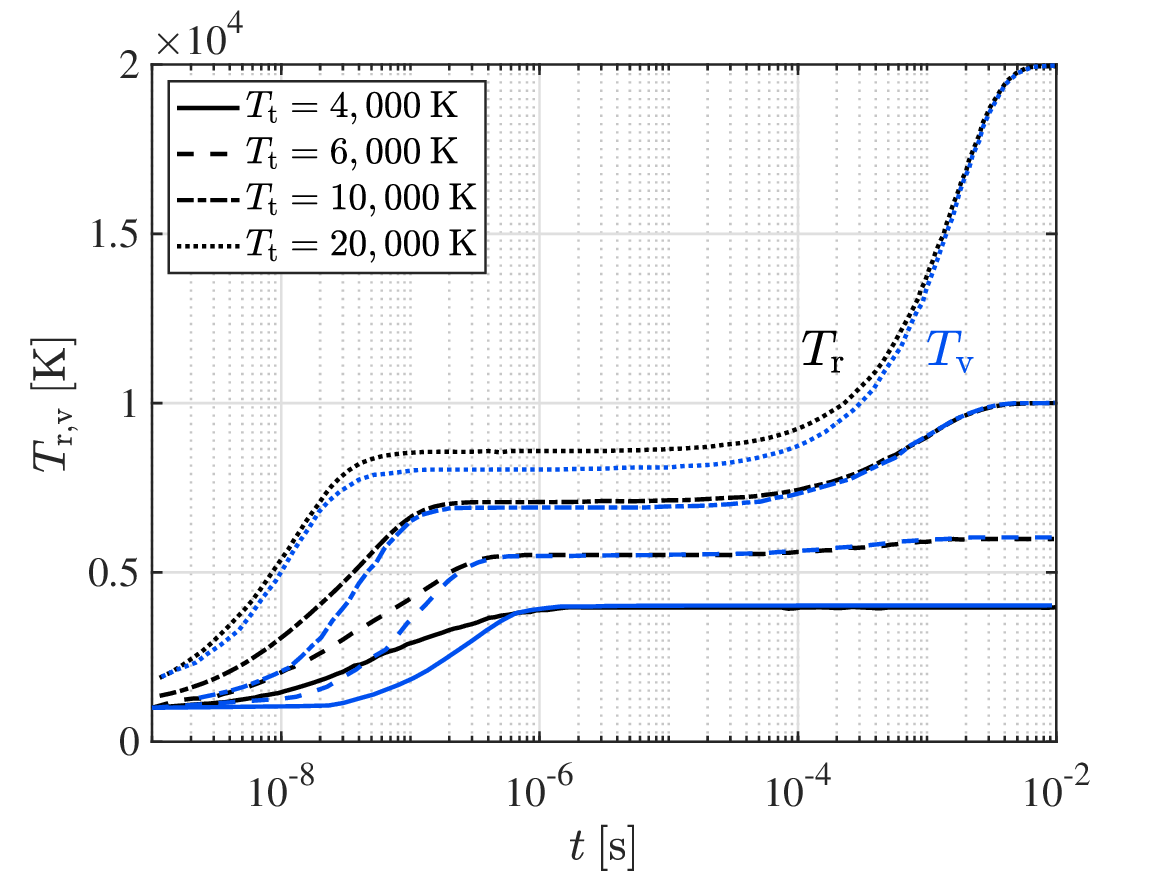}
        \caption{M = $\rm H_2$}
    \end{subfigure}
    \hfill
    \begin{subfigure}[b]{0.32\textwidth}
        \centering
        \includegraphics[width=\textwidth,trim={1.3cm 0cm 1.3cm 0cm},clip]{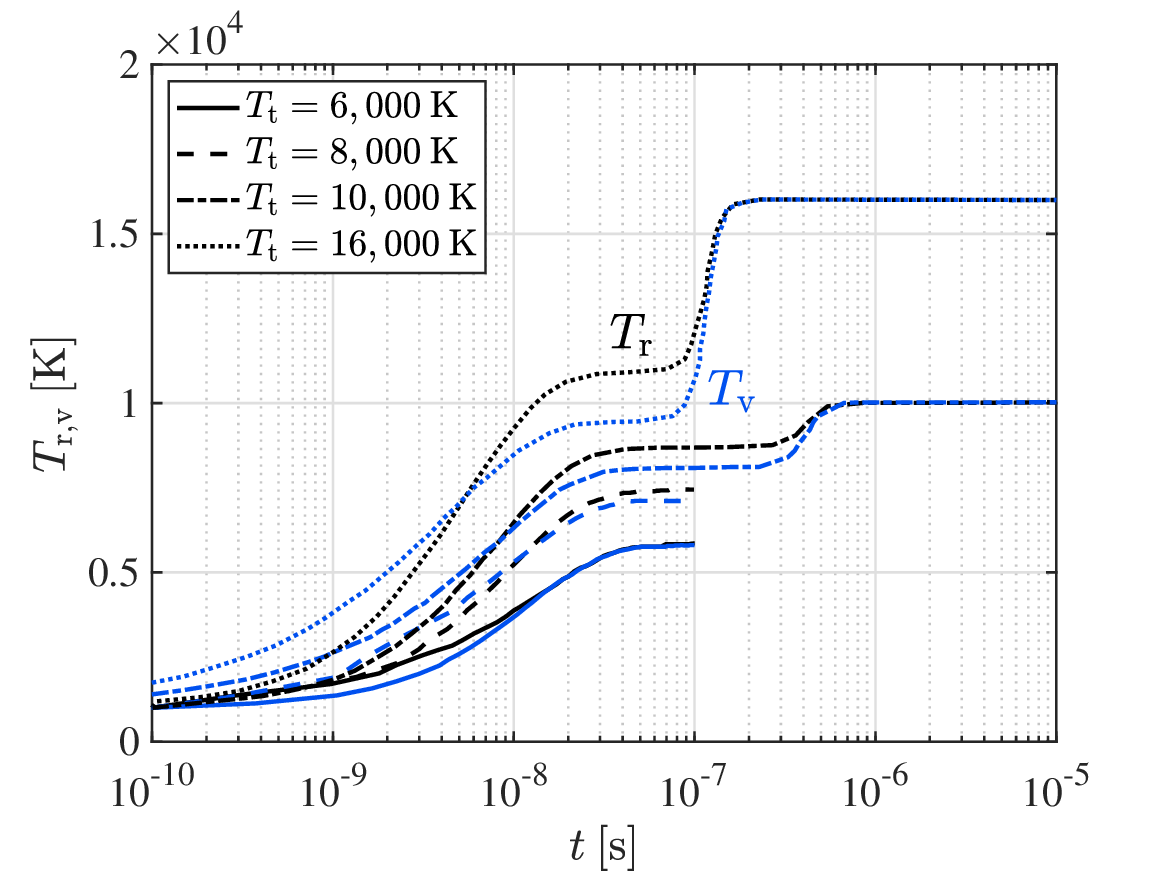}
        \caption{M = H}
    \end{subfigure}
    \hfill
    \begin{subfigure}[b]{0.32\textwidth}
        \centering
        \includegraphics[width=\textwidth,trim={1.3cm 0cm 1.3cm 0cm},clip]{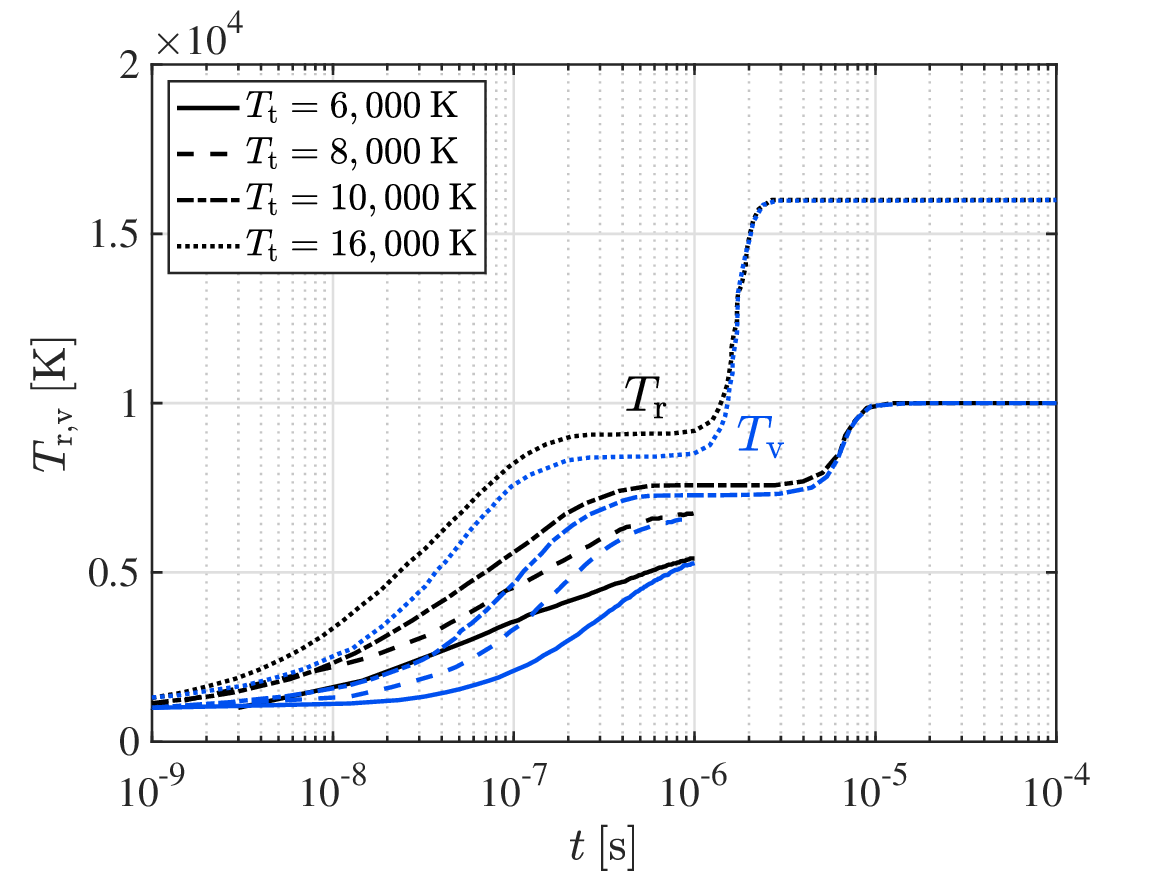}
        \caption{M = He}
    \end{subfigure}
    \caption{Rotational and vibrational temperature profiles for the master equation calculations of Kim and Boyd~\cite{Kim2012} (M = $\rm H_2$), Kim~\cite{Kim2015} (M = H and He, $T_{\rm t}$ = 6,000 K and 8,000 K), and Kim et al.~\cite{Kim2009} (M = H and He, $T_{\rm t}$ = 10,000 K and 16,000 K).}
    \label{fig:Trv_master}
\end{figure*}

To validate the analysis presented in the previous sections, the predictions of the QSS and pre-QSS theories are compared to the results of the master equation calculations for the dissociation of $\rm H_2$ by Kim and Boyd~\cite{Kim2012} (M = $\rm H_2$) and Kim~\cite{Kim2015} (M = H and He). In these computational studies, state-specific rate constants were evaluated using the QCT method. For the case of M = $\rm H_2$ by Kim and Boyd~\cite{Kim2012}, the complete set of state-specific rate constants for both target and projectile $\rm H_2$ molecules were evaluated using the QCT method along with response surfaces designed by the ordinary Kriging model. For each third-body, the full set of master equations was used to simulate a 0-D isothermal and isochoric reactor in a heating environment where $T_{\rm t}$ and $\tilde{n}_{\rm H}$ were held constant. $\rm H_2$ molecules were allowed to dissociate, and the evolution to equilibrium of the initial rovibrational distribution of $\rm H_2$ (characterized by $T_{\rm r,0} = T_{\rm v,0} < T_{\rm t}$) was observed. The resulting number density and internal temperature profiles from these calculations are reproduced in Fig.~\ref{fig:n_master} and \ref{fig:Trv_master}, respectively\footnote{Number densities, rovibrational temperatures, and rate constants for these studies were only reported as plots. Therefore, these values have all been obtained via digitizations.}. For the 10,000 K and 16,000 K cases with M = H and He, rovibrational temperatures are obtained from the earlier study by Kim et al.~\cite{Kim2009}, as temperatures were reported over a wider range of simulation times. For all plotted cases, rovibrational temperatures were initialized at $T_{\rm r,0} = T_{\rm v,0}$ = 1,000 K, and the number density of $\rm H_2$ was initialized at $1.0\times10^{18}$ $\rm cm^{-3}$. For the cases with M = H and He, the number density of the third-body was additionally set to $5.0\times10^{17}$ $\rm cm^{-3}$. For the higher temperature cases with all third-bodies, the results of Fig.~\ref{fig:Trv_master} highlight distinct plateaus in the $T_{\rm r}$ and $T_{\rm v}$ profiles between the initial value at 1,000 K and the final equilibrium values at $T_{\rm t}$. These plateaus correspond to the non-recombining QSS limit, and the regions before and after the plateaus correspond to the pre-QSS and recombining regions, respectively.

\begin{figure*}[hbt!]
    \centering
    \begin{subfigure}[b]{0.34\textwidth}
        \centering
        \includegraphics[width=\textwidth,trim={0cm 1cm 1cm 0.2cm},clip]{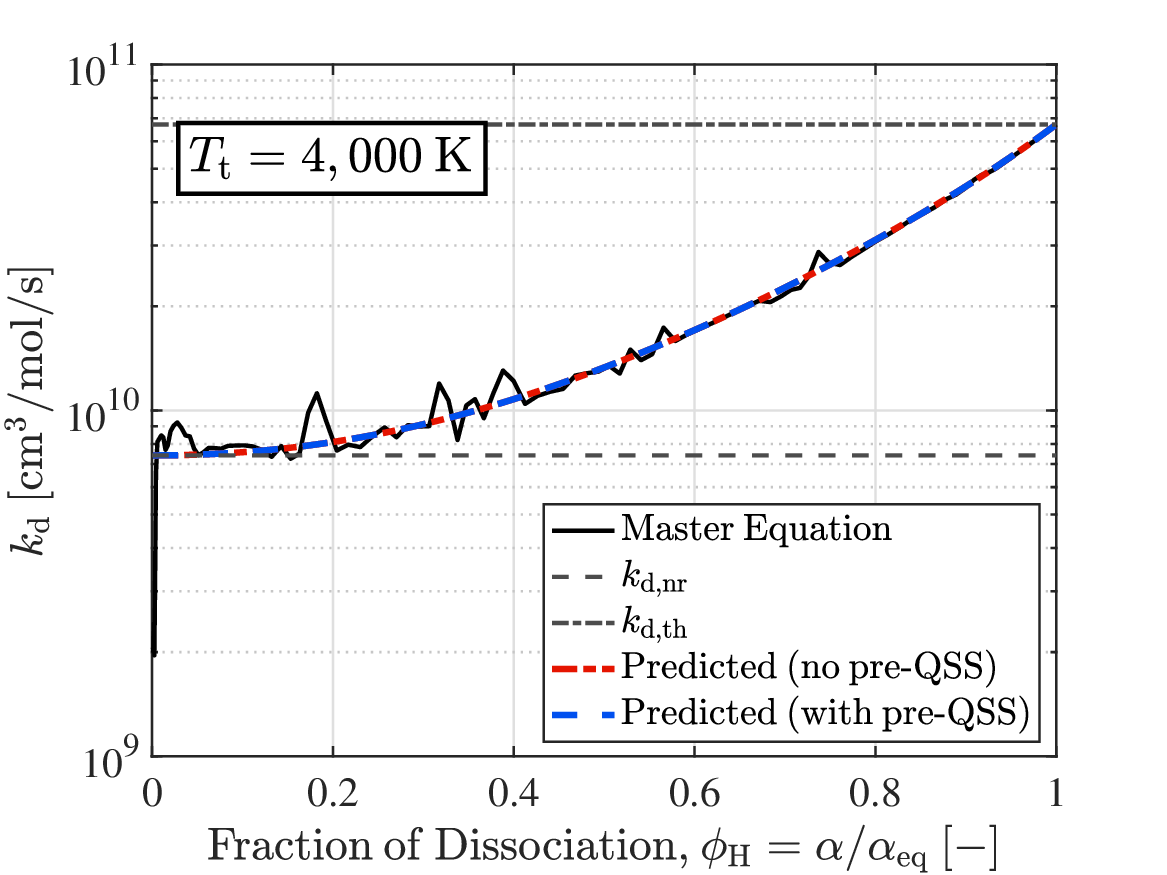}
    \end{subfigure}
    \hfill
    \begin{subfigure}[b]{0.32\textwidth}
        \centering
        \includegraphics[width=\textwidth,trim={1.3cm 1cm 1cm 0.2cm},clip]{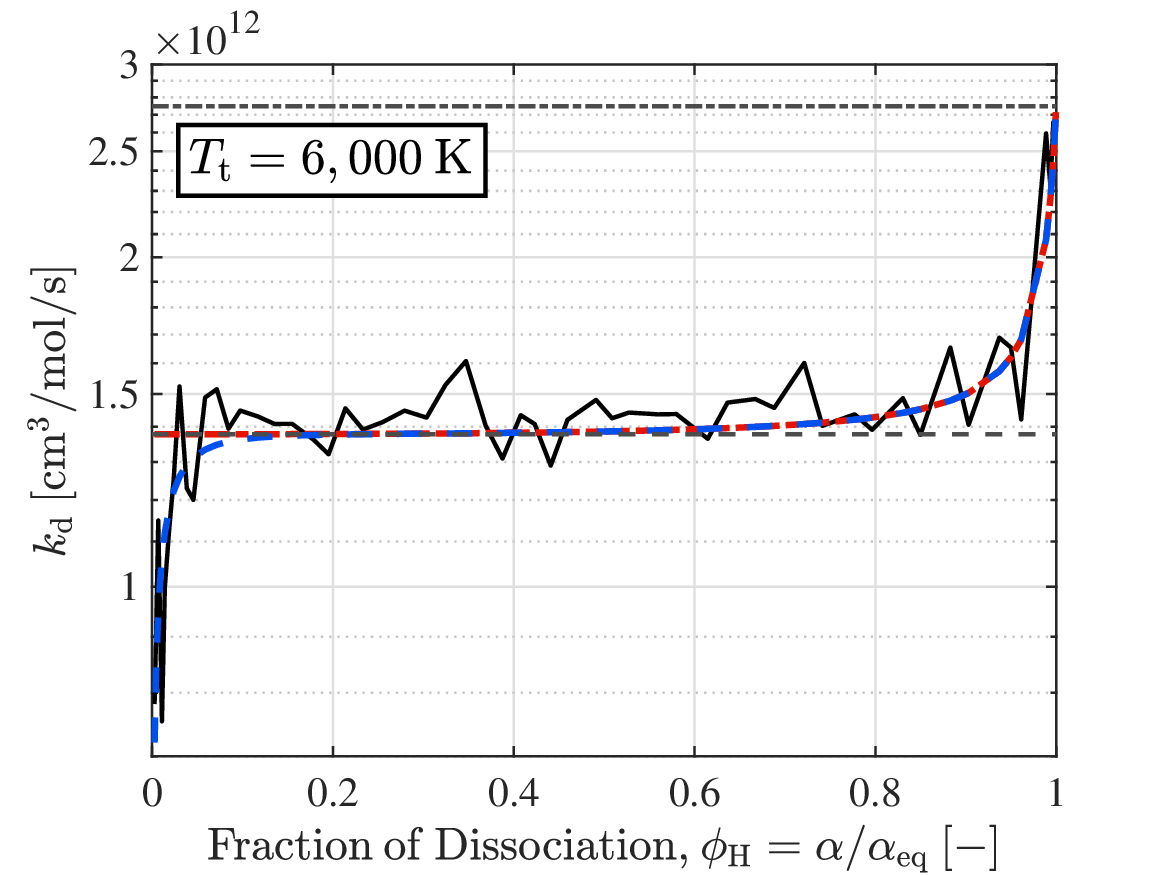}
    \end{subfigure}
    \hfill
    \begin{subfigure}[b]{0.32\textwidth}
        \centering
        \includegraphics[width=\textwidth,trim={1.2cm 1cm 1cm 0.2cm},clip]{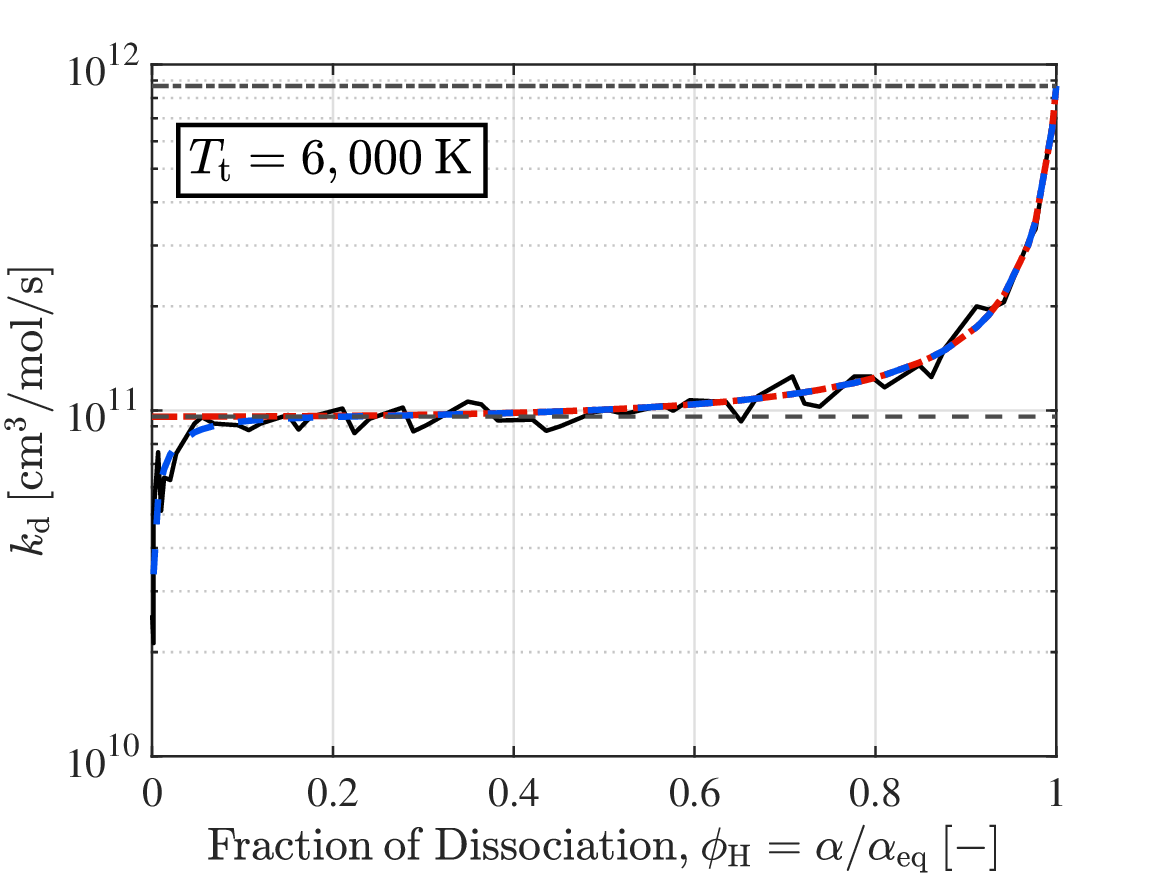}
    \end{subfigure}
    \begin{subfigure}[b]{0.34\textwidth}
        \centering
        \includegraphics[width=\textwidth,trim={0cm 1cm 1cm 0.2cm},clip]{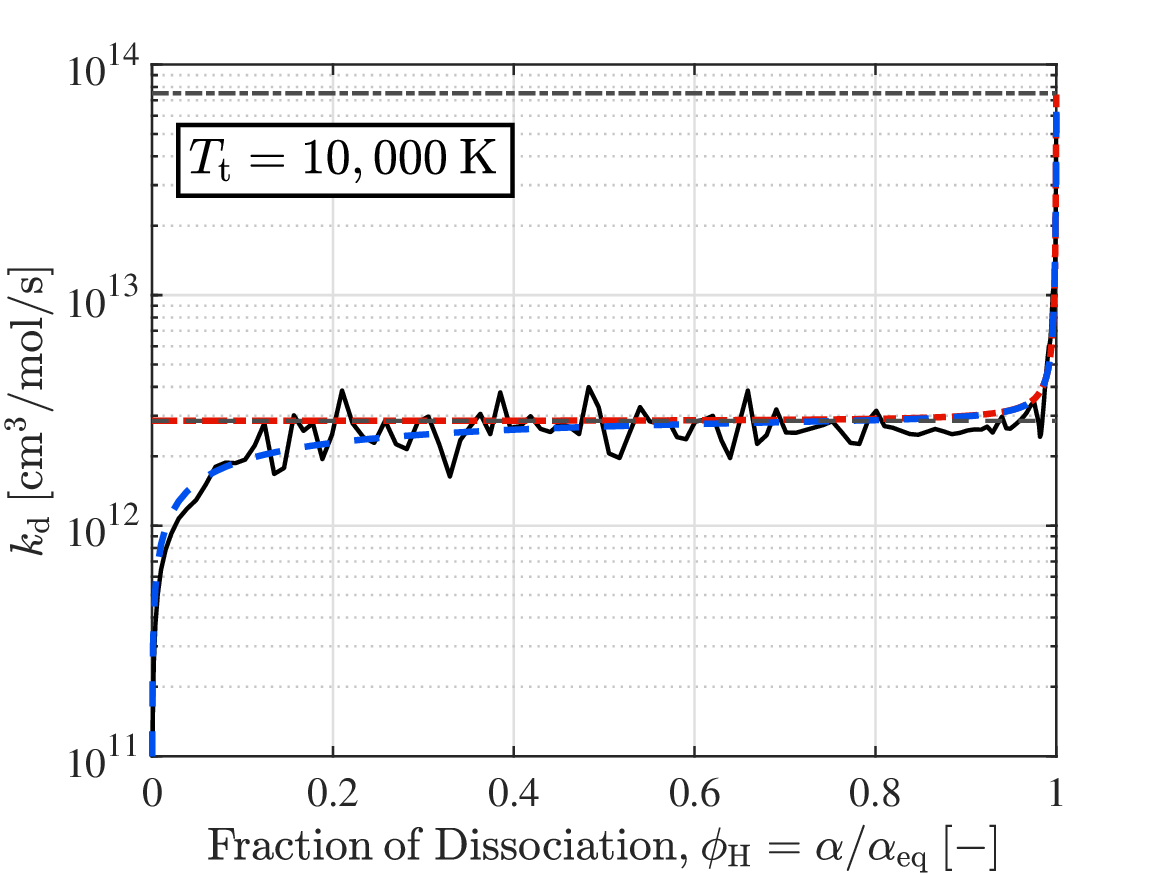}
    \end{subfigure}
    \hfill
    \begin{subfigure}[b]{0.32\textwidth}
        \centering
        \includegraphics[width=\textwidth,trim={1.2cm 1cm 1cm 0.2cm},clip]{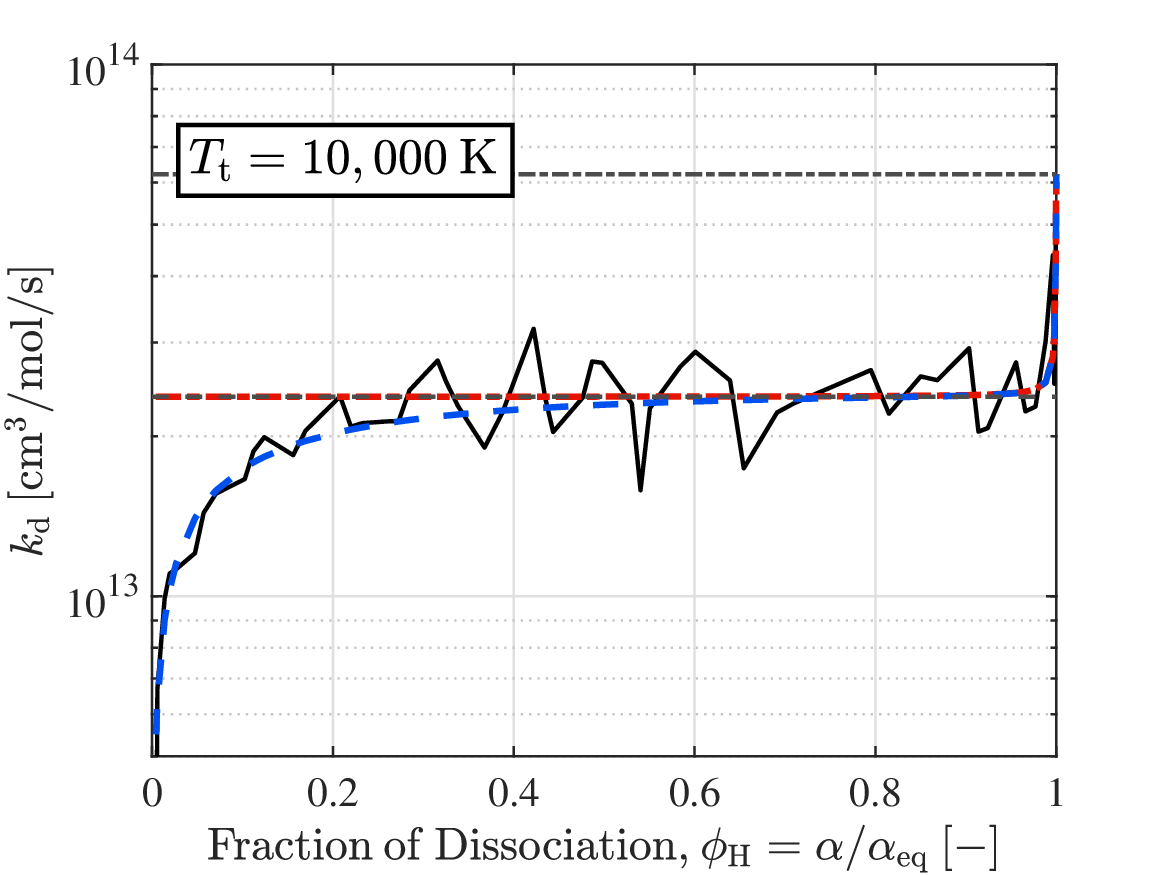}
    \end{subfigure}
    \hfill
    \begin{subfigure}[b]{0.32\textwidth}
        \centering
        \includegraphics[width=\textwidth,trim={1.2cm 1cm 1cm 0.2cm},clip]{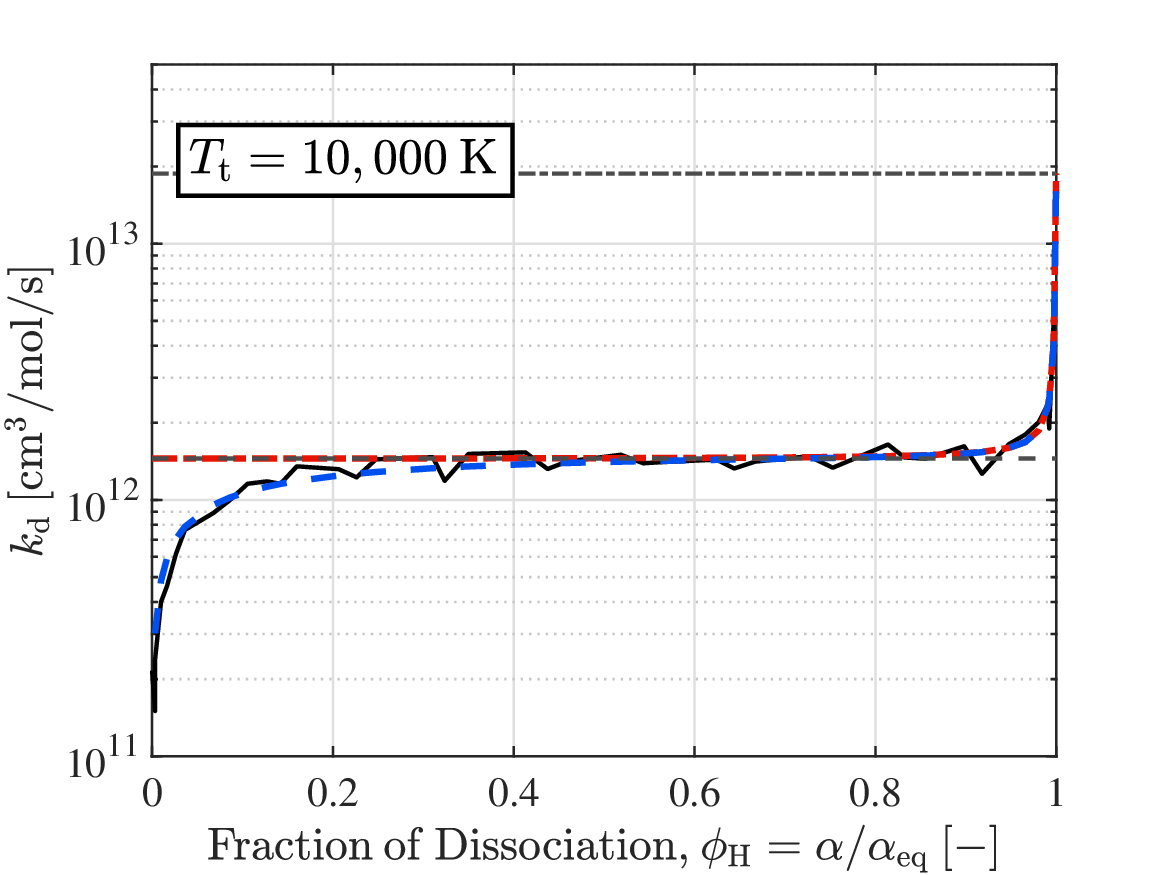}
    \end{subfigure}
    \begin{subfigure}[b]{0.34\textwidth}
        \centering
        \includegraphics[width=\textwidth,trim={0cm 0cm 1cm 0.2cm},clip]{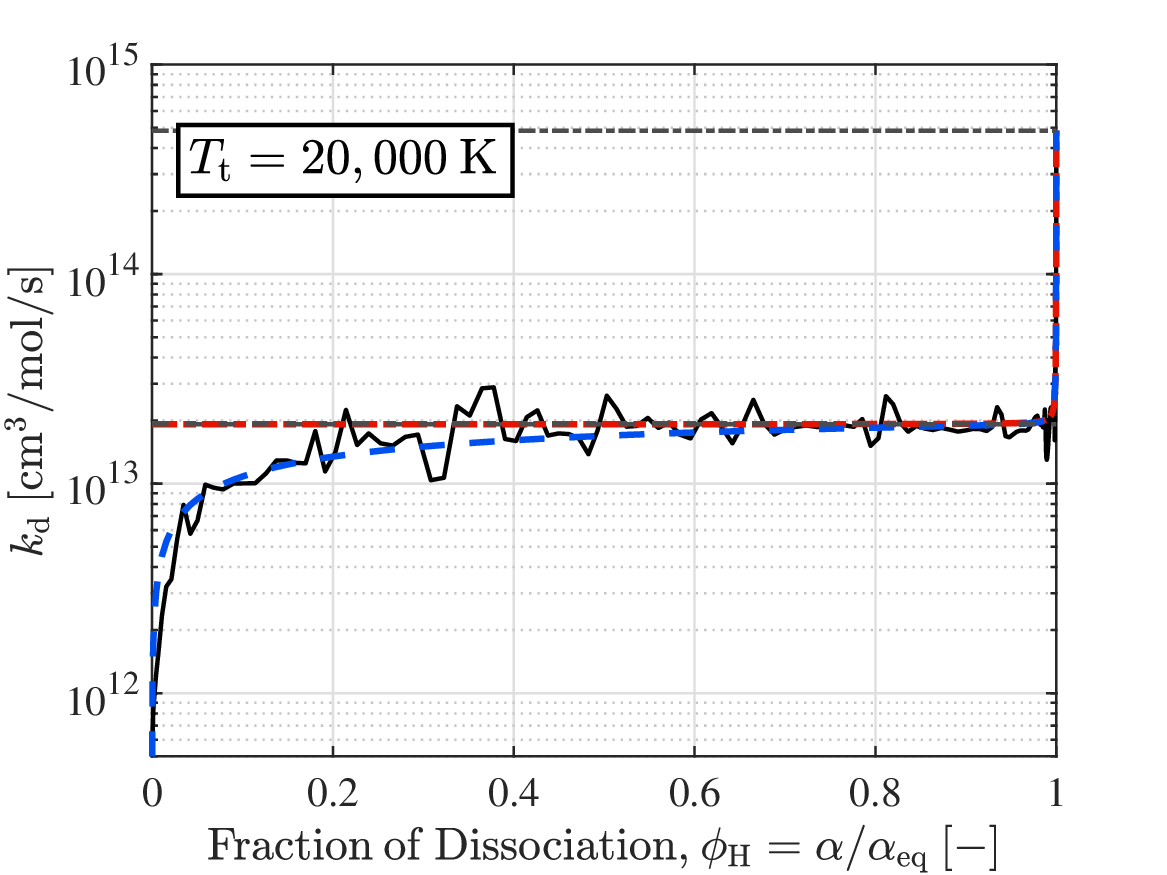}
        \caption{M = $\rm H_2$}
    \end{subfigure}
    \hfill
    \begin{subfigure}[b]{0.32\textwidth}
        \centering
        \includegraphics[width=\textwidth,trim={1.2cm 0cm 1cm 0.2cm},clip]{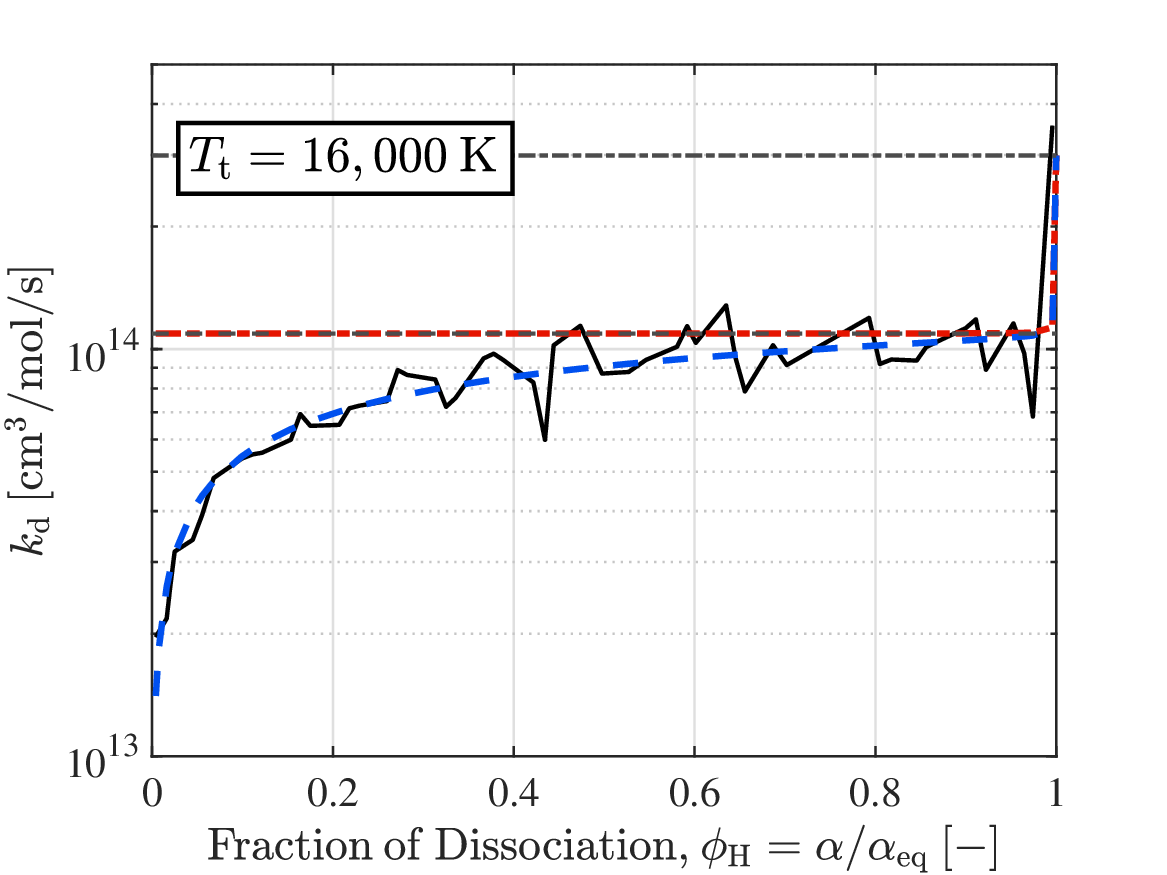}
        \caption{M = H}
    \end{subfigure}
    \hfill
    \begin{subfigure}[b]{0.32\textwidth}
        \centering
        \includegraphics[width=\textwidth,trim={1.2cm 0cm 1cm 0.2cm},clip]{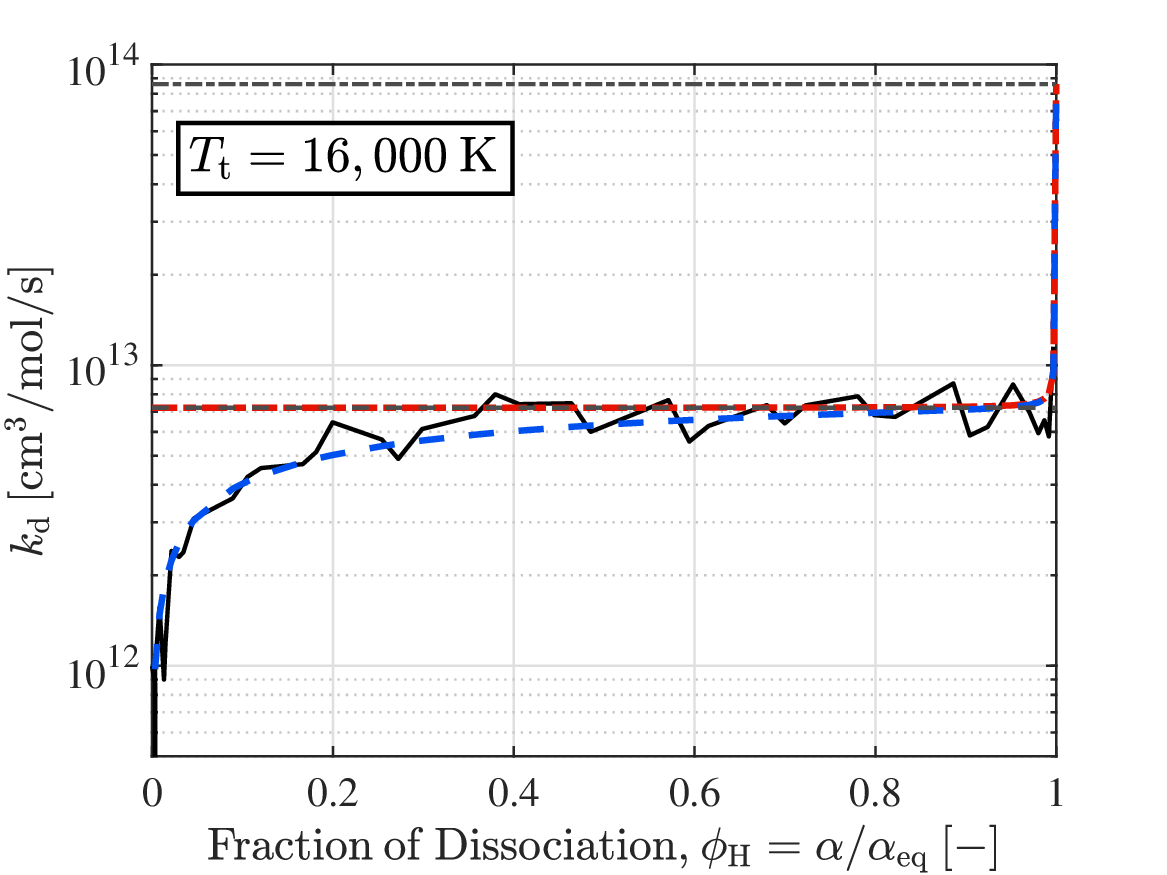}
        \caption{M = He}
    \end{subfigure}
    \caption{Aggregate dissociation rate constants for the master equation calculations of Kim and Boyd~\cite{Kim2012} (M = $\rm H_2$, left) and Kim~\cite{Kim2015} (M = H, middle, and M = He, right) at increasing temperatures (top to bottom). Solid black lines correspond to the master equation calculations, while the dashed blue and dash-dotted red lines correspond to the predictions from the QSS theory with and without the pre-QSS correction, respectively.}
    \label{fig:kd_master}
\end{figure*}

For several of these master equation cases, the corresponding aggregate $k_{\rm d}$ is calculated and shown as solid black lines in Fig.~\ref{fig:kd_master} \footnote{For the case of M=$\rm H_2$, a modified equilibrium constant, $K_{\rm eq,mod}=2K_{\rm eq}$, was required to reproduce the equilibrium number densities plotted by Kim and Boyd~\cite{Kim2012}. Additionally, for the case of M=H, $k_{\rm d}$ values were extracted from the number density profiles assuming $n_{\rm M}$ was fixed and equal to $n_{\rm H,0}$.}. These values are computed by substituting $k_{\rm d,th}$ into Eq.~\eqref{eqn:A2DEB} to compute $k_{\rm r}$, after which $k_{\rm r}$ and the corresponding number densities from Fig.~\ref{fig:n_master} and their derivatives are substituted into Eq.~\eqref{eqn:dnA} to compute $k_{\rm d}$. For the case of M = $\rm H_2$, $k_{\rm d,th}$ values are taken directly from Kim and Boyd~\cite{Kim2012}. However, in the cases of M = H and He, $k_{\rm d,th}$ values were not reported by Kim~\cite{Kim2015}. Therefore, $k_{\rm d,th}$ values are instead taken from the recent QCT study by Vargas et al.~\cite{Vargas2023}. The master equation-extracted $k_{\rm d}$ values show similar trends for all third-bodies. Namely, the rate constants start at a low value, increase to reach a plateau, and then equilibrate to the thermal limit. As with the $T_{\rm r}$ and $T_{\rm v}$ plots from Fig.~\ref{fig:Trv_master}, the initial increase corresponds to the pre-QSS period, the plateau corresponds to the non-recombining QSS limit, and the final increase corresponds to the recombining region. For the lower temperature cases, the gradual evolution of $k_{\rm d}$ from the non-recombining QSS to the thermal limit is apparent; at higher temperatures this transition occurs rapidly near $\phi_{\rm H}=1$. Consistent with the QSS theory discussed previously, these results show that $k_{\rm d}$ only reaches $k_{\rm d,th}$ when $\phi_{\rm H}=1$. In other words, thermal and chemical equilibria are reached simultaneously. At the other end near $\phi_{\rm H}=0$, the pre-QSS region is negligible for the lower temperature cases.  As $T_{\rm t}$ is increased, the fraction of dissociation that occurs in the pre-QSS region increases.

To compare to the master equation results, predicted $k_{\rm d}$ values from the QSS as well as the pre-QSS theories are computed. These are shown in Fig.~\ref{fig:kd_master} as dash-dotted red and dashed blue lines, respectively. For the QSS predicted values, $k_{\rm d}$ is computed using Eq.~\eqref{eqn:kdnrth}; for the QSS values with the pre-QSS correction, Eq.~\eqref{eqn:kdprenrth} is used instead with $\epsilon=0$. The $\eta$ values used in the pre-QSS expressions are computed for each third-body and temperature case using a least squares fit to the master equation results. As expected, the $k_{\rm d}$ values from the QSS theory without the pre-QSS correction are able to reproduce the results of the master equations everywhere outside of the pre-QSS region. Namely, the $k_{\rm d,nr}$ values reflect the plateaus in the master equation results, and the transition from the non-recombining QSS to the thermal limit is captured correctly for all third-bodies and temperatures by the QSS theory. Additionally, with the simple pre-QSS correction of Eq.~\eqref{eqn:kdpreQSS4}, the $k_{\rm d}$ values are also able to reproduce the results of the master equation calculations in the pre-QSS regions.

\subsection{Fraction of Dissociation in the Pre-QSS and QSS Regions}
\label{sec:fracdiss}

\begin{figure}[hbt!]
    \centering
    \begin{subfigure}[b]{0.4\textwidth}
        \centering
        \includegraphics[width=\textwidth,trim={0cm 0cm 1.3cm 0cm},clip]{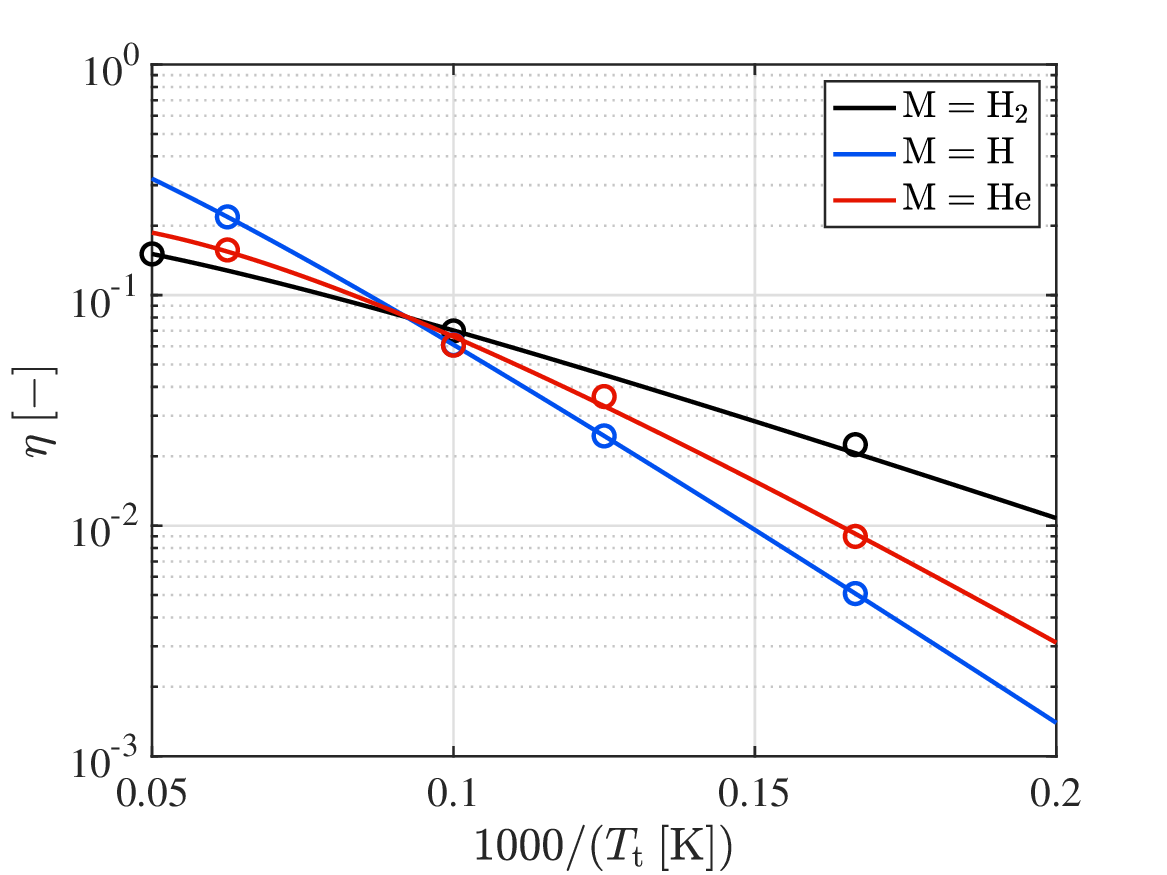}
    \end{subfigure}
    \caption{Extracted $\eta$ values for individual master equation cases (symbols) and associated Arrhenius fits (lines).}
    \label{fig:etafit}
\end{figure}

To estimate the fraction of dissociation that occurs in the pre-QSS for any temperature, fits of $\eta(T_{\rm t})$ for each third-body are first computed. The definition of $\eta(T_{\rm t})\equiv k_{\rm d,nr}/(2(\lambda_1-\lambda_0))$ gives that $\eta$ is effectively the ratio of two rate constants, as $\lambda_0$ and $\lambda_1$ both correspond to eigenvalues of the rate matrix $\bm{\tilde{M}}$. Therefore, a modified Arrhenius form is used for these fits. The parameters for the Arrhenius fits are determined by fitting $\eta(T_{\rm t})$ as a function of temperature (in K) across each of the master equation cases discussed in the previous section. For M = $\rm H_2$, H, and He, these fits are given by $\eta_{\rm M = H_2}(T_{\rm t}) = 6.489\times10^{1}T_{\rm t}^{-0.50}\exp(-2.221\times10^{4}/T_{\rm t})$, $\eta_{\rm M = H}(T_{\rm t}) = 1.897\times10^{3}T_{\rm t}^{-0.66}\exp(-4.236\times10^{4}/T_{\rm t})$, and $\eta_{\rm M = He}(T_{\rm t}) = 1.668\times10^{6}T_{\rm t}^{-1.41}\exp(-4.035\times10^{4}/T_{\rm t})$, respectively. Figure~\ref{fig:etafit} compares these fits (in solid lines) to the master equation-extracted $\eta$ values (in open symbols). The Arrhenius fits are in excellent agreement with the master equation-extracted $\eta$ values. In fact, using these Arrhenius fits in Eq.~\eqref{eqn:kdprenrth} instead of the individual master equation-extracted $\eta$ values produced effectively identical predictions of the pre-QSS dissociation rate constants for the master equation cases discussed in the previous section.

\begin{figure*}[hbt!]
    \centering
    \begin{subfigure}[b]{0.345\textwidth}
        \centering
        \includegraphics[width=\textwidth,trim={0cm 0cm 1cm 0.2cm},clip]{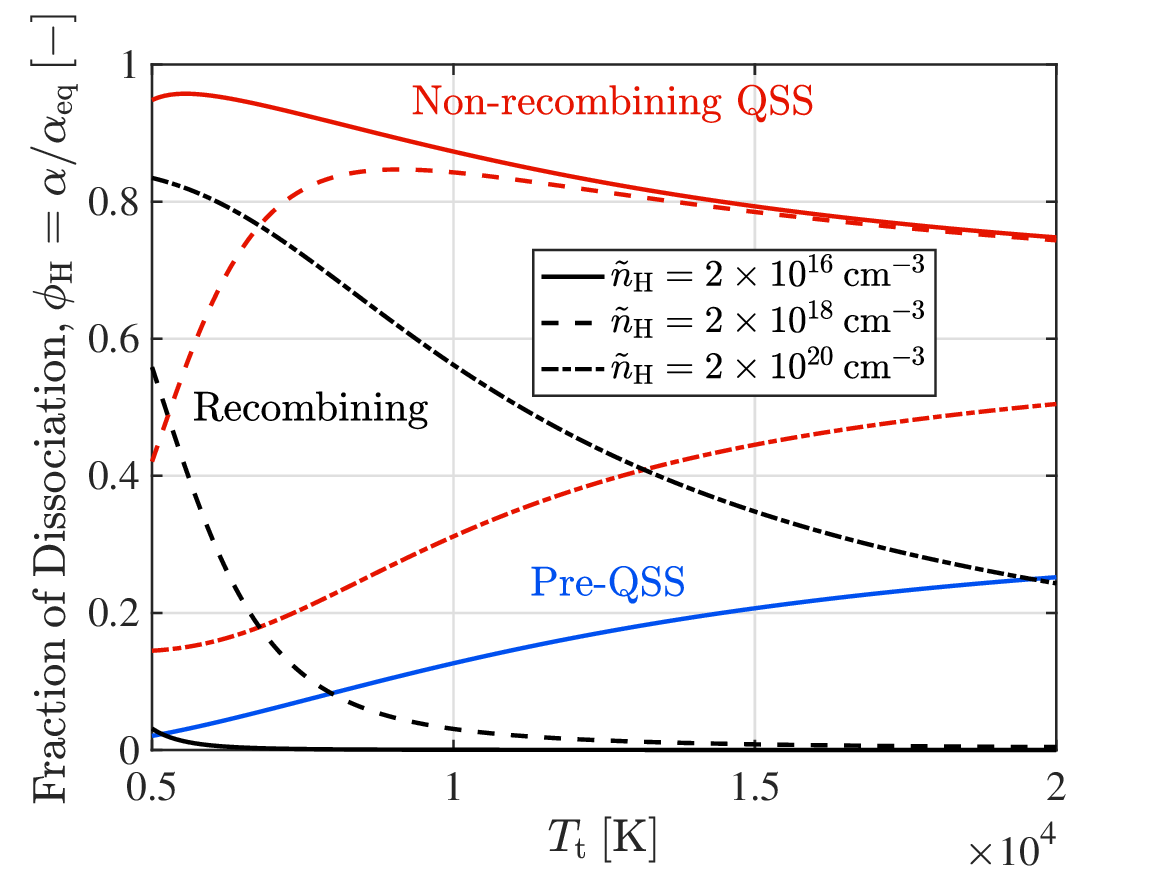}
        \caption{M = $\rm H_2$}
    \end{subfigure}
    \hfill
    \begin{subfigure}[b]{0.32\textwidth}
        \centering
        \includegraphics[width=\textwidth,trim={1.35cm 0cm 1cm 0.2cm},clip]{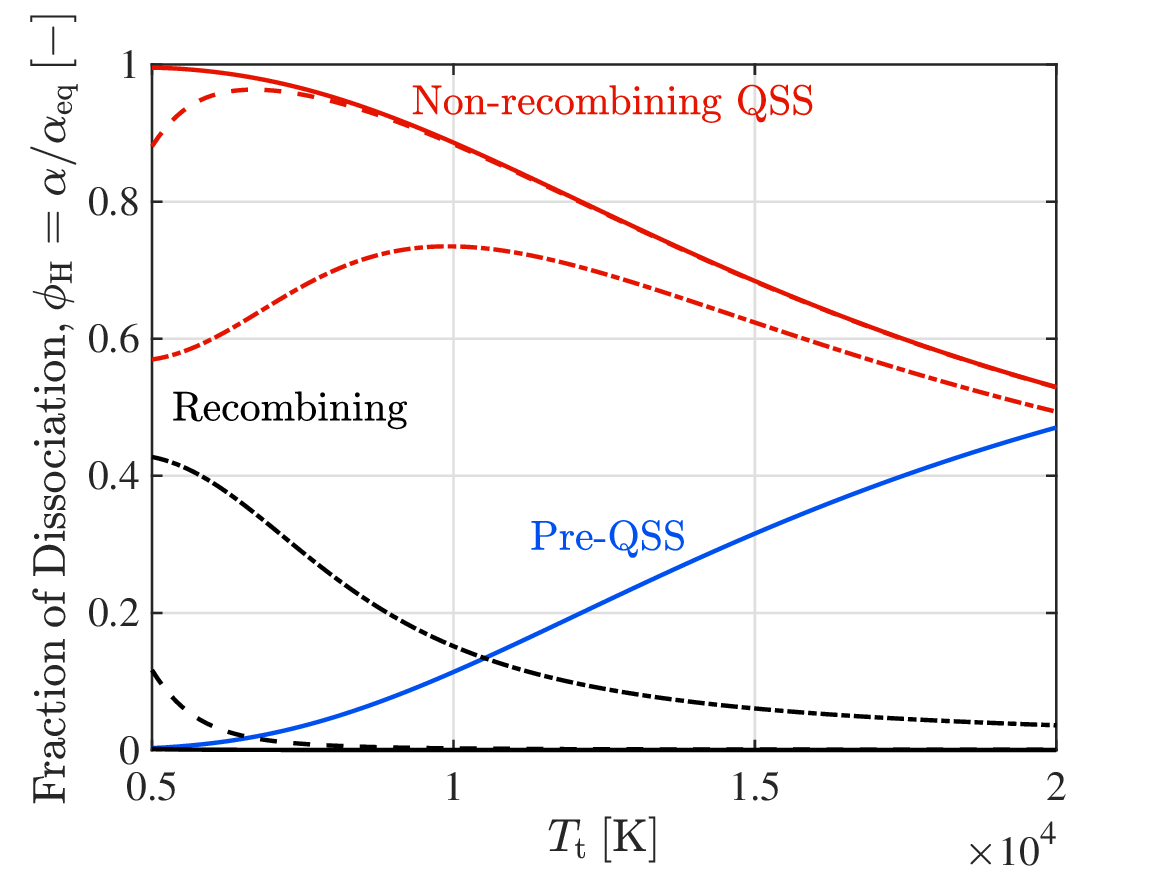}
        \caption{M = H}
    \end{subfigure}
    \hfill
    \begin{subfigure}[b]{0.32\textwidth}
        \centering
        \includegraphics[width=\textwidth,trim={1.35cm 0cm 1cm 0.2cm},clip]{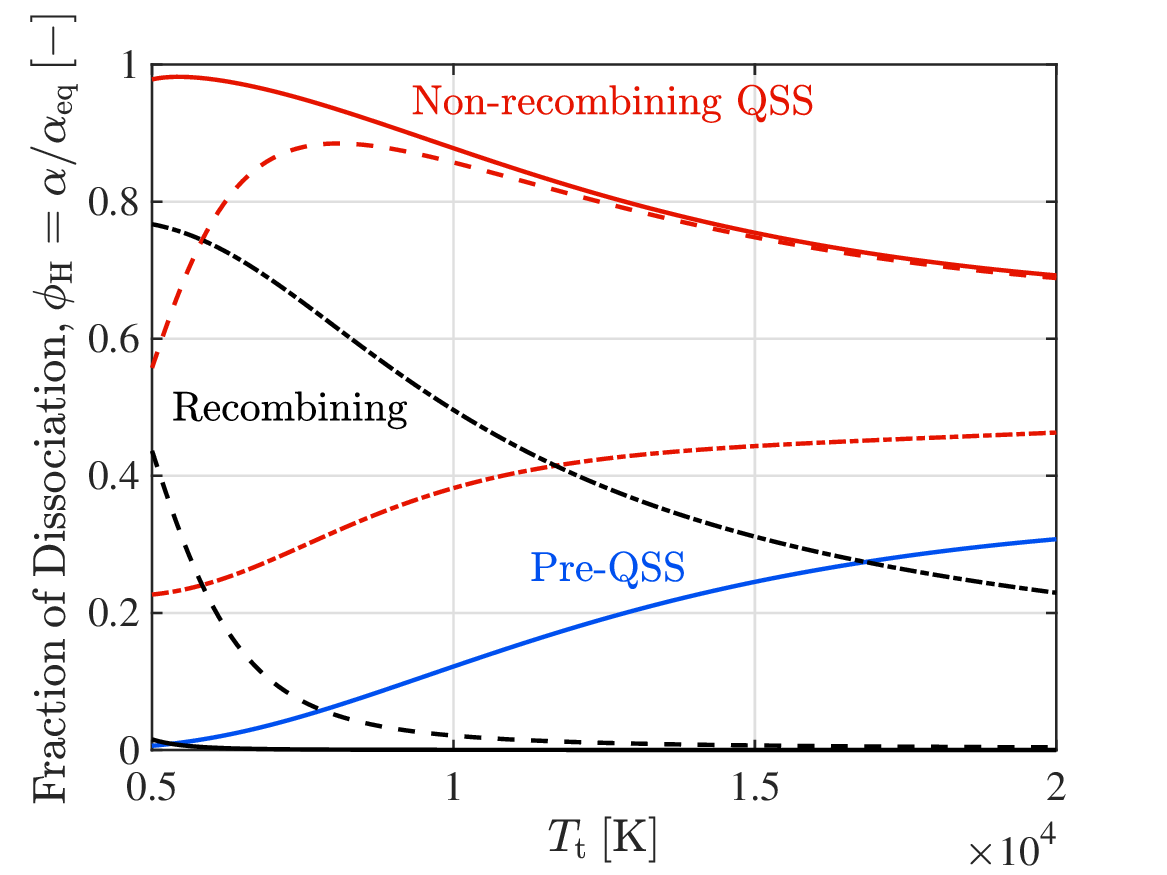}
        \caption{M = He}
    \end{subfigure}
    \caption{Fraction of dissociation that occurs in the pre-QSS (blue lines), non-recombining QSS (red lines), and recombining (black lines) regions for varying number densities and temperatures.}
    \label{fig:fracdiss}
\end{figure*}

Using these $\eta(T_{\rm t})$ fits, the fraction of dissociation that occurs in the pre-QSS, non-recombining QSS, and recombining regions as a function of $T_{\rm t}$ and $\tilde{n}_{\rm H}$ is estimated using Eq.~\eqref{eqn:phiA2chi_nr}, \eqref{eqn:alphanorm}, and \eqref{eqn:fracpreQSS}. In these equations, the exact value of $\delta$ and hence the cutoffs between the pre-QSS, non-recombining QSS, and recombining regions is arbitrary. As will be shown in section~\ref{sec:review}, the uncertainties in the rate constants from the literature are approximately bounded by a factor of two. In light of these uncertainties, a $\delta$ value of 0.25 is used for the present analysis. For the M = $\rm H_2$ case, continuous fits of $k_{\rm d,nr}$ and $k_{\rm d,th}$ are obtained by fitting the reported values from Kim and Boyd~\cite{Kim2012} to a modified Arrhenius form. For the M = H and He cases, fits of $k_{\rm d,nr}$ and $k_{\rm d,th}$ are taken from Kim et al.~\cite{Kim2009} and Vargas et al.~\cite{Vargas2023}, respectively. The results of these calculations are shown in Fig.~\ref{fig:fracdiss}. The selected temperature and number density ranges correspond to the range of conditions relevant for post-shock hypersonic entry flows as well as the shock tube experiments discussed later in section~\ref{sec:highexp}. The $\tilde{n}_{\rm H}=2.0\times10^{18}$ $\rm cm^{-3}$ case in particular corresponds to the master equation calculations discussed in the previous section. These plots highlight that for most temperatures and number densities, the majority of dissociation for all three third-bodies occurs near the non-recombining QSS limit where $k_{\rm d} \approx k_{\rm d,nr}$. As the temperature decreases and/ or as the number density increases, the fraction of dissociation that occurs in the recombining region increases. Luckily, as discussed in section~\ref{sec:QSSsrc}, the source terms for the entire QSS region can be captured by a one-temperature fit of $k_{\rm d,nr}(T_{\rm t})$, as the dependence on number densities for the transition of $k_{\rm d}$ between the non-recombining and thermal limits is captured implicitly by Eq.~\eqref{eqn:dnA_3}. Following Eq.~\eqref{eqn:fracpreQSS}, the fraction of dissociation that occurs in the pre-QSS region is only a function of temperature. This is consistent with the empirical observations from the $\rm O_2$ + O master equation calculations of Venturi et al.~\cite{Venturi2020}. While the fraction of dissociation that occurs in the pre-QSS is negligible at low temperatures, it increases monotonically with temperature, and in the case of M = H, becomes comparable to the non-recombining contribution at 20,000 K.

\subsection{Predicted Number Densities and Rovibrational Energies}
\label{sec:prednum}

\begin{figure*}[hbt!]
    \centering
    \begin{subfigure}[b]{0.345\textwidth}
        \centering
        \includegraphics[width=\textwidth,trim={0cm 0cm 1cm 0cm},clip]{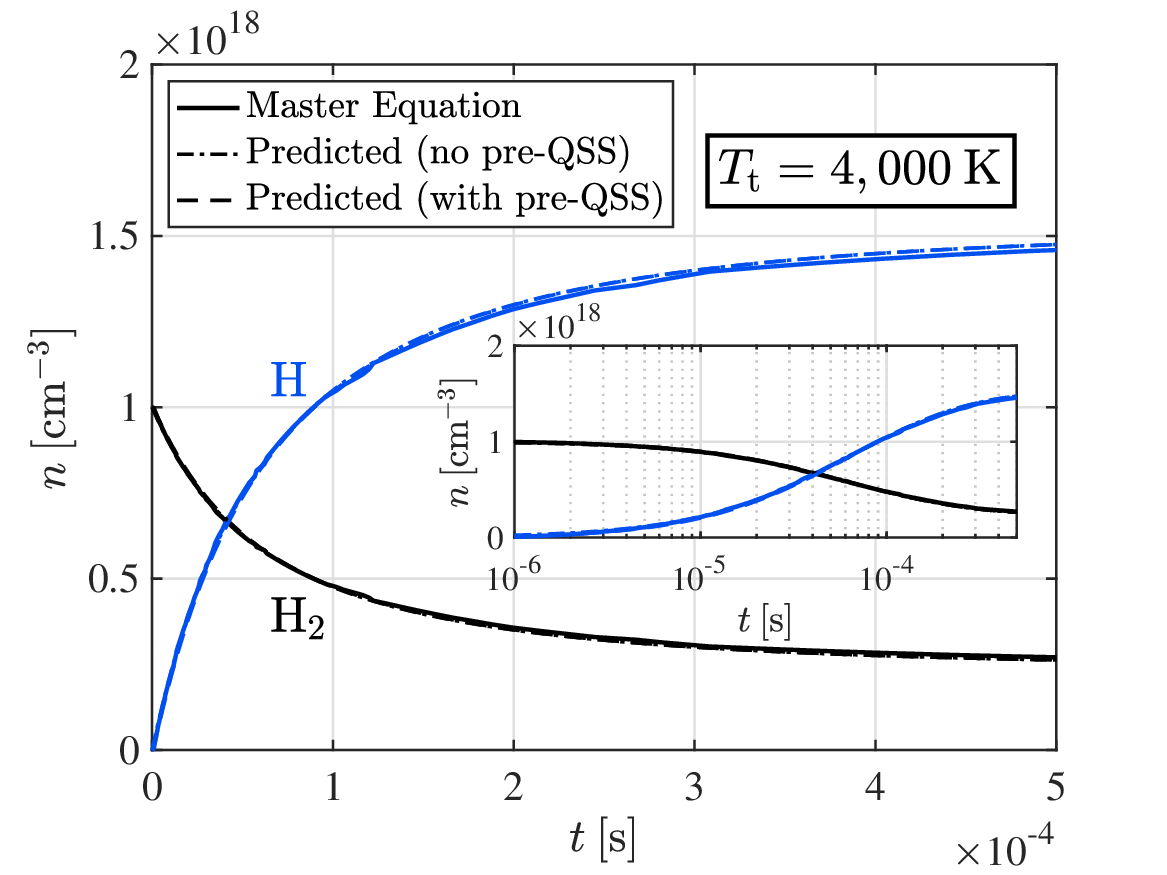}
    \end{subfigure}
    \hfill
    \begin{subfigure}[b]{0.32\textwidth}
        \centering
        \includegraphics[width=\textwidth,trim={1.3cm 0cm 1cm 0cm},clip]{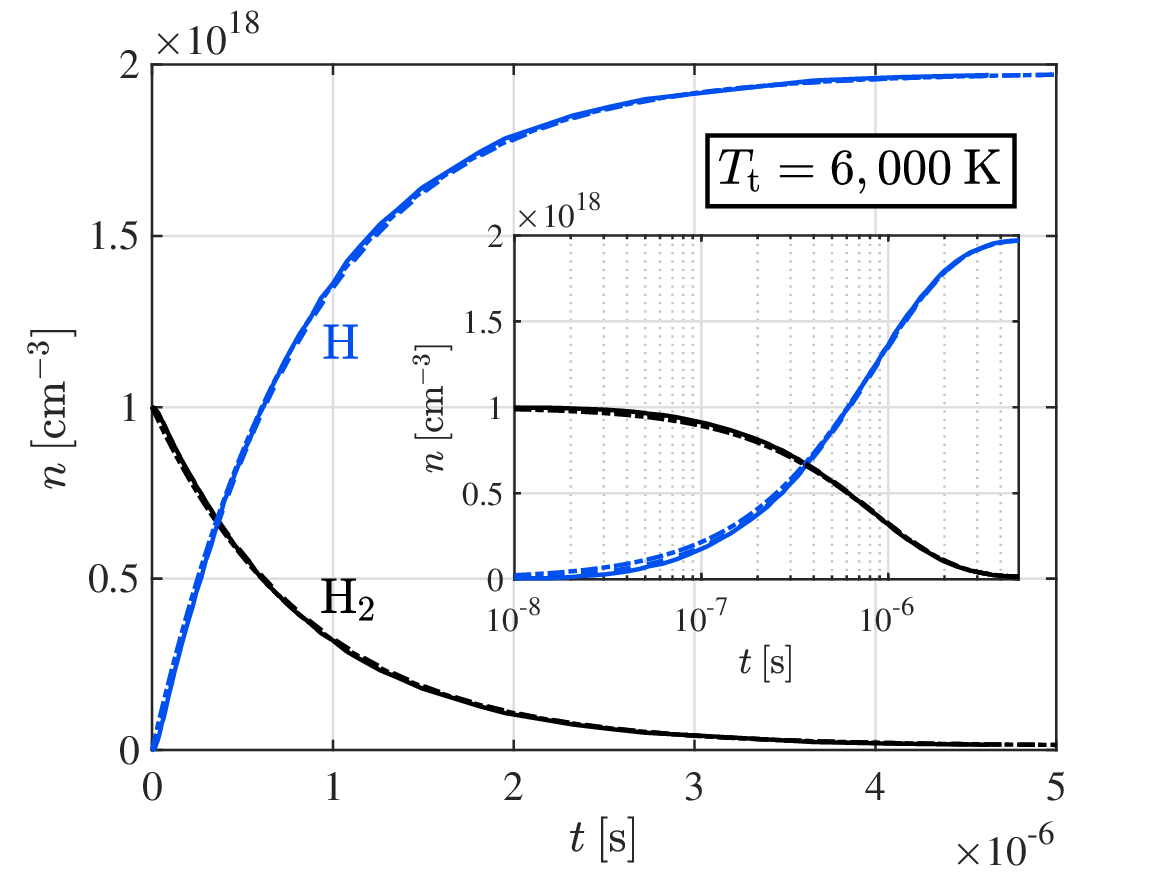}
    \end{subfigure}
    \hfill
    \begin{subfigure}[b]{0.32\textwidth}
        \centering
        \includegraphics[width=\textwidth,trim={1.3cm 0cm 1cm 0cm},clip]{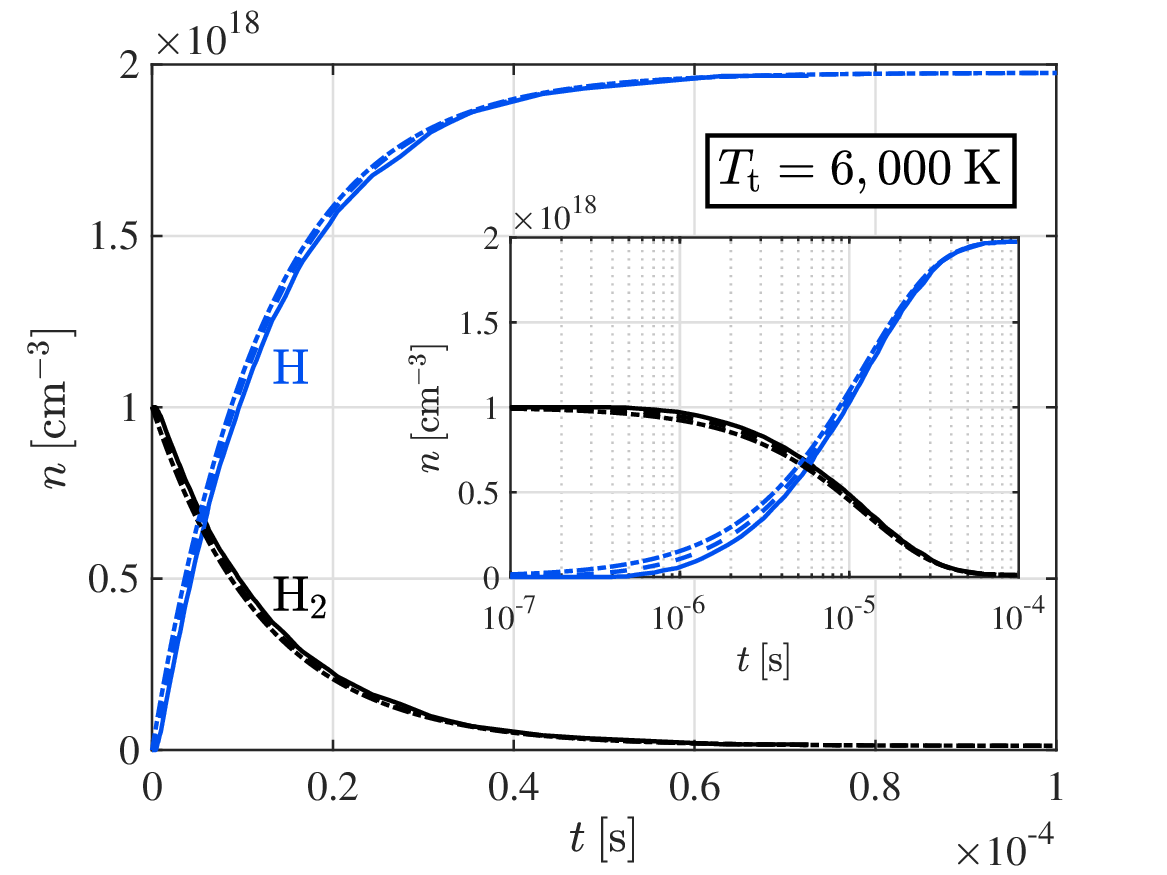}
    \end{subfigure}
    \begin{subfigure}[b]{0.345\textwidth}
        \centering
        \includegraphics[width=\textwidth,trim={0cm 0cm 1cm 0cm},clip]{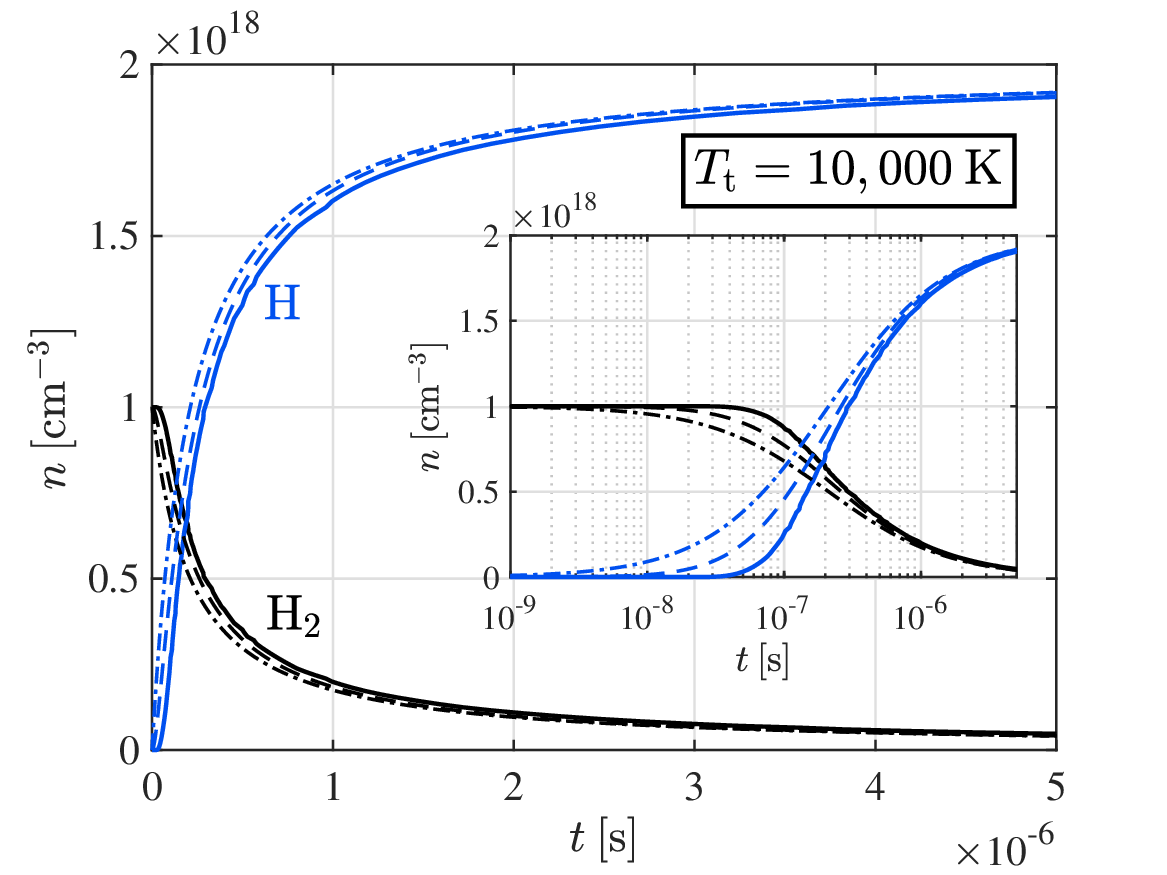}
    \end{subfigure}
    \hfill
    \begin{subfigure}[b]{0.32\textwidth}
        \centering
        \includegraphics[width=\textwidth,trim={1.3cm 0cm 1cm 0cm},clip]{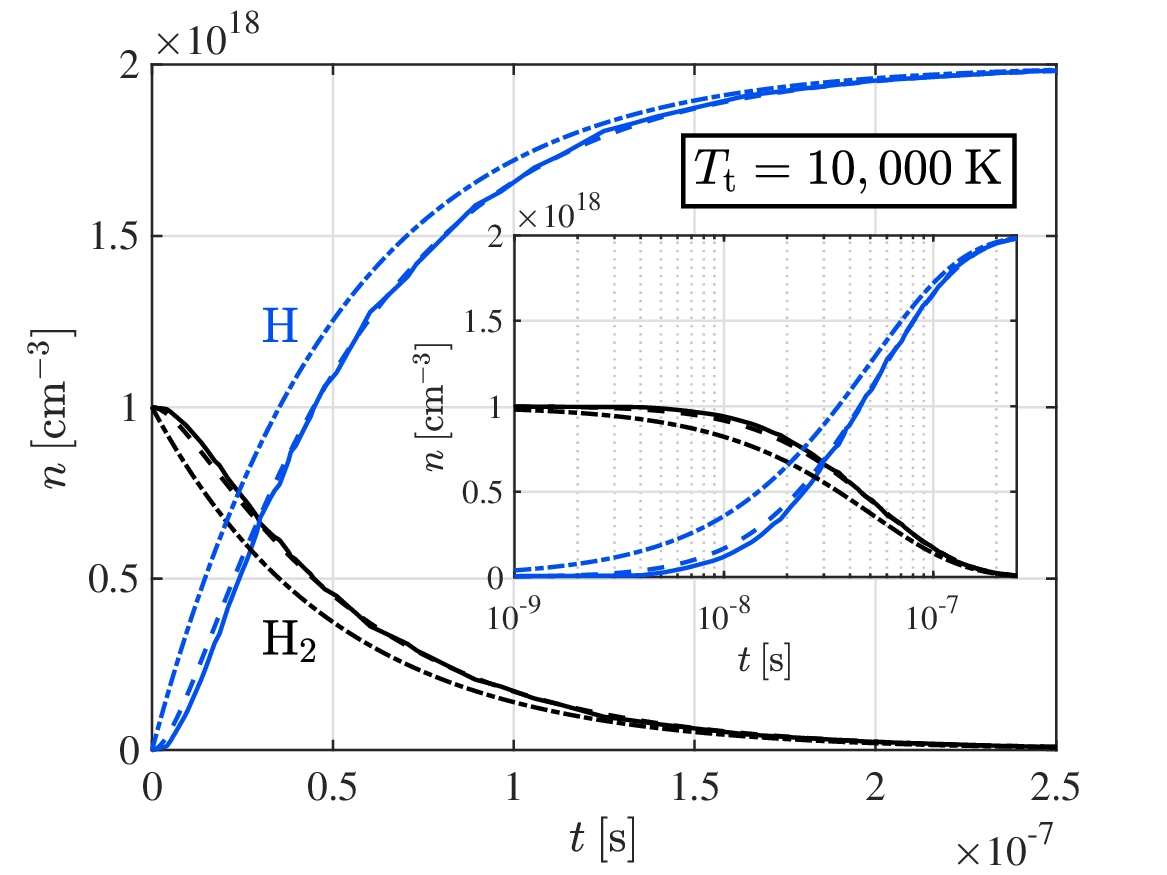}
    \end{subfigure}
    \hfill
    \begin{subfigure}[b]{0.32\textwidth}
        \centering
        \includegraphics[width=\textwidth,trim={1.3cm 0cm 1cm 0cm},clip]{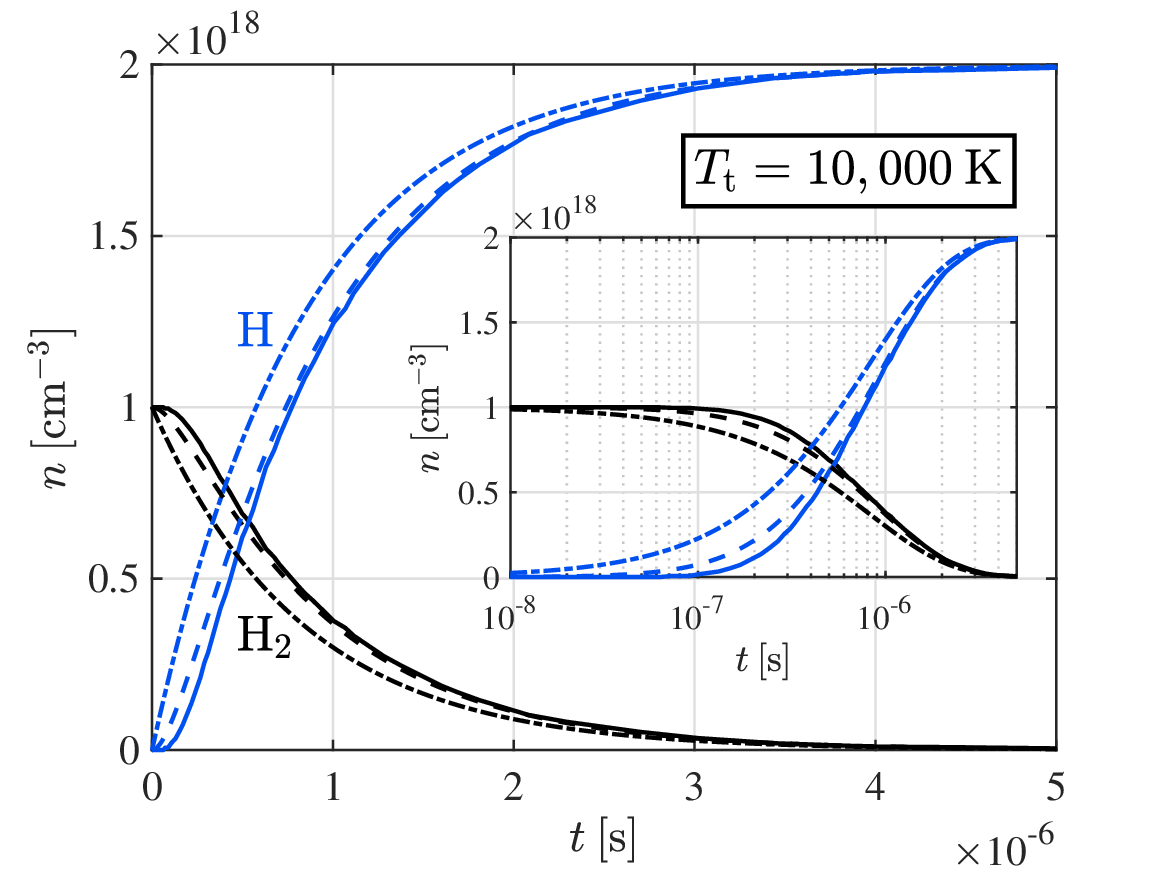}
    \end{subfigure}
    \begin{subfigure}[b]{0.345\textwidth}
        \centering
        \includegraphics[width=\textwidth,trim={0cm 0cm 1cm 0cm},clip]{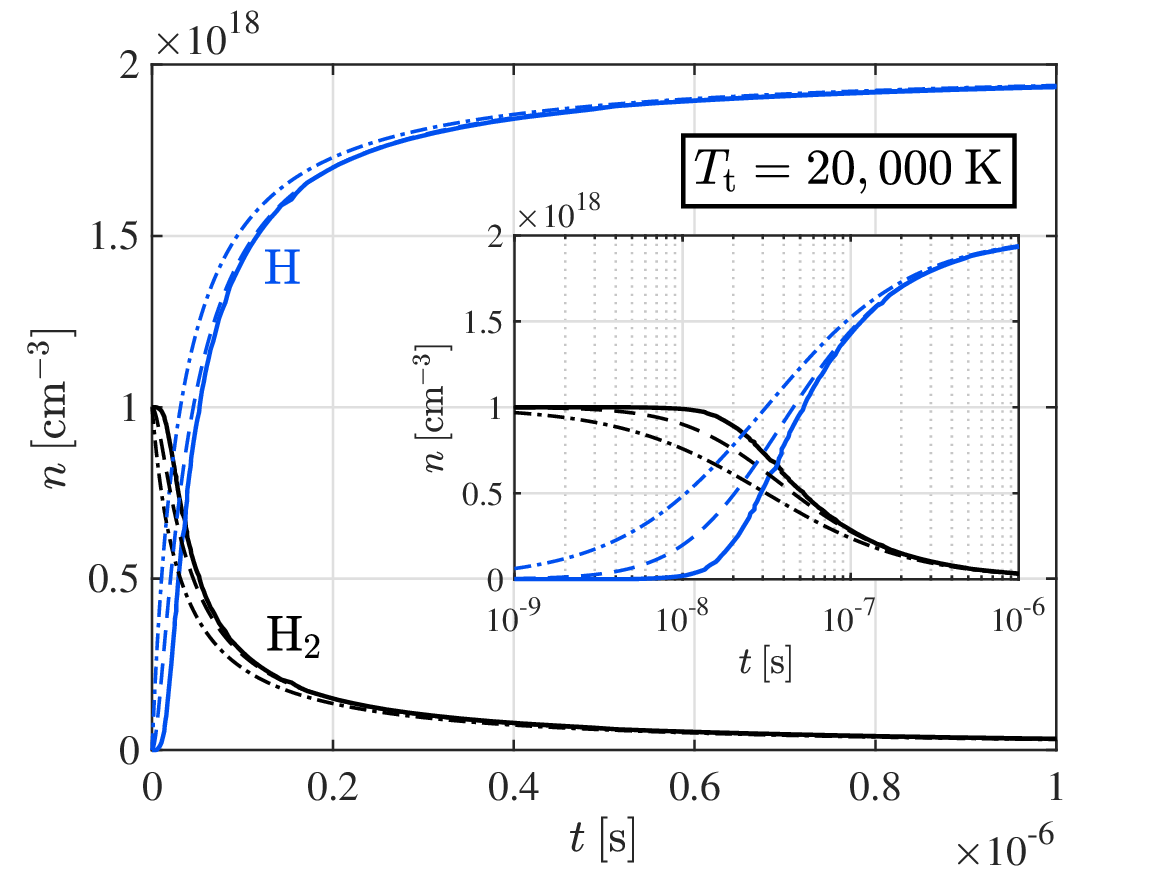}
        \caption{M = $\rm H_2$}
    \end{subfigure}
    \hfill
    \begin{subfigure}[b]{0.32\textwidth}
        \centering
        \includegraphics[width=\textwidth,trim={1.3cm 0cm 1cm 0cm},clip]{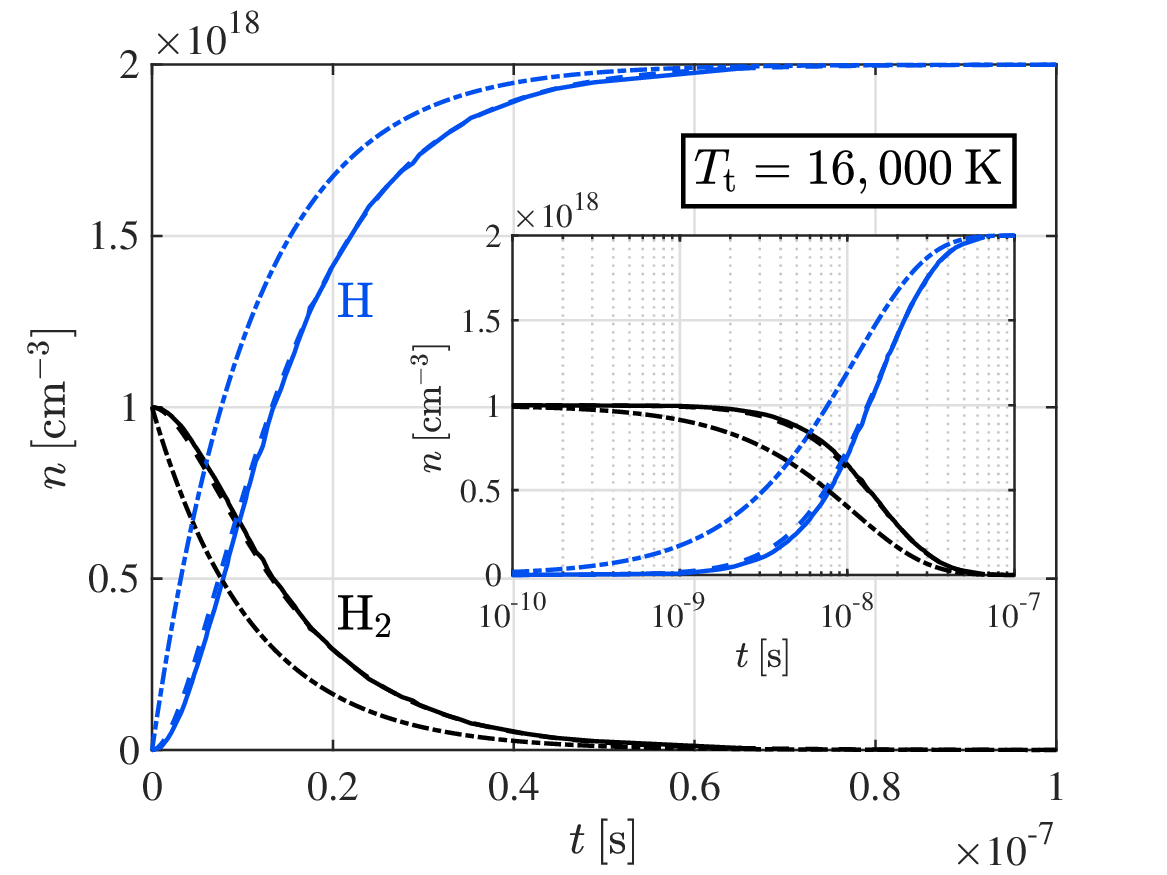}
        \caption{M = H}
    \end{subfigure}
    \hfill
    \begin{subfigure}[b]{0.32\textwidth}
        \centering
        \includegraphics[width=\textwidth,trim={1.3cm 0cm 1cm 0cm},clip]{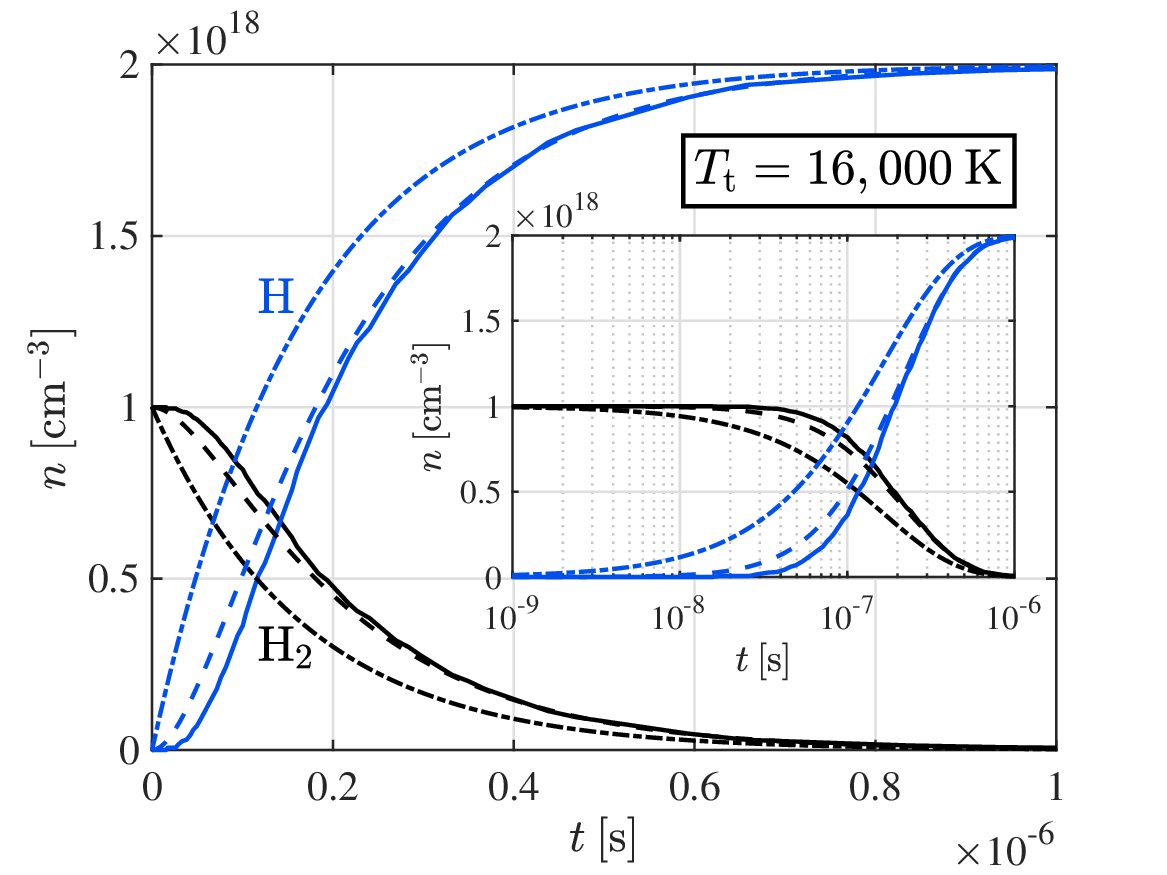}
        \caption{M = He}
    \end{subfigure}
    \caption{Number density profiles for the master equation calculations of Kim and Boyd~\cite{Kim2012} (M = $\rm H_2$, left) and Kim~\cite{Kim2015} (M = H, middle, and M = He, right) at increasing temperatures (top to bottom). Solid lines correspond to the master equation calculations, while the dashed and dash-dotted lines correspond to the predictions from the QSS theory with and without the pre-QSS correction, respectively.}
    \label{fig:prednum}
\end{figure*}

The QSS source term expressions with and without the pre-QSS correction (Eq.~\eqref{eqn:dnA_preQSS} and \eqref{eqn:dnA_3}, respectively) are numerically integrated, and the resulting number density profiles are compared to those of the master equation calculations in Fig.~\ref{fig:prednum}. For the pre-QSS expression, $\epsilon$ is set to $10^{-3}$ and the $\eta(T_{\rm t})$ fits obtained in the previous section are used. For the lowest temperature cases (4,000 K and 6,000 K), both sets of the predicted profiles reproduce the master equation results accurately for all three third-bodies. This serves as a verification of the source term equivalence discussed in section~\ref{sec:QSSsrc}, as Fig.~\ref{fig:kd_master} shows that the aggregate $k_{\rm d}$ varies considerably between the $k_{\rm d,nr}$ and $k_{\rm d,th}$ limits for these low temperature cases, and yet integrating Eq.~\eqref{eqn:dnA_3} and \eqref{eqn:dnA_preQSS} (which are only functions of $k_{\rm d,nr}$ and $k_{\rm d,pre-QSS}$, respectively) still yields the correct number density profiles. For the higher temperature cases ($T_{\rm t}\geq$ 10,000 K), a larger difference appears between the predicted profiles with and without the pre-QSS correction. In particular, the QSS expression without the pre-QSS correction results in dissociation that is too fast when compared to the master equation calculations. However, with the pre-QSS correction, the predicted profiles are in much better agreement with the master equation results, especially for the M = H and He cases where there is excellent agreement for the higher temperature cases.

\begin{figure*}[hbt!]
    \centering
    \begin{subfigure}[b]{0.35\textwidth}
        \centering
        \includegraphics[width=\textwidth,trim={0cm 0cm 1.3cm 0cm},clip]{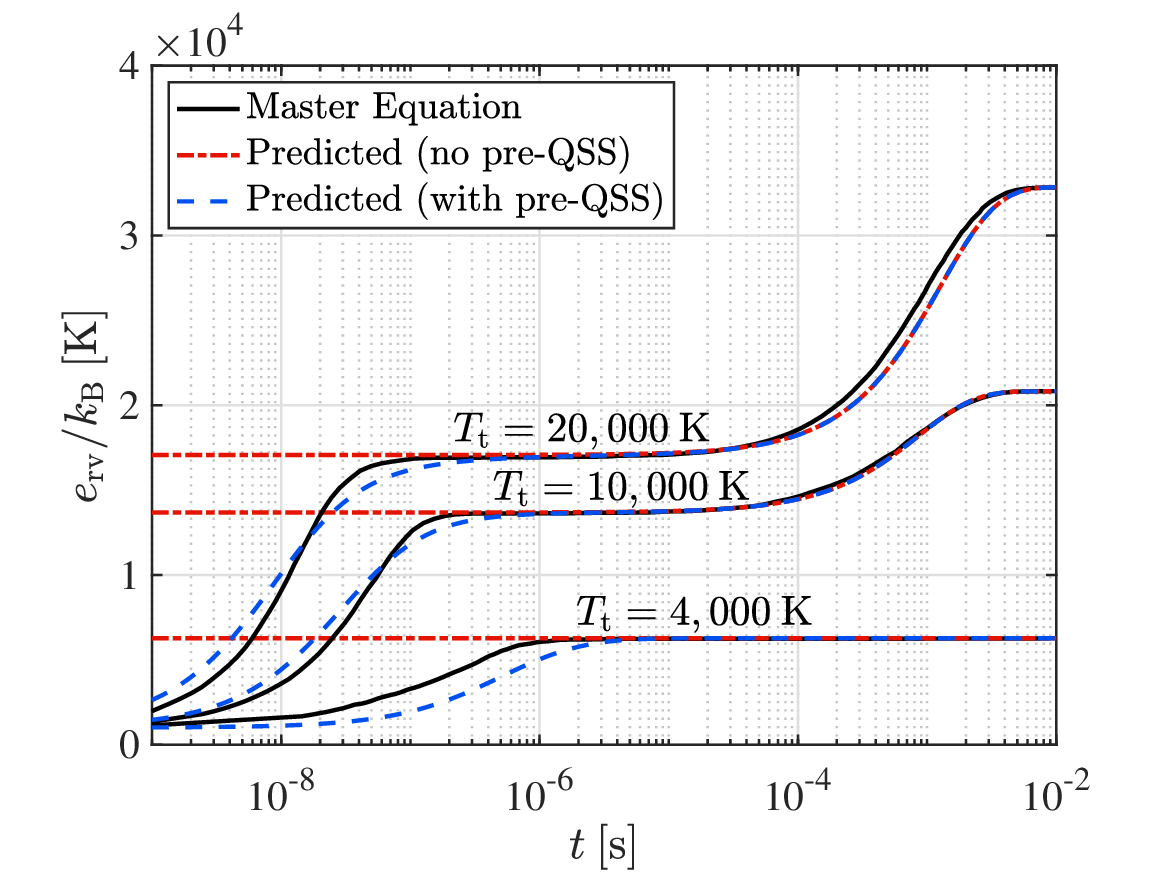}
        \caption{M = $\rm H_2$}
    \end{subfigure}
    \hfill
    \begin{subfigure}[b]{0.315\textwidth}
        \centering
        \includegraphics[width=\textwidth,trim={1.8cm 0cm 1.3cm 0cm},clip]{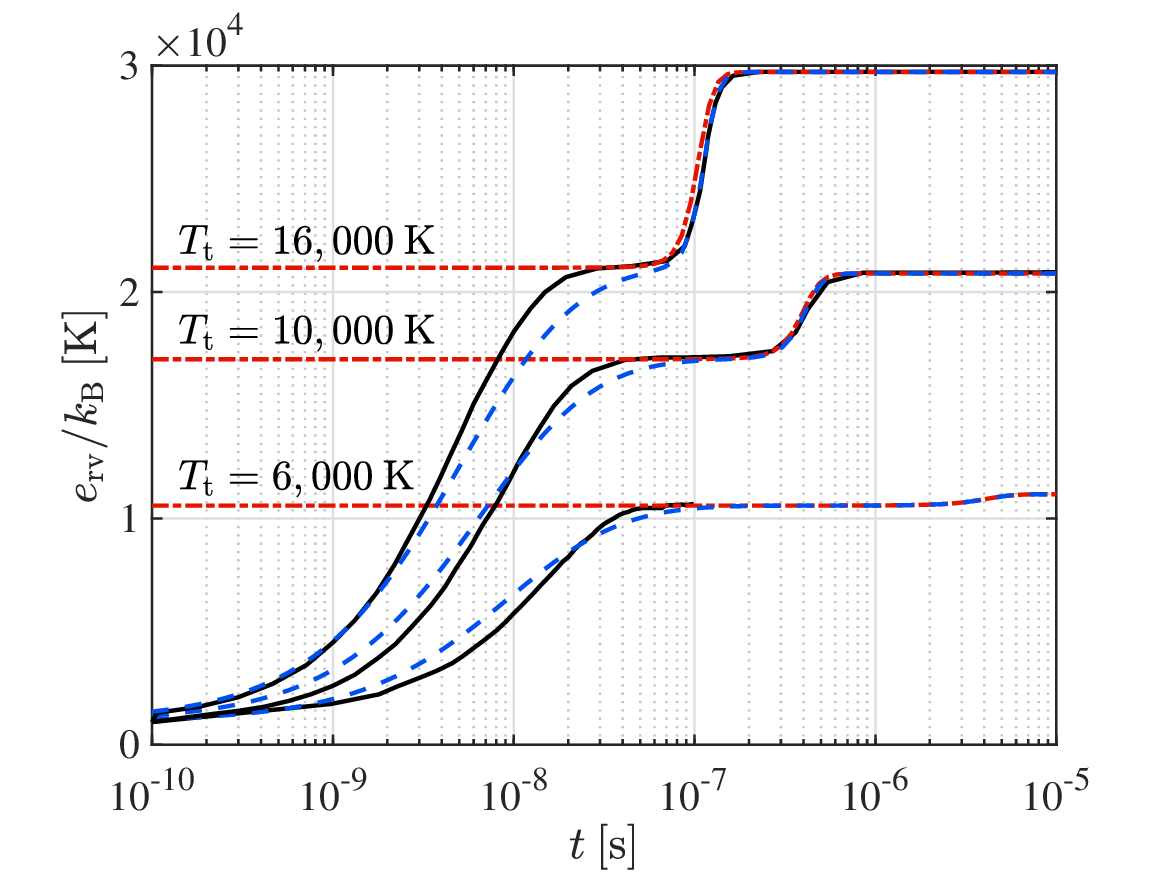}
        \caption{M = H}
    \end{subfigure}
    \hfill
    \begin{subfigure}[b]{0.315\textwidth}
        \centering
        \includegraphics[width=\textwidth,trim={1.8cm 0cm 1.3cm 0cm},clip]{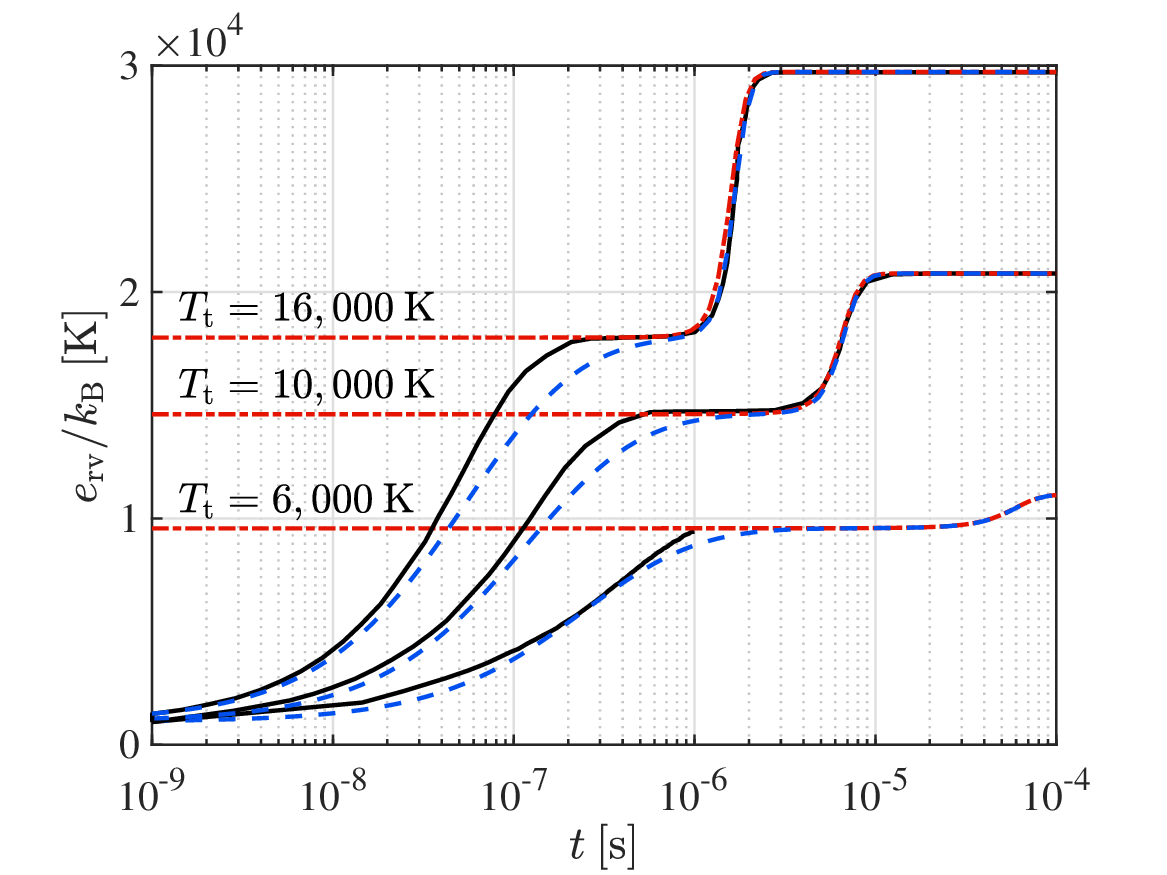}
        \caption{M = He}
    \end{subfigure}
    \caption{Average rovibrational energy profiles for the master equation calculations of Kim and Boyd~\cite{Kim2012} (M = $\rm H_2$), Kim~\cite{Kim2015} (M = H and He, $T_{\rm t}$ = 6,000 K and 8,000 K), and Kim et al.~\cite{Kim2009} (M = H and He, $T_{\rm t}$ = 10,000 K and 16,000 K) compared to the predictions of the present work. Solid black lines correspond to the master equation calculations, while the dashed blue and dash-dotted red lines correspond to the predictions from the QSS theory with and without the pre-QSS correction, respectively.}
    \label{fig:prederv}
\end{figure*}

Figure~\ref{fig:prederv} shows the corresponding average rovibrational energy profiles for the same simulation cases. For the master equation calculations, the reported $T_{\rm r}$ and $T_{\rm v}$ values from Fig.~\ref{fig:Trv_master} are used with Eq.~\eqref{eqn:er}-\eqref{eqn:erv_sum} to compute $e_{\rm rv}$ values. For the predicted values, $T_{\rm r,nr}$ and $T_{\rm v,nr}$ are first estimated from Fig.~\ref{fig:Trv_master}, then are used in Eq.~\eqref{eqn:erv_nr2} to compute $e_{\rm rv,nr}$. Then, values of $e_{\rm rv}$ are computed using either Eq.~\eqref{eqn:erv_preQSS2} or Eq.~\eqref{eqn:erv_QSS} with the predicted number densities from Fig.~\ref{fig:prednum} for the cases with and without the pre-QSS correction, respectively. Overall, the predicted profiles with the pre-QSS correction reproduce the trends seen in the master equation calculations, without any refitting of $\eta(T_{\rm t})$. In the recombining regions for all third-bodies, there is excellent agreement in the transition of $e_{\rm rv}$ between the $e_{\rm rv,nr}$ and $e_{\rm rv,th}$ limits.

The results of Fig.~\ref{fig:prednum} and \ref{fig:prederv} show that the relatively simple pre-QSS corrected expressions of Eq.~\eqref{eqn:kdpreQSS4} and \eqref{eqn:erv_preQSS2} are sufficient for capturing the majority of the non-equilibrium behavior of the full master equation calculations for $\rm H_2$ dissociation. Importantly, the rate constant expression only requires fits for $k_{\rm d,nr}(T_{\rm t})$ and $\eta(T_{\rm t})$, and is ultimately only a function of $T_{\rm t}$ and the fraction of dissociation, $\phi_{\rm H}=n_{\rm H}/n_{\rm H,eq}$. In CFD calculations, $T_{\rm t}$ and $n_{\rm H}$ are already computed via the energy and species' continuity equations. Therefore, the non-equilibrium dissociation behavior (including QSS and pre-QSS effects) may be captured with just a one-temperature formulation that only considers bulk species. This approach offers benefits over other approaches from the literature.

The most widely used class of non-equilibrium dissociation models in the literature is known broadly as ``multi-temperature'' models. Examples of such models include the two-temperature formulation by Park~\cite{Park1988,Park1989,Park1989NonequilibriumHA}, as well as the more recent QCT-informed modified Marrone-Treanor (MMT)~\cite{Chaudhry2019,Chaudhry2020} and Singh-Schwartzentruber models~\cite{Singh2020_1,Singh2020_2}. Multi-temperature models require the solution of additional energy equations (one per independently tracked energy mode/ temperature) to describe the coupling between internal energy relaxation and dissociation. As a consequence, multi-temperature models (at a minimum) require inputs for relaxation time-scales and energy feedback factors, as well as multi-temperature fits of rate constants. These three model inputs are inherently coupled and cannot be estimated independently. In particular, they must be evaluated in a consistent manner to recover QSS dissociation rates accurately. By contrast, this coupling is already captured by design in the one-temperature formulation of the present work, as the expressions for $k_{\rm d}$ (Eq.~\eqref{eqn:kdpreQSS4}) and $e_{\rm rv}$ (Eq.~\eqref{eqn:erv_preQSS2}) are derived from the same non-equilibrium internal energy distribution solution.

A more general class of non-equilibrium dissociation models from the literature are based on the multi-group/ coarse-graining approach outlined by Liu et al.~\cite{Liu2015}. In this approach, individual rovibrational states are binned into a number of ``groups''. Typically, this is done by assuming that the states within a given group follow a Boltzmann distribution about a locally defined internal group temperature. Then, the master equations are solved for the grouped states instead of the entire set of rovibrational states. In the limit where each ``group'' consists of a single rovibrational state, the full master equation/ state-to-state approach is recovered. Conversely, in the limit where all rovibrational states are binned together into one group, the multi-temperature approach is recovered. While this approach has been used successfully in the past to reproduce the dissociation dynamics of full master equation calculations with a fraction of the species~\cite{Magin2012,Munafo2014,Sahai2017,Sahai2019,Venturi2020}, there are a few limitations. Firstly, similar to the multi-temperature approach, the coarse-graining approach requires the solution of additional continuity and internal energy conservation equations for each group of states. Even if relatively few groups are required, the associated computational cost is still larger than the one-temperature approach. Secondly, finding the optimal grouping strategy to capture the correct physics while minimizing computational cost is not a trivial task~\cite{Sahai2017}. This is not a problem for the present formulation, as only bulk species need to be tracked explicitly.

\section{Review of Rate Constants for $\rm \bf H_2$ Dissociation}
\label{sec:review}

While the master equation studies of Kim and Boyd~\cite{Kim2012} and Kim~\cite{Kim2015}/ Kim et al.~\cite{Kim2009} discussed in the previous section provide a good validation of the QSS and pre-QSS theories and framework, they only represent one set of calculations for determining $\rm H_2$ dissociation rate constants. Even detailed master equation studies such as those by Kim and Boyd~\cite{Kim2012} and Kim~\cite{Kim2015}/ Kim et al.~\cite{Kim2009} can have associated uncertainties due to potential inaccuracies in the underlying PESs or QCT calculations. As a consequence, accurate estimates of rate constants should rely on compilations of data from as many sources as possible. To implement the rate constant models proposed in section~\ref{sec:preQSS}, only fits of $k_{\rm d,nr}(T_{\rm t})$ and $\eta(T_{\rm t})$ for each third-body are required. Since $\eta(T_{\rm t})$ can only be computed by fitting directly to master equation results, the expressions presented previously in section~\ref{sec:fracdiss} already represent the best available estimates. However, estimates of $k_{\rm d,nr}(T_{\rm t})$ can be obtained from a much wider variety of sources. Therefore, in this section, rate constants for the dissociation of $\rm H_2$ from both experimental and computational studies are reviewed for each relevant third-body. Then, from this review, new fits of $k_{\rm d,nr}(T_{\rm t})$ that are valid from 200 - 20,000 K are proposed.

The available data sets in the literature can be split into three primary categories: high temperature (2,000 - 8,000 K) rate constants extracted from shock tube experiments (reviewed in section~\ref{sec:highexp}), low temperature ($\lesssim$ 350 K) rate constants extracted from discharge-flow tube experiments (reviewed in section~\ref{sec:lowexp}), and computational studies (reviewed in section~\ref{sec:comp}). A comparison of the rate constants from these sources along with new fits of the rate constants for each third-body are presented in section~\ref{sec:rates}.

\subsection{High Temperature/ Shock Tube Experiments}
\label{sec:highexp}

Table~\ref{tab:highexp} summarizes the relevant shock tube studies in the literature for $\rm H_2$ dissociation. While there were differences in the measurement techniques and measured quantities used to determine the degree of dissociation in these studies, the overall method used to compute rate constants was the same. Namely, measurements were taken behind incident shocks of varying strengths and compositions, for mixtures composed of $\rm H_2$ and a heavy noble gas (i.e., Ar, Xe, or Kr). Then, corresponding inviscid 1-D shock calculations were performed with varying rate constants assuming the linear mixture rule (see Appendix~\ref{sec:mixtures}) until the calculations could reproduce the measured results. For most studies, this was done by assuming that the dissociation rate constants followed the modified Arrhenius form given by
\begin{equation}
    k_{\rm d} = A_{\rm d} T^{n_{\rm d}} \exp \left( - \frac{\theta_{\rm d, H_2}}{T} \right),
\end{equation}
or alternatively
\begin{equation}
    k_{\rm r} = A_{\rm r} T^{n_{\rm r}}
\end{equation}
for recombination rate constants. Here, the scalars $A_{\rm d,r}$ and $n_{\rm d,r}$ are fitting parameters specific to each third-body. Because these experiments were conducted behind strong incident shocks, the $\rm H_2$ dissociation process was dominant over recombination. Nevertheless, rate constants were often still reported in the recombination direction instead of dissociation by assuming $k_{\rm r}=k_{\rm d}/K_{\rm eq}$.

\begin{table*}[hbt!]
\footnotesize
\caption{\label{tab:highexp}Experimental Sources for High Temperature Rate Constants}
\begin{ruledtabular}
\begin{tabular}{lllll}
Source & Measurement Method & M & $T_{\rm t}$ [K] & $P$ [atm] \\\hline
Gardiner and Kistiakowsky, 1961~\cite{Gardiner1961} & X-ray densitometry & $\rm H_2$, H, Xe & 3,000 - 4,500 & 3.6e-1 - 6.2e-1 \\
Sutton, 1962~\cite{Sutton1962}                      & light interferometry & $\rm H_2$\footnotemark[1], H\footnotemark[2], Ar\footnotemark[1] & 2,800 - 4,500 & 2.6e-2 - 1.3e-1 \\
Rink, 1962~\cite{Rink1962}                          & X-ray densitometry &$\rm H_2$, H, Ar, Xe, Kr & 2,800 - 5,000 & 1.2e-2 - 6.7e-2 \\
Patch, 1962~\cite{Patch1962}                        & UV absorption ($\rm H_2$) &$\rm H_2$, H, Ar & 2,950 - 5,330 & 2.5e-3 - 1.1e-1 \\
Jacobs, Geidt, and Cohen, 1967~\cite{Jacobs1967}    & IR emission (trace HCl) &$\rm H_2$, H, Ar & 2,900 - 4,700 & 1.4e0 - 1.8e0 \\
Myerson and Watt, 1968~\cite{Myerson1968}           & VUV absorption (H) &$\rm H_2$, Ar & 2,290 - 3,790 & 2.0e-1 - 1.0e0 \\
Hurle, Jones, and Rosenfeld, 1969~\cite{Hurle1969}  & Na or Cr line-reversal &$\rm H_2$, H\footnotemark[2], Ar & 2,500 - 7,000 & 6.6e-3 - 2.6e-2 \\
Breshears and Bird, 1973~\cite{Breshears1972}       & laser schlieren &$\rm H_2$, H, Ar, Xe & 3,500 - 8,000 & 3.9e-3 - 6.6e-2 \\
\end{tabular}
\end{ruledtabular}
\footnotetext[1]{Rate constants were obtained from the fits by Baulch et al.~\cite{Baulch2005}.}
\footnotetext[2]{Rate constants were obtained via digitizations of the published plots.}
\end{table*}

For all of the shock tube studies, one major source of uncertainty arises from the fact that the rate constants were fitted simultaneously for all three third-bodies ($\rm H_2$, H, and the chosen noble gas). Therefore, an underprediction of a rate constant for one third-body would result in an overprediction of the rate constant for a different third-body. Additionally, due to the limited temperature ranges, Gardiner and Kistiakowsky~\cite{Gardiner1961}, Rink~\cite{Rink1962}, Patch~\cite{Patch1962}, and Jacobs et al.~\cite{Jacobs1967} reported that the temperature exponential term in the Arrhenius fits could not be determined reliably. As a consequence, these four studies all assumed a value of $n_{\rm r} = -1$ for their reported $k_{\rm r}$ fits with all third-bodies. Therefore, while the magnitude of the reported rate constants from these four shock tube studies may be meaningful, the temperature dependencies may not be as accurate across larger temperature ranges. This will be discussed further in the context of other low temperature and computational data sets in section~\ref{sec:rates}.

\begin{figure}[hbt!]
    \centering
    \begin{subfigure}[b]{0.45\textwidth}
        \centering
        \includegraphics[width=\textwidth,trim={0cm 0 0.8cm 0cm},clip]{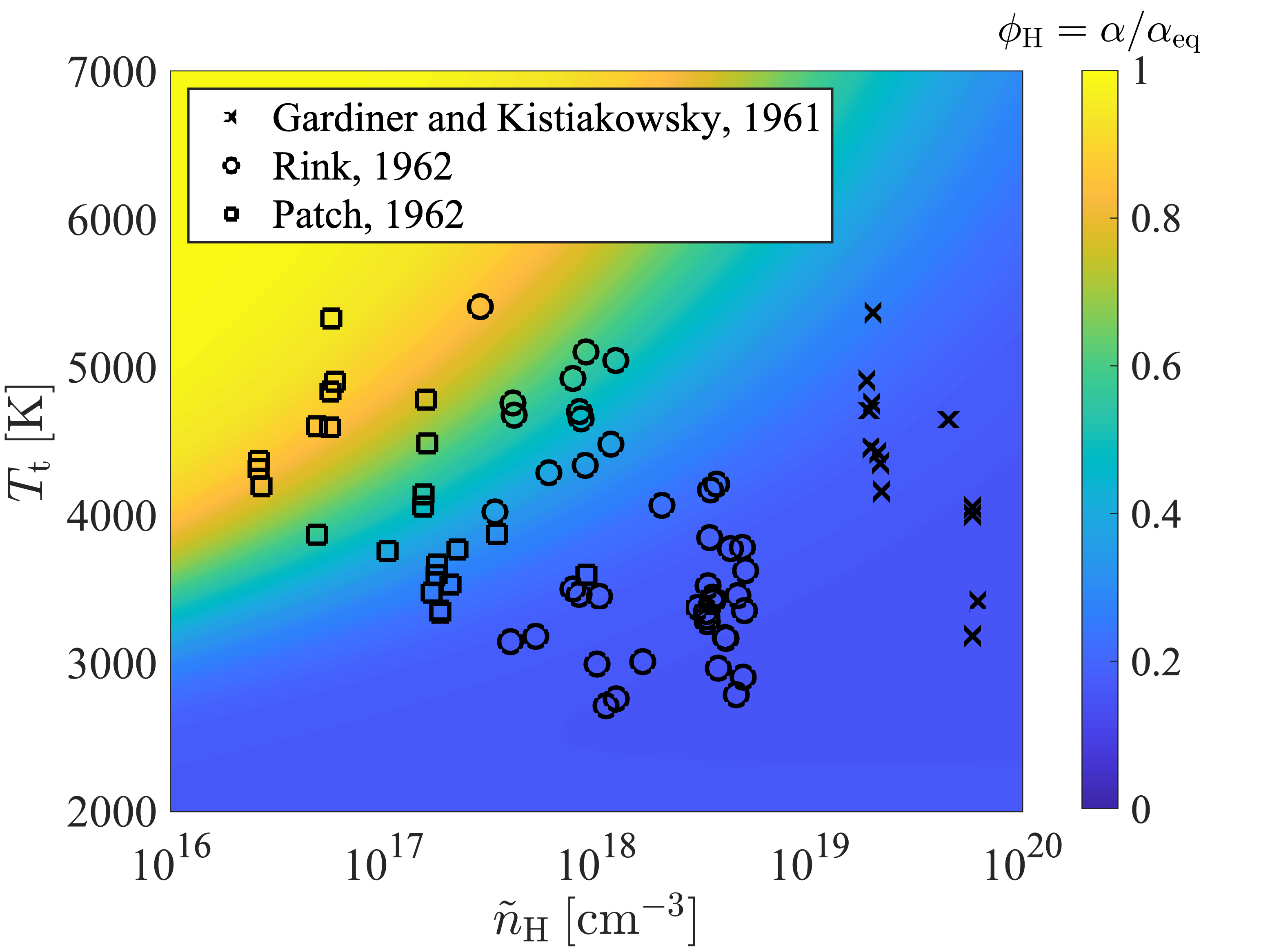}
    \end{subfigure}
    \caption{Contours of the estimated fraction of dissociation, $\phi_{\rm H} = \alpha/ \alpha_{\rm eq}$, that is reached before $k_{\rm d}$ exceeds the $k_{\rm d,nr}$ limit for M = $\rm H_2$.}
    \label{fig:alphacontours}
\end{figure}

Based on the QSS and pre-QSS theories presented earlier in section~\ref{sec:QSS} and \ref{sec:preQSS}, a relevant question is what impact non-equilibrium effects may have had on the reported rate constants from these shock tube studies. Based on Fig.~\ref{fig:fracdiss}, it is clear that pre-QSS effects were likely negligible even at the highest temperatures (7,000 - 8,000 K) investigated in the shock tube studies. However, the estimated fraction of dissociation that occurred in the non-recombining versus recombining regions varies significantly based on the number density conditions. To illustrate this point, Fig.~\ref{fig:alphacontours} shows contours of the estimated fraction of dissociation that occurred in the non-recombining limit, computed using Eq.~\eqref{eqn:phiA2chi_nr} and \eqref{eqn:alphanorm} for the case of M = $\rm H_2$. These limits are computed with $k_{\rm d,nr}$ and $k_{\rm d,th}$ taken from Kim and Boyd~\cite{Kim2012}, and with $\delta$ = 0.25. Overlaid are the frozen $T_{\rm t}$ and $\tilde{n}_{\rm H}$ values from the shot conditions in the studies by Gardiner and Kistiakowsky~\cite{Gardiner1961}, Rink~\cite{Rink1962}, and Patch~\cite{Patch1962}, computed using the CEA code~\cite{Gordon1996}. For the sake of clarity, only the shots from these three studies are plotted, however, the shot conditions from any of the other studies in Table~\ref{tab:highexp} could be plotted similarly. From Fig.~\ref{fig:alphacontours}, it is clear that the fraction of dissociation that occurred in the non-recombining limit varied drastically for different shot conditions. For example, for some of the highest temperature cases by Patch, 95\% of the dissociation is estimated to have occurred in the non-recombining limit; however, for the lowest temperature cases by Gardiner and Kistiakowski, this is just 15\%.

Luckily, in all of the considered shock tube studies, it was assumed that dissociation and recombination rate constants were related via macroscopic detailed balance. As discussed in section~\ref{sec:QSSsrc}, and in particular as given by Eq.~\eqref{eqn:dnA_3}, this implies that the reported rate constant values from the shock tube studies correspond to $k_{\rm d,nr}$ (and $k_{\rm d,nr}/K_{\rm eq}$ where recombination rate constants were reported instead), regardless of the temperature and number density conditions in which the experiments were conducted.

\subsection{Low Temperature/ Discharge-Flow Tube Experiments}
\label{sec:lowexp}
Table~\ref{tab:lowexp} summarizes the discharge-flow tube studies considered in the present work for the determination of rate constants at low temperatures ($T_{\rm t} \lesssim $ 350 K). These experiments relied on a conventional dynamical flow tube setup, wherein atomic hydrogen was first generated, then continually pumped along a tube with measurements taken as H atom concentrations decayed with distance. The only exception to this is the study by Lynch et al.~\cite{Lynch1976}, which instead relied on a static method with H atom decay measured instead as a function of time. It should be noted that there are a number of earlier discharge-flow tube studies in the literature (prior to 1960) that show a wide range of possible recombination rates near $T_{\rm t} \approx $ 300 K with approximately a magnitude of disagreement for the third-bodies M = $\rm H_2$ and Ar. However, the previous reviews by Cohen and Westburg~\cite{Cohen1983} and Baulch et al.~\cite{Baulch2005} suggest that the rate constants reported by these earlier investigations are in doubt, and therefore they are not considered in the present work.

\begin{table*}[hbt!]
\footnotesize
\caption{\label{tab:lowexp}Experimental Sources for High Temperature Rate Constants}
\begin{ruledtabular}
\begin{tabular}{lllll}
Source & Measurement Method & M & $T_{\rm t}$ [K] & $P$ [atm] \\\hline
Larkin, 1968~\cite{Larkin1968} & calorimetric probe & $\rm H_2$ & 291 & 8.4e-3 \\
                               &                    & Ar & 213 - 349  & 6.0e-3 - 7.1e-3 \\
Bennett and Blackmore, 1968~\cite{Bennett1968} & ESR spectroscopy\footnotemark[1] & $\rm H_2$, H\footnotemark[2] & 300 & 1.3e-3 - 1.3e-2 \\
Bennett and Blackmore, 1970~\cite{Bennett1970} & ESR spectroscopy\footnotemark[1] & $\rm H_2$ & 300 & 1.8e-3 - 7.7e-3 \\
Bennett and Blackmore, 1971~\cite{Bennett1971} & ESR spectroscopy\footnotemark[1] & $\rm H_2$, He, Ar & 298 & 6.5e-2 - 3.3e-1 \\
Trainor, Ham, and Kaufman, 1973~\cite{Trainor1973} & calorimetric probe &$\rm H_2$, He, Ar & 77 - 298 & 2.6e-3 - 2.0e-2 \\
Lynch, Schwab, and Michael, 1976~\cite{Lynch1976} & H Lyman-$\alpha$ absorption & $\rm H_2$, He, Ne, Ar, Kr & 298 & 6.5e-1 - 2.0e0 \\
Mitchell and LeRoy, 1977~\cite{Mitchell1977} & capillary flow-meter & He\footnotemark[3] & 296.5 & 7.9e-3 - 2.4e-2  \\
\end{tabular}
\end{ruledtabular}
\footnotetext[1]{Electron Spin Resonance spectroscopy.}
\footnotetext[2]{Only an upper bound value was provided.}
\footnotetext[3]{Both lower and upper bound values were reported.}
\end{table*}

Rate constants are available from the listed sources for $\rm H_2$ and the noble gases He, Ar, and Kr as third-bodies.  Unfortunately, due to low temperature limitations, rates with H as a third-body could not reliably be determined using discharge-flow tube systems. Bennett and Blackmore~\cite{Bennett1968} proposed a room temperature recombination rate for M = H, but only as an upper bound. However, this value has been called into question by Baulch et al.~\cite{Baulch1972}.

As was done implicitly in the previous reviews by Cohen and Westburg~\cite{Cohen1983} and Baulch et al.~\cite{Baulch2005}, it is assumed that at the low temperatures encountered in the discharge-flow tube studies ($T_{\rm t} \lesssim$ 350 K), there is no distribution dependence for the reported rate constants. As a consequence, the recombination rate constants reported in these studies are interpreted directly as estimates of $k_{\rm d,nr}/K_{\rm eq}$. Additionally, it is assumed that recombination is a third-order reaction. The validity of this assumption is discussed in Appendix~\ref{sec:2body}.

\subsection{Computational Studies}
\label{sec:comp}

Table~\ref{tab:comp} summarizes the computational studies considered in the present work. Each of these studies followed a similar methodology to that already discussed previously in section~\ref{sec:kdextract} for the works of Kim and Boyd~\cite{Kim2012} and Kim~\cite{Kim2015}/ Kim et al.~\cite{Kim2009}. Namely, $(J,\nu)$ state-specific relaxation and reaction rates were first computed using QCT calculations on detailed ab-initio PESs. Once converged state-specific rates were obtained, aggregate rate constants were computed. The specific method used to compute these aggregate rate constants differs slightly between the different studies.

\begin{table*}[hbt!]
\footnotesize
\caption{\label{tab:comp}Computational Sources for Rate Constants}
\begin{ruledtabular}
\begin{tabular}{llll}
Source & M & $T_{\rm t}$ [K] & PES \\\hline
Schwenke, 1990~\cite{Schwenke1990} & $\rm H_2$ & 1,000 - 5,000 & Schwenke, 1988~\cite{Schwenke1988} \\
                                   & H         & 3,000 - 5,000 & Schwenke, 1988~\cite{Schwenke1988} \\
Martin, Schwarz, and Mandy, 1996~\cite{Martin1996} & H\footnotemark[1] & 450 - 45,000 & Liu, Siegbahn, Truhlar, and Horowitz, 1978~\cite{Liu1973,Siegbahn1978,Truhlar1978,Truhlar1979}\\
Furudate, Fujita, and Abe, 2006~\cite{Furudate2006} & $\rm H_2$\footnotemark[2]\footnotemark[3] & 5,000 - 50,000 & Schwenke, 1988~\cite{Schwenke1988} \\
Kim, Kwon, and Park, 2009~\cite{Kim2009} & H  & 2,000 - 16,000 & Boothroyd, Keogh, Martin, and Peterson, 1996~\cite{Boothroyd1996} \\
                                         & He & 2,000 - 16,000 & Boothroyd, Martin, and Peterson, 2003~\cite{Boothroyd2003} \\ 
Kim, Kwon, and Park, 2010~\cite{Kim2010} & $\rm H_2$ & 4,000 - 30,000 & Schwenke, 1988~\cite{Schwenke1988} \\
Kim and Boyd, 2012~\cite{Kim2012} & $\rm H_2$\footnotemark[2] & 1,000 - 30,000 & Schwenke, 1988~\cite{Schwenke1988} \\
\end{tabular}
\end{ruledtabular}
\footnotetext[1]{The sum of dissociation via both collision-induced and tunneling contributions is considered in the present work.}
\footnotetext[2]{Rate constants were obtained via digitizations of the published plots.}
\footnotetext[3]{$k_{\rm d,nr}$ was computed using two methods. Only the first (equivalent to Eq.~\eqref{eqn:kdnr}) is considered in the present work.}
\end{table*}

For the studies by Furudate et al.~\cite{Furudate2006}, Kim and Boyd~\cite{Kim2012}, and both studies by Kim et al.~\cite{Kim2009,Kim2010}, aggregate dissociation rate constants were computed by using the same QSS formulation outlined in section~\ref{sec:QSS}. As a consequence, the reported rate constants correspond directly to values of $k_{\rm d,nr}$. For the study by Furudate et al.~\cite{Furudate2006}, two sets of rate constants were reported: one based directly on QCT calculations, and one where the same underlying state-specific rate constants were modified by an arbitrary scaling factor to match the state-specific rate constants reported by Sharma~\cite{Sharma1994}. These were labeled as the ``uncorrected'' and ``corrected'' rate constants, respectively. Both sets of rate constants are presented in section~\ref{sec:rates}.

For the study by Martin et al.~\cite{Martin1996}, recombination terms were not included in their formulation of the master equations, so their ``steady-state'' dissociation rate constants also correspond to $k_{\rm d,nr}$. The authors reported two limiting dissociation rate constants: a ``low density'' rate constant for $n_{\rm H}\lesssim1$ $\rm cm^{-3}$ and a ``high density'' rate constant for $n_{\rm H}\gtrsim10^4$ $\rm cm^{-3}$, with a smooth transition in between. This dependence on $n_{\rm H}$ appeared because radiative transitions were included in their calculations. For $n_{\rm H}\lesssim10^4$ $\rm cm^{-3}$, radiative transitions were non-negligible and served to depopulate the excited ${\rm H_2}(J,\nu)$ state distribution, thereby lowering $k_{\rm d,nr}$. However, the values of $n_{\rm H}$ that are relevant for hypersonic entry calculations are generally several magnitudes greater than $n_{\rm H}\approx10^4$ $\rm cm^{-3}$. Therefore, only the ``high density'' fit of $k_{\rm d,nr}$ is considered in the present work.

Finally, for the study by Schwenke~\cite{Schwenke1990}, a different approach was taken in which isothermal simulations were first performed for a $\rm H_2$/ H mixture using the master equations. Then, using several different approaches, corresponding phenomenological rate constants were extracted and compared. Of the tested methods, the most meaningful was determined to be the one in which $\rm H_2$ and H number densities from the master equation calculations were substituted into the aggregate source term equations, Eq.~\eqref{eqn:dnA2} and \eqref{eqn:dnA}, assuming the linear mixture rule (see Appendix~\ref{sec:mixtures}) and assuming that $k_{\rm d}/k_{\rm r}=K_{\rm eq}$. This method was described as the ``small master equation'' approach. As shown in section~\ref{sec:QSS}, this approach is equivalent to fitting $k_{\rm d,nr}$ values using Eq.~\eqref{eqn:dnA_3}. Because the maximum temperature simulated by Schwenke~\cite{Schwenke1990} was 5,000 K, pre-QSS effects (as estimated by Fig.~\ref{fig:fracdiss}) were negligible. Hence, the rate constants reported by Schwenke can also be interpreted directly as values of $k_{\rm d,nr}$.

Of the four studies that computed rate constants for M = $\rm H_2$, only Kim and Boyd~\cite{Kim2012} considered the complete set of state-to-state rate coefficients for both target and projectile $\rm H_2$ molecules. The other three studies assumed that the internal state distribution of the projectile $\rm H_2$ followed a Boltzmann distribution defined by either by the translational temperature (Schwenke~\cite{Schwenke1990} and Furudate et al.~\cite{Furudate2006}) or a separate non-equilibrium internal temperature (Kim et al.~\cite{Kim2010}). As discussed by Kim and Boyd~\cite{Kim2012}, the reported dissociation rate constants from Kim et al.~\cite{Kim2010} were still similar to those of Kim and Boyd~\cite{Kim2012}.

Outside of the master equation studies discussed above, there are a number of earlier studies in the literature that also relied on master equation calculations to compute $\rm H_2$ dissociation rate constants. However, these earlier studies generally made several more simplifying assumptions. These assumptions include the local equilibrium/ Boltzmann treatment of the rotational mode~\cite{Shui1971_Ar,Shui1971,Shui1973,Roberge1982,Esposito1999}, the calculation of state-specific rate constants using simple scaling laws instead of trajectory calculations~\cite{Dove1972,Lepp1983}, or the explicit treatment of only para-$\rm H_2$ (as opposed to both ortho and para-$\rm H_2$)~\cite{Blais1979,Haug1987,Dove1979,Dove1987}. Because the validity of these assumptions is questionable, these earlier studies are not considered for the review in the present work.

\subsection{Comparison and Fits of Rate Constant Data}
\label{sec:rates}

\begin{figure*}[hbt!]
    \centering
    \begin{subfigure}[b]{0.92\textwidth}
        \centering
        \includegraphics[width=\textwidth,trim={3.5cm 1.2cm 4cm 1cm},clip]{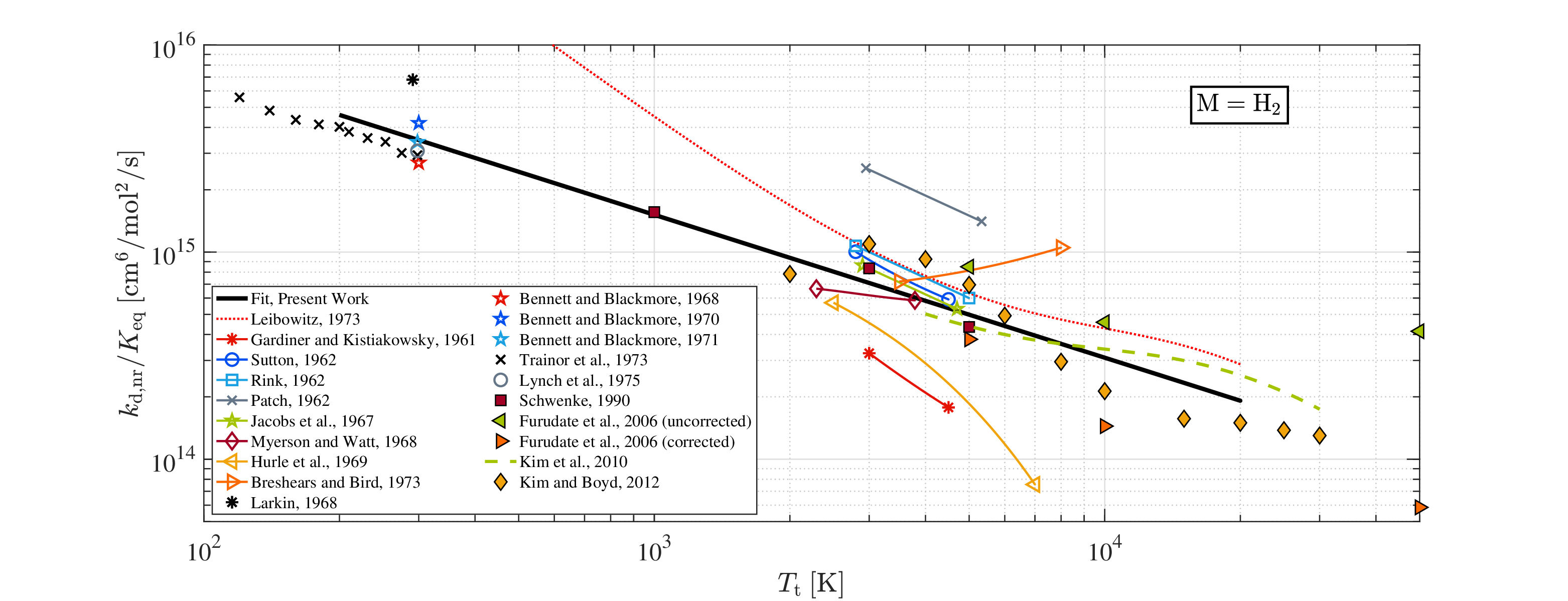}
    \end{subfigure}
    \begin{subfigure}[b]{0.92\textwidth}
        \centering
        \includegraphics[width=\textwidth,trim={3.5cm 1.2cm 4cm 1cm},clip]{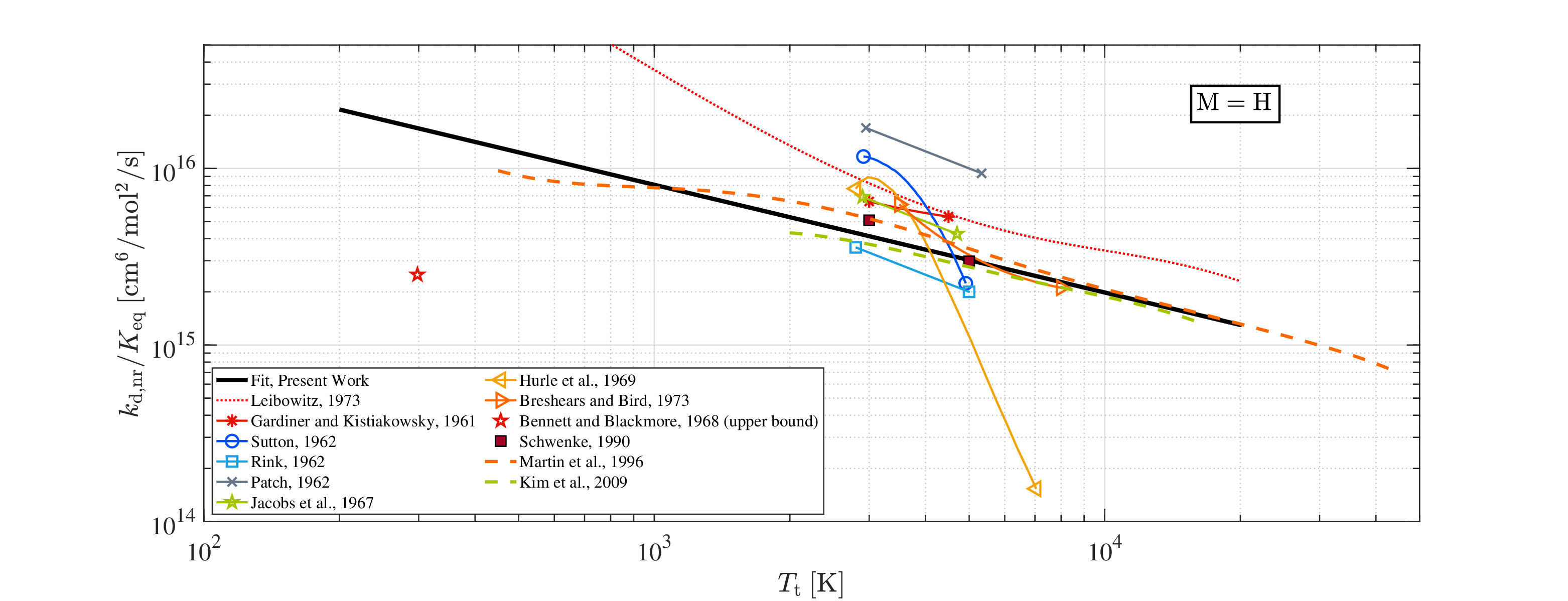}
    \end{subfigure}
    \begin{subfigure}[b]{0.92\textwidth}
        \centering
        \includegraphics[width=\textwidth,trim={3.5cm 0 4cm 1cm},clip]{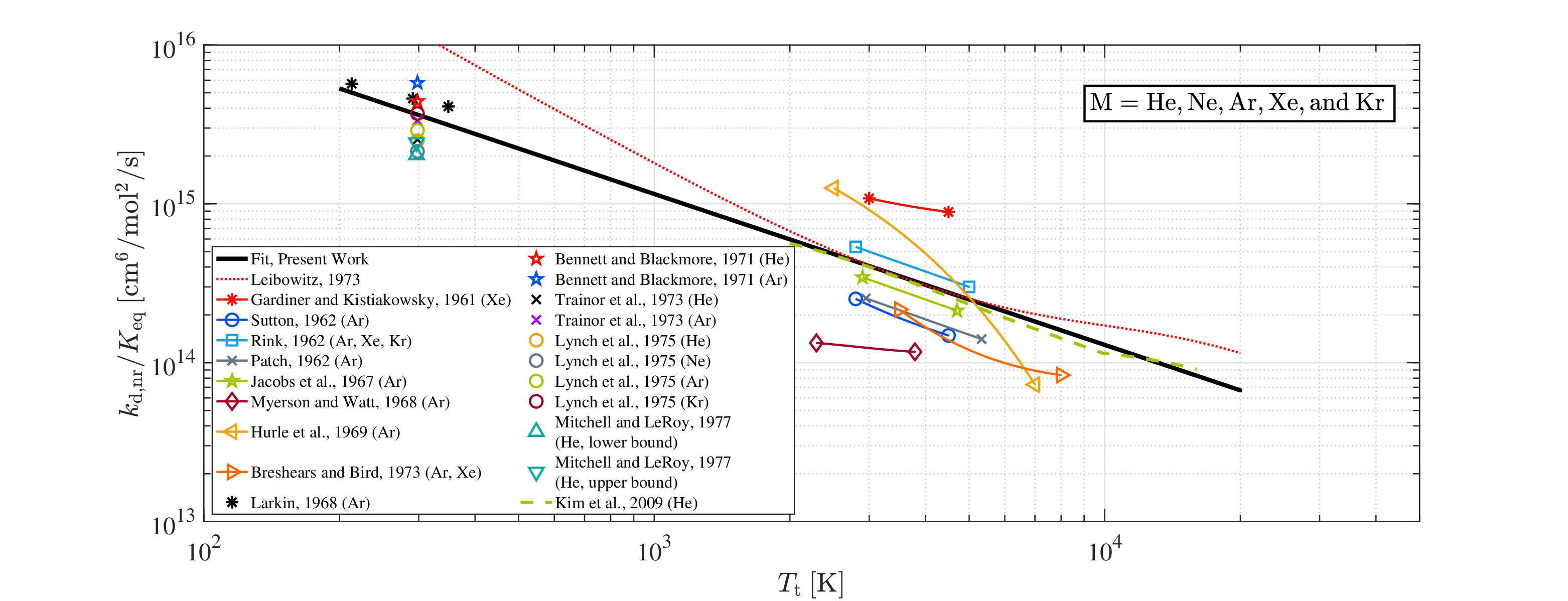}
    \end{subfigure}
    \caption{Review of rate constant data for M = $\rm H_2$ (top), M = H (middle), and M = \{noble gas\} (bottom). Open symbols (including x's and *'s) correspond to discharge-flow tube studies, solid lines with open symbols correspond to shock tube studies, and both closed symbols and dashed lines correspond to computational studies. Solid black lines and dotted red lines correspond to the rate constant fits of the present work and Leibowitz~\cite{Leibowitz1973_1}, respectively.}
    \label{fig:kdnrKeq}
\end{figure*}

Figure~\ref{fig:kdnrKeq} shows the values of $k_{\rm d,nr}/K_{\rm eq}$ from all of the sources discussed in the previous sections. The rate constant data is plotted as $k_{\rm d,nr}/K_{\rm eq}$ instead of $k_{\rm d,nr}$, as this significantly reduces the total change in magnitude of the plotted values between the low and high temperature regimes, making comparisons between different data sets easier. However, these plots should not be mistaken as plots of $k_{\rm r}$, as $k_{\rm d,nr}/K_{\rm eq}\neq k_{\rm r}$ (see section~\ref{sec:QSS}). Additionally, $k_{\rm d,nr}/K_{\rm eq}$ values are only plotted as functions of $T_{\rm t}$, as both $k_{\rm d,nr}$ and $K_{\rm eq}$ are functions of $T_{\rm t}$ alone. For the cases with a noble gas as a third-body, i.e., M = He, Ne, Ar, Xe, and Kr, the rate constants have all been plotted together. While there have been some studies in the literature~\cite{Shui1971,Whitlock1974} that have indicated that the rate constants for $\rm H_2$ dissociation are not identical between different noble gas third-bodies (especially between He and the heavier noble gases), the scatter of the rate constant data is seemingly larger than any of these differences. Therefore, it is assumed that any differences in dissociation rate constants for varying noble gas third-bodies are negligible.

There are two key observations that can be made about the rate constant data shown in Fig.~\ref{fig:kdnrKeq}. Firstly, while there is considerable scatter in both the discharge-flow tube and shock tube data, the computational data from the master equation studies are generally in good agreement with each other and roughly lie in the middle of the scatter of the experimental data. The only exception to this is the computational data of Furudate et al.~\cite{Furudate2006} (M = $\rm H_2$), where there is disagreement between even the ``corrected'' and ``uncorrected'' versions of their rate constants. These two rate constants roughly bracket the rate constants from the other computational studies.

A second observation is that the temperature dependencies of most of the shock tube data are similar to that of the computational data between 2,000 and 5,000 K. Recall that Gardiner and Kistiakowsky~\cite{Gardiner1961}, Rink~\cite{Rink1962}, Patch~\cite{Patch1962}, and Jacobs et al.~\cite{Jacobs1967} assumed a value of $n_{\rm r} = -1$ for their reported $k_{\rm r}$ fits, as discussed previously in section~\ref{sec:highexp}. The other shock tube studies did not however, and report rate constants with similar magnitudes and trends. There are a couple of notable exceptions. For M = $\rm H_2$, Breshears and Bird~\cite{Breshears1972} report a rate constant that increases with temperature, which is inconsistent with all of the other studies. For M = H, Hurle et al.~\cite{Hurle1969} argued that the extreme temperature dependence in their data (and that of Sutton~\cite{Sutton1962}), along with the low-temperature, ``upper bound'' estimate by Bennett and Blackmore~\cite{Bennett1968}, suggested a non-monotonic trend in the M = H rate constant with a maximum value between approximately 2,000 and 3,000 K. This argument is not supported by the computational results of Martin et al.~\cite{Martin1996} and Kim et al.~\cite{Kim2009}, which predict monotonically decreasing trends with temperature. Shui~\cite{Shui1973} suggested that the high-temperature results of Hurle et al.~\cite{Hurle1969} may be in error due to the reliance of their measurements on the assumption that the vibrational mode of $\rm H_2$ was in thermal equilibrium with the translational mode. From section~\ref{sec:kdextract}, it is clear that for the higher temperatures in the experiments by Hurle et al.~\cite{Hurle1969} (approaching 7,000 K), an assumption of thermal equilibrium would be invalid in the non-recombining QSS limit where $T_{\rm v}$ < $T_{\rm t}$. Ultimately, this suggests that due to the lack of reliable low-temperature data for M = H, the temperature dependence of the M = H fit must be determined almost entirely from the computational studies of Martin et al.~\cite{Martin1996} and Kim et al.~\cite{Kim2009}.

Based on the results of the review, new rate constant fits are proposed and plotted along with the rate constants of Leibowitz~\cite{Leibowitz1973_1} in Fig.~\ref{fig:kdnrKeq}. The Leibowitz rate constants are the most widely used in the ice and gas giant entry flow literature~\cite{Palmer2014,Higdon2018,Liu2021,Hansson2021,Carroll2023_conv,Carroll2023,Coelho2023,Rataczak2024}, and are based on an extrapolation of the rate constants of Jacobs et al.~\cite{Jacobs1967}. A more detailed discussion of the Leibowitz rate constants is provided in Appendix~\ref{sec:Leib}. The rate constant fits were chosen to lie in the middle of the scatter of both the low-temperature and high-temperature experimental data where available, while still following the larger temperature dependencies predicted by the computational studies. The proposed rate constant fits are summarized in Table~\ref{tab:fits}. These fits are for $k_{\rm d,nr}/K_{\rm eq}$ instead of $k_{\rm d,nr}$, as this was found to provide a better fit to the rate constant data. The M = H fit below 450 K and the M = \{noble gas\} fit above 16,000 K should be used with some caution, as the fits in these regions are extrapolations that go beyond the temperature ranges of the reviewed data.

\begin{table}[hbt!]
\footnotesize
\caption{\label{tab:fits}Fits of $k_{\rm d,nr}$/$ K_{\rm eq}$}
\begin{tabular}{ll}
\hline \hline
M & $k_{\rm d,nr}/K_{\rm eq}$ $\rm [cm^6/mol^2/s] $\\\hline
$\rm H_2$          & $1.779\times10^{17}T_{\rm t}^{-0.69}$ \\
H                  & $5.461\times10^{17}T_{\rm t}^{-0.61}$ \\
He, Ne, Ar, Xe, Kr & $8.168\times10^{17}T_{\rm t}^{-0.95}$ \\
\hline \hline
\end{tabular}
\end{table}

\begin{figure*}[hbt!]
    \centering
    \begin{subfigure}[b]{0.92\textwidth}
        \centering
        \includegraphics[width=\textwidth,trim={4cm 1.2cm 4cm 0.5cm},clip]{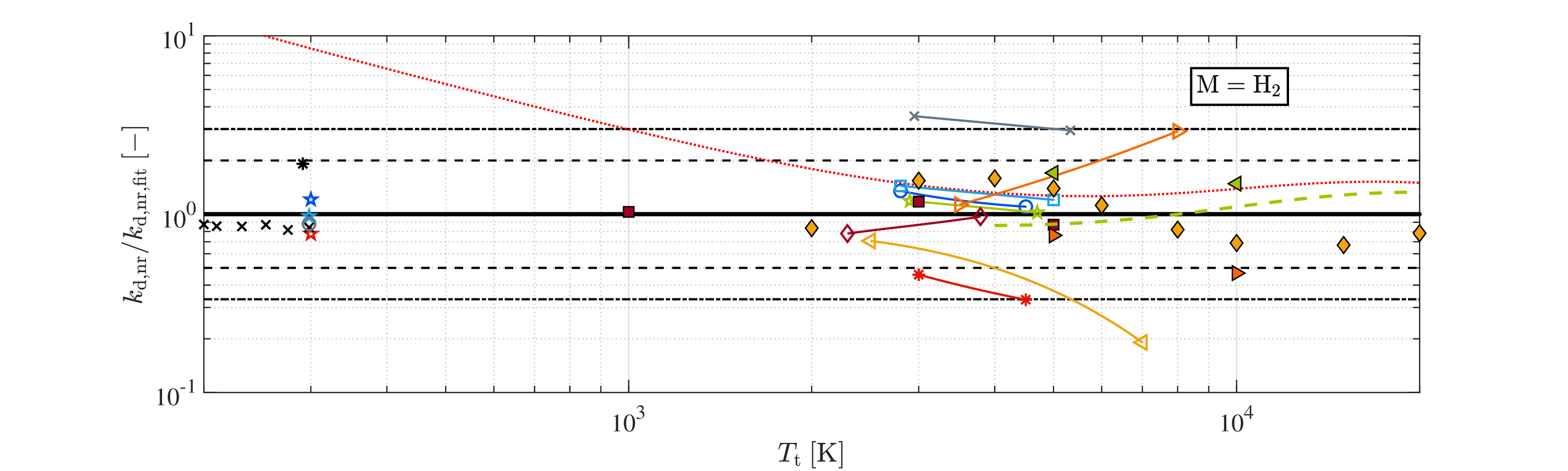}
    \end{subfigure}
    \begin{subfigure}[b]{0.92\textwidth}
        \centering
        \includegraphics[width=\textwidth,trim={4cm 1.2cm 4cm 0.5cm},clip]{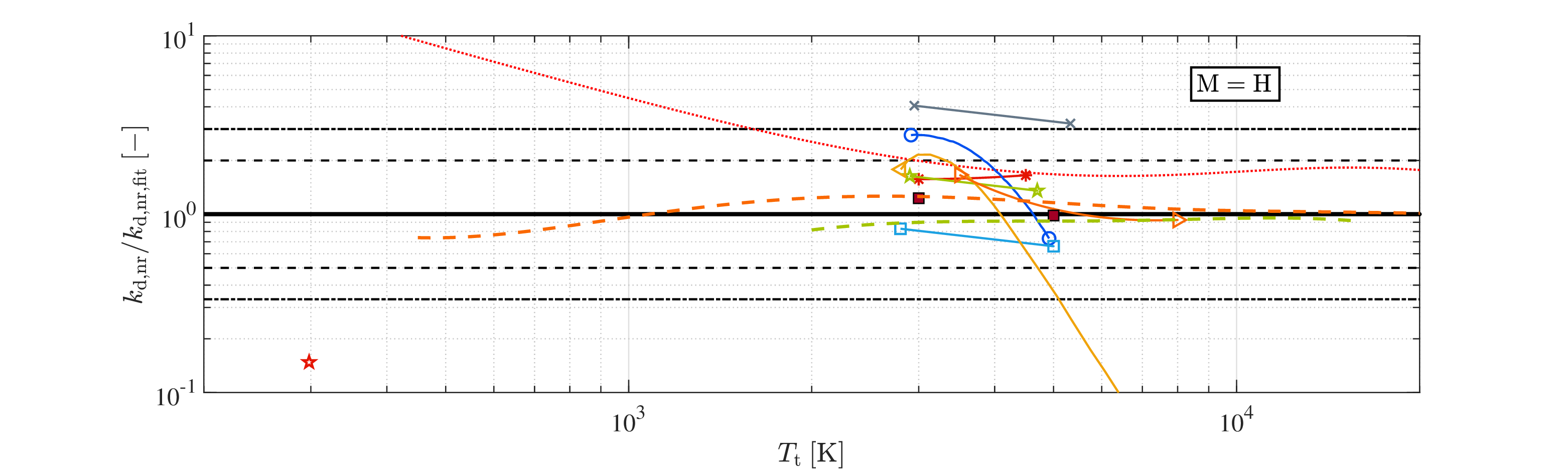}
    \end{subfigure}
    \begin{subfigure}[b]{0.92\textwidth}
        \centering
        \includegraphics[width=\textwidth,trim={4cm 0 4cm 0.5cm},clip]{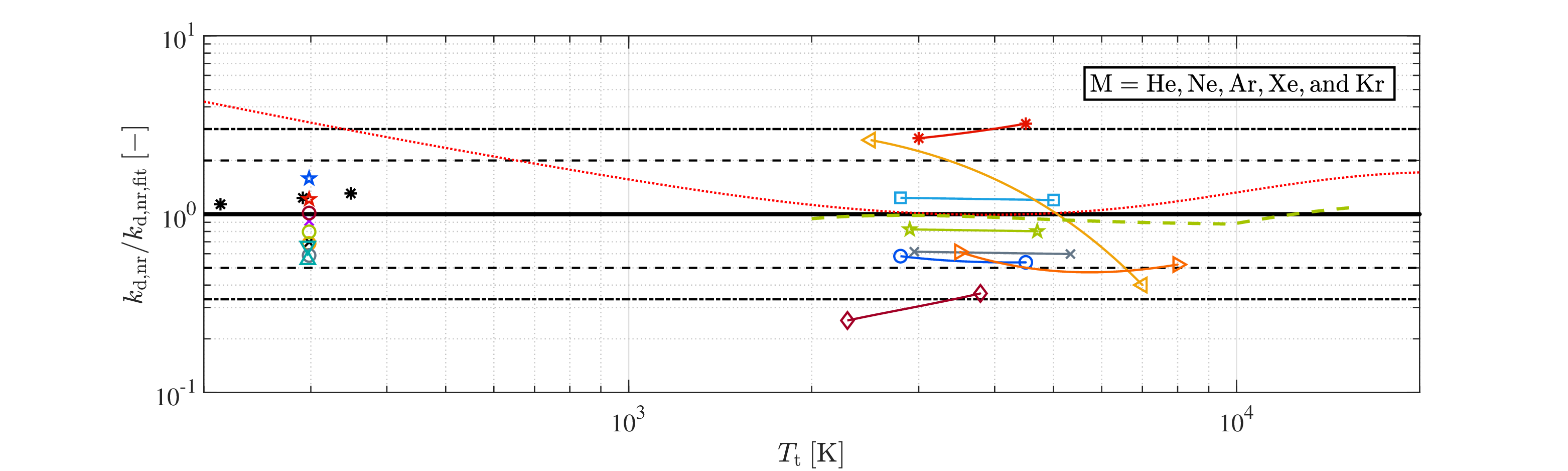}
    \end{subfigure}
    \caption{Rate constant data for M = $\rm H_2$ (top), M = H (middle), and M = \{noble gas\} (bottom) normalized by the fits of the present work. The dashed and dash-dotted black lines correspond to the fits scaled by a factor of two and three, respectively. For all other lines and symbols, the legends are the same as Fig.~\ref{fig:kdnrKeq}.}
    \label{fig:kdnrKeqnorm}
\end{figure*}

Figure~\ref{fig:kdnrKeqnorm} shows the same rate constant data as Fig.~\ref{fig:kdnrKeq}, but now normalized by the fits of the present work. The majority of the reviewed data falls within a factor of two of the proposed fits, and in general, the fits reproduce the reviewed data more accurately than the rate constants of Leibowitz~\cite{Leibowitz1973_1}. This is especially true in the lower temperature ranges, where the Leibowitz rate constants severely overpredict the reviewed data. This overprediction is not surprising, as the shock tube study of Jacobs et al.~\cite{Jacobs1967} (upon which the Leibowitz rate constants are based) only measured rates to a minimum temperature of 2,900 K. The aerothermal heating uncertainty study for Saturn and Uranus entry probes by Palmer et al.~\cite{Palmer2014} assumed an uncertainty factor of $\pm$ 1 order of magnitude on the rate constants for $\rm H_2$ dissociation with all third-bodies. The more recent uncertainty quantification study for ice giant aerocapture applications by Rataczak et al.~\cite{Rataczak2024} instead assumed uncertainty bounds of 0.25 and 3 times the baseline rate constants. Both studies used the rate constants of Leibowitz~\cite{Leibowitz1973_1} as a baseline. The results of Fig.~\ref{fig:kdnrKeq} highlight that, with the fits of the present work, an uncertainty factor of two is more appropriate than the wider uncertainty bounds used in either of these previous uncertainty quantification studies.

Figure~\ref{fig:kdnrKeqnorm} also shows that none of the rate constants from any one of the shock tube studies consistently over or underpredicts the fits for all three third-bodies. For example, in the case of Patch~\cite{Patch1962}, the reported rate constants overpredict the fits for the M = $\rm H_2$ and H cases, but underpredict the fit for the M = \{noble gas\} case. The only exception are the results of Myerson and Watt~\cite{Myerson1968}, but only because they did not report a rate constant for the M = H case. This result suggests that while the total dissociation rates measured by the shock tube studies may be accurate, the reported third-body efficiencies may not be, possibly due to the simultaneous third-body fitting procedure that was used in all of the shock tube studies.

\section{Conclusion}
\label{sec:conclusions}

In this work, a detailed description of the non-equilibrium dissociation of diatomic molecules was presented, then applied to the case of $\rm H_2$ dissociation. The master equation formulation was used to analyze general non-equilibrium rate constant expressions in the thermal equilibrium limit, the QSS regime, and the pre-QSS regime. In the QSS regime (up to and including thermal equilibrium), it was found that the chemical source term could be described entirely as a function of the steady dissociation rate constant in the absence of recombination, $k_{\rm d,nr}$, which in turn is only a function of the translational temperature, $T_{\rm t}$. To account for pre-QSS effects, a simple modification of the QSS rate constant expression was proposed through the use of an additional rate constant term, $\eta(T_{\rm t})$. The complete expression, which is ultimately only a function of $T_{\rm t}$ and the fraction of dissociation, $\phi_{\rm H}$, was able to reproduce the results of detailed master equation simulations of a 0-D isothermal and isochoric reactor for the case of $\rm H_2$ dissociation with the third-bodies $\rm H_2$, H, and He. Finally, an extensive review of $k_{\rm d,nr}(T_{\rm t})$ for $\rm H_2$ dissociation was performed by considering both experimental and computational data sources. From this review, fits of $k_{\rm d,nr}/K_{\rm eq}$ that could describe the majority of the reviewed rate constants from the literature within a factor of two were proposed for each relevant third-body.

Practically, if pre-QSS effects can be neglected, the use of the QSS source term expression of Eq.~\eqref{eqn:dnA_3} along with the fits of $k_{\rm d,nr}/K_{\rm eq}$ from Table~\ref{tab:fits} are recommended. By implementing the fits of $k_{\rm d,nr}/K_{\rm eq}$ as ``recombination'' rate constants, the dissociation rate constants, $k_{\rm d,nr}$, can be computed by applying macroscopic detailed balance, where $K_{\rm eq}$ can be computed readily using the NASA9 polynomial fits of McBride et al.~\cite{McBride2002}. If pre-QSS effects are non-negligible, the same approach can be applied, but instead through the use of the pre-QSS corrected source term (Eq.~\eqref{eqn:dnA_preQSS}) and rate constant (Eq.~\eqref{eqn:kdpreQSS4} with $\epsilon=10^{-3}$ and the fits of $\eta(T_{\rm t})$ from section~\ref{sec:fracdiss}) expressions.

There are two caveats that should be considered when using the rate constant expressions proposed in the present work. First, for the simulation of mixtures, the analysis presented is only strictly valid when the dissociation of the diatomic molecule of interest, e.g., $\rm H_2$, occurs in an infinitely dilute bath composed solely of one third-body. However, in most practical CFD simulations and shock tube experiments, dissociation does not occur in an infinitely dilute bath, but instead occurs in a multi-component mixture. If the linear mixture rule is assumed such that the chemical source terms can be computed as just a linear sum over each of the third-bodies in the mixture, then, the same expressions proposed in this work can be used directly. This is discussed in more detail in Appendix~\ref{sec:mixtures}.

Separately, it is not clear if the proposed expressions are sufficient for the accurate simulation of recombination-dominated flows at high temperatures. As highlighted in the recent study by Macdonald~\cite{Macdonald2024}, recombining flows are characterized by internal state distributions with an overpopulation of excited states (relative to a Boltzmann distribution). This is the opposite of the QSS behavior in dissociation-dominated flows, which is characterized by an underpopulation of excited states~\cite{Singh2017,Singh2020}. As implied by Eq.~\eqref{eqn:A2kd_sum}, this difference in rovibrational distributions will lead to different behaviors of the aggregate dissociation rate constant, $k_{\rm d}$. This means that even if the recombination rate constant is always given by $k_{\rm r}=k_{\rm d,th}/K_{\rm eq}$, the chemical source term will not be accurately predicted without considering the impact of this difference in dissociation rates. The further analysis of this topic will be the subject of future work.

\begin{acknowledgments}
This work is supported by the NASA entry systems modeling project under grant number 80NSSC21K1751.
\end{acknowledgments}

\section*{Author Declarations}
\subsection*{Conflict of Interest}
The authors have no conflicts to disclose.

\subsection*{Author Contributions}
{\bf Alex T. Carroll:} Conceptualization (equal), Methodology (equal), Data curation (lead), Formal analysis (lead), Investigation (lead), Software (lead), Validation (lead), Visualization (lead), Writing - original draft (lead), Writing - review and editing (equal).
{\bf Jacob Wolmer:} Data curation (supporting), Investigation (supporting), Visualization (supporting).
{\bf Guillaume Blanquart:} Conceptualization (equal), Methodology (equal), Formal analysis (supporting), Writing - review and editing (equal), Funding acquisition (lead), Supervision (lead).
{\bf Aaron M. Brandis:} Writing - review and editing (supporting), Supervision (supporting).
{\bf Brett A. Cruden:} Writing - review and editing (supporting), Supervision (supporting).

\section*{Data Availability Statement}
The data that support the findings of this study are available from the corresponding author upon reasonable request.

\appendix

\section{Dissociation in Multi-Component Mixtures}
\label{sec:mixtures}

Throughout the present work, only dissociation that occurs in a single-component mixture, i.e., dissociation in an infinitely dilute bath composed solely of one third-body, has been considered. However, for most practical simulations and experiments of interest, it is necessary to consider dissociation that occurs in a multi-component mixture. Considering again the general diatom $\rm A_2$, for a mixture, the source term for A can be written as 
\begin{equation}
    \frac{d n_{\rm A}}{dt} = \sum_{\rm M} \left[ 2n_{\rm M} n_{\rm A_2} \sum_{\nu=0}^{\nu_{\rm max}} \sum_{J=0}^{J_{\rm max}(\nu)} \frac{n_{\rm A_2 \it (J,\nu)}}{n_{\rm A_2}} k_{\rm M}(J,\nu \rightarrow c) - 2 n_{\rm M} n_{\rm A}^2\sum_{\nu=0}^{\nu_{\rm max}} \sum_{J=0}^{J_{\rm max}(\nu)} k_{\rm M}(c \rightarrow J,\nu) \right],
    \label{eqn:masterA_mix}
\end{equation}
where $k_{\rm M}(J,\nu \rightarrow c)$ and $k_{\rm M}(c \rightarrow J,\nu)$ are the state-specific dissociation and recombination rate constants with the third-body M. Because the recombination term in Eq.~\eqref{eqn:masterA_mix} is not a function of the internal state distribution of $\rm A_2$, it is the sum of the recombination terms of all third-bodies considered independently. This is not true for the dissociation term in Eq.~\eqref{eqn:masterA_mix}, as $n_{{\rm A_2}(J,\nu)}$ is now determined through collisions with all third-bodies in the mixture. Stated differently, the mixture-dependent $n_{{\rm A_2}(J,\nu)}$ in Eq.~\eqref{eqn:masterA_mix} is not necessarily the same as the distributions obtained in the single-component cases, denoted here as $n_{{\rm A_2}(J,\nu),{\rm M}}$.

The most common method for relating $n_{{\rm A_2}(J,\nu)}$ to $n_{{\rm A_2}(J,\nu),{\rm M}}$ is through the use of the linear mixture rule. In the linear mixture rule, it is assumed that the source term for A can be computed as
\begin{equation}
    \frac{d n_{\rm A}}{dt} = \sum_{\rm M} \frac{d n_{\rm A}}{dt} \biggr|_{\rm M} = \sum_{\rm M} \left[ 2n_{\rm M} n_{\rm A_2} k_{\rm d,M} - 2 n_{\rm M} n_{\rm A}^2 k_{\rm r,M} \right].
    \label{eqn:LMR}
\end{equation}
Here, $d n_{\rm A}/dt|_{\rm M}$, $k_{\rm d,M}$, and $k_{\rm r,M}$ are the source term and aggregate dissociation and recombination rate constants for the single-component cases. To get Eq.~\eqref{eqn:LMR} from the general expression of Eq.~\eqref{eqn:masterA_mix}, it is sufficient to assume that $n_{{\rm A_2}(J,\nu)}$ is related to $n_{{\rm A_2}(J,\nu),{\rm M}}$ by
\begin{equation}
    n_{{\rm A_2}(J,\nu)} = \frac{\sum_{\rm M} n_{{\rm A_2}(J,\nu){\rm ,M}} n_{\rm M} k_{\rm M}(J,\nu \rightarrow c)}{\sum_{\rm M} n_{\rm M} k_{\rm M}(J,\nu \rightarrow c)}.
    \label{eqn:LMR_jv}
\end{equation}
In other words, $n_{{\rm A_2}(J,\nu)}$ is a linear combination of $n_{{\rm A_2}(J,\nu),{\rm M}}$ weighted by the state-specific branching ratios.

As discussed by Boyd~\cite{Boyd1977}, Dove et al.~\cite{Dove1984}, and Carruthers and Teitelbaum~\cite{Carruthers1988}, the range of validity of the linear mixture rule is an open question. Unfortunately, to the authors' knowledge, there are currently no detailed/ master equation simulation data available for $\rm H_2$/ H/ He mixtures. Therefore, the recommended practice remains to use Eq.~\eqref{eqn:LMR} for the simulation of mixtures.

\section{Recombination as a Second or Third-Order Process}
\label{sec:2body}

Throughout this work, it has been implicitly assumed that recombination is a third-order process. This assumption has been made in the master equations and the subsequent QSS formulation presented in sections~\ref{sec:noneq} and \ref{sec:applicationH2}, as well as in the interpretation of rate constant data from the literature for $\rm H_2$ dissociation throughout section~\ref{sec:review}. In particular, it has been assumed that recombination occurs as the direct inverse process of dissociation, and hence that recombination rates can be characterized entirely by applying the micro-reversibility relation of Eq.~\eqref{eqn:A2microdiss}.

Under certain pressure and temperature conditions however, recombination can occur primarily as a second-order process. This can be seen by considering recombination that occurs through a series of two-body collisions, as opposed to a ``direct'' three-body collision. As reviewed by Mirahmadi and Pérez-Ríos~\cite{Mirahmadi2022}, this ``indirect'' approach treats recombination in two steps. First, two bodies collide to form an intermediate complex,
\begin{equation}
    \rm A + A \leftrightarrow A_2^*.
    \label{eqn:ET_1}
\end{equation}
Then, this intermediate complex is stabilized into a bound state/ recombined molecule via a collision with a third-body,
\begin{equation}
    \rm A_2^* + M \leftrightarrow A_2 + M.
    \label{eqn:ET_2}
\end{equation}
Here, $\rm A_2$ is a generic diatom as before, and $\rm A_2^*$ is an intermediate complex/ quasibound state. These quasibound states are bound classically, but have a finite lifetime quantum mechanically due to tunneling. If the backwards direction of reaction~\eqref{eqn:ET_1} is dominant, recombination will occur primarily as a third-order process. Conversely, if the forwards direction of reaction~\eqref{eqn:ET_2} is dominant, recombination will occur primarily as a second-order process.

In detailed chemical models, the pressure dependence of a recombination reaction rate is often modeled with the Lindemann form~\cite{Lindemann1922} or the more general Troe form~\cite{Gilbert1983}.
In the Lindemann expression, low-pressure and high-pressure rate constants, $k_{\rm 0}$ and $k_{\infty}$ respectively, are first defined. Then, the rate constant at any intermediate pressure is computed as
\begin{equation}
    k = \frac{k_0 n_{\rm M}}{P_{\rm r}+1},
    \label{eqn:Lind}
\end{equation}
where
\begin{equation}
    P_{\rm r} \equiv \frac{k_{0}n_{\rm M}}{k_{\infty}}
    \label{eqn:Pr}
\end{equation}
is the reduced pressure. For an infinitely dilute mixture with the third-body, M, $n_{\rm M}=P/(k_{\rm B}T_{\rm t})$. For the recombination of a diatom, the high-pressure and low-pressure limits correspond to the limits dominated by two-body and three-body collisions, respectively. For $\rm H_2$, $k_{\rm 0}$ can be estimated by using the rate constant fits from section~\ref{sec:rates}. $k_{\infty}$ on the other hand can be estimated by using the upper-bound value of the bimolecular collision rate for homonuclear collisions,
\begin{equation}
    k_{\infty} \approx \frac{1}{2}\sigma_{\rm H_2}\sqrt{\frac{8k_{\rm B}T_{\rm t}}{\pi \mu}},
\end{equation}
where $\sigma_{\rm H_2}=\pi d_{\rm H}^2$ is the hard-sphere collision cross section with $d_{\rm H}\approx 3 \rm \mathring{A}$~\cite{Jasper2014}, and $\mu=m_{\rm H}/2=8.37\times10^{-28}$ kg is the reduced mass. Of all of the discharge-flow tube studies considered in section~\ref{sec:lowexp}, the experiments by Lynch et al.~\cite{Lynch1976} at $T_{\rm t}=$ 298 K and $P=2$ atm exhibit the largest values of $P_{\rm r}$ and hence the furthest departures from the low-pressure limit. However, even at this condition, $P_{\rm r}$ as estimated by Eq.~\eqref{eqn:Pr} is still approximately $10^{-3}$. Therefore, from Eq.~\eqref{eqn:Lind}, $k\approx k_0 n_{\rm M}$, and the choice to interpret the rate constants from the discharge-flow tube studies as third-order is appropriate.

Outside of the discharge-flow tube studies, there have been a number of computational studies~\cite{Pack1972,Whitlock1972,Whitlock1974,Orel1987,Schwenke1988,Esposito2009,Forrey2013} that have directly computed the third-order recombination rate constant of $\rm H_2$ by using the resonance complex theory of Roberts et al.~\cite{Roberts1969}. In this theory, it is assumed that recombination occurs exclusively through the indirect pathway. Then, a small number of select quasibound states are considered to be in equilibrium with the continuum (i.e., reaction~\eqref{eqn:ET_1}), and the downward flux of these quasibound states (i.e., reaction~\eqref{eqn:ET_2}) are assumed to be predominantly responsible for the recombination rate. Unfortunately, the accuracy of the phenomenological PESs used in several of these studies is questionable, and as pointed out by Schwenke~\cite{Schwenke1988,Schwenke1990}, the reported rate constants from these studies are not always consistent with the high-temperature experimental and computational rate constants reviewed in sections~\ref{sec:highexp} and \ref{sec:comp}, respectively. For these reasons, the studies based on the resonance complex theory are not considered for the review in the present work.

\section{Rate Constants of Leibowitz}
\label{sec:Leib}

Three different sets of $\rm H_2$ dissociation rate constants have been proposed by Leibowitz, namely the models in Leibowitz~\cite{Leibowitz1973_1}, Leibowitz et al.~\cite{Leibowitz1973_2}, and Leibowitz and Kuo~\cite{Leibowitz1976}. The reported rate constants are summarized below in Table~\ref{tab:leib}. In all three versions, the rate constants for the third-bodies M = ${\rm H_2}$ and H are evaluated from the M = He case using a constant multiplicative factor.

\begin{table*}[hbt!]
\footnotesize
\caption{\label{tab:leib}Rate Constants of Leibowitz in $\rm cm^3/mol/s$}
\begin{ruledtabular}
\begin{tabular}{llll}
    & M = He & M = $\rm H_2$ & M = H \\\hline
Leibowitz, 1973~\cite{Leibowitz1973_1} & $4.17\times10^{18}T_{\rm t}^{-1}\exp(-51952/T_{\rm t})$\footnotemark[1] & $2.5k_{\rm d,M=He}$ & $20.0k_{\rm d,M=He}$ \\
Leibowitz et al. 1973~\cite{Leibowitz1973_2} & $1.1\times10^{18}T_{\rm t}^{-1}[1-\exp(-1.5\times10^8/T_{\rm t}^2)]\exp(-52340/T_{\rm t})$\footnotemark[2] & $2.5k_{\rm d,M=He}$ & $14.0k_{\rm d,M=He}$ \\
Leibowitz and Kuo, 1976~\cite{Leibowitz1976} & $4.3\times10^{18}T_{\rm t}^{-1}[1-\exp(-1.5\times10^8/T_{\rm t}^2)]\exp(-52340/T_{\rm t})$\footnotemark[2]\footnotemark[3] & $2.5k_{\rm d,M=He}$ & $14.0k_{\rm d,M=He}$ \\
\end{tabular}
\end{ruledtabular}
\footnotetext[1]{The term inside the exponential is mistakenly reported as a positive value in Leibowitz~\cite{Leibowitz1973_1}.}
\footnotetext[2]{The leading $T_{\rm t}^{-1}$ term is mistakenly reported as just $T_{\rm t}$ in both Leibowitz et al.~\cite{Leibowitz1973_2} and Leibowitz and Kuo~\cite{Leibowitz1976}.}
\footnotetext[3]{The term $[1-\exp(-1.5\times10^8/T_{\rm t}^2)]$ is mistakenly reported as $[1-\exp(1.5\times10^8/T_{\rm t}^2)]$ in Leibowitz and Kuo~\cite{Leibowitz1976}.}
\end{table*}

It is important to note that the three different sets of rate constants are not based on the same source of rate constant data. For the first set, Leibowitz~\cite{Leibowitz1973_1} cites the shock tube study of Jacobs et al.~\cite{Jacobs1967}. For the second set, Leibowitz et al.~\cite{Leibowitz1973_2} cites the shock tube study of Hurle et al.~\cite{Hurle1969}, adding that the rate constants were selected to be ``consistent with shock tube measurements below 7,000 K, reach a maximum at 20,000 K, and then decay with increasing temperature''. Finally, for the third set by Leibowitz and Kuo~\cite{Leibowitz1976}, the same temperature dependence and third-body efficiencies are used as in the second set by Leibowitz et al.~\cite{Leibowitz1973_2}, but the rate constants are all increased by a factor of four. Leibowitz and Kuo~\cite{Leibowitz1976} do not provide an explanation for this change, and simply cite the two previous sets of rate constants by Leibowitz~\cite{Leibowitz1973_1} and Leibowitz and Kuo~\cite{Leibowitz1973_2} as the source for these rate constants.

\begin{figure*}[hbt!]
    \centering
    \begin{subfigure}[b]{0.34\textwidth}
        \centering
        \includegraphics[width=\textwidth,trim={0cm 0cm 1.3cm 0cm},clip]{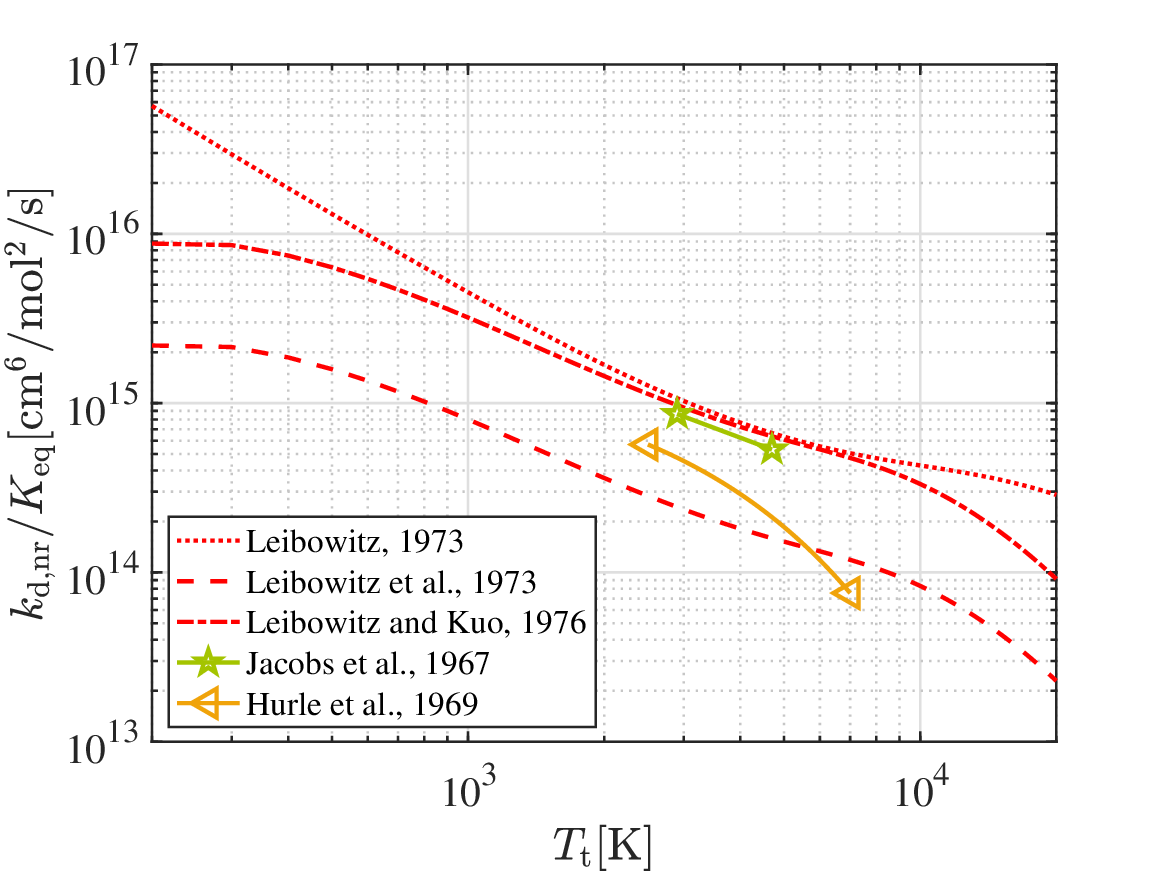}
        \caption{M = $\rm H_2$}
    \end{subfigure}
    \hfill
    \begin{subfigure}[b]{0.32\textwidth}
        \centering
        \includegraphics[width=\textwidth,trim={1cm 0cm 1.3cm 0cm},clip]{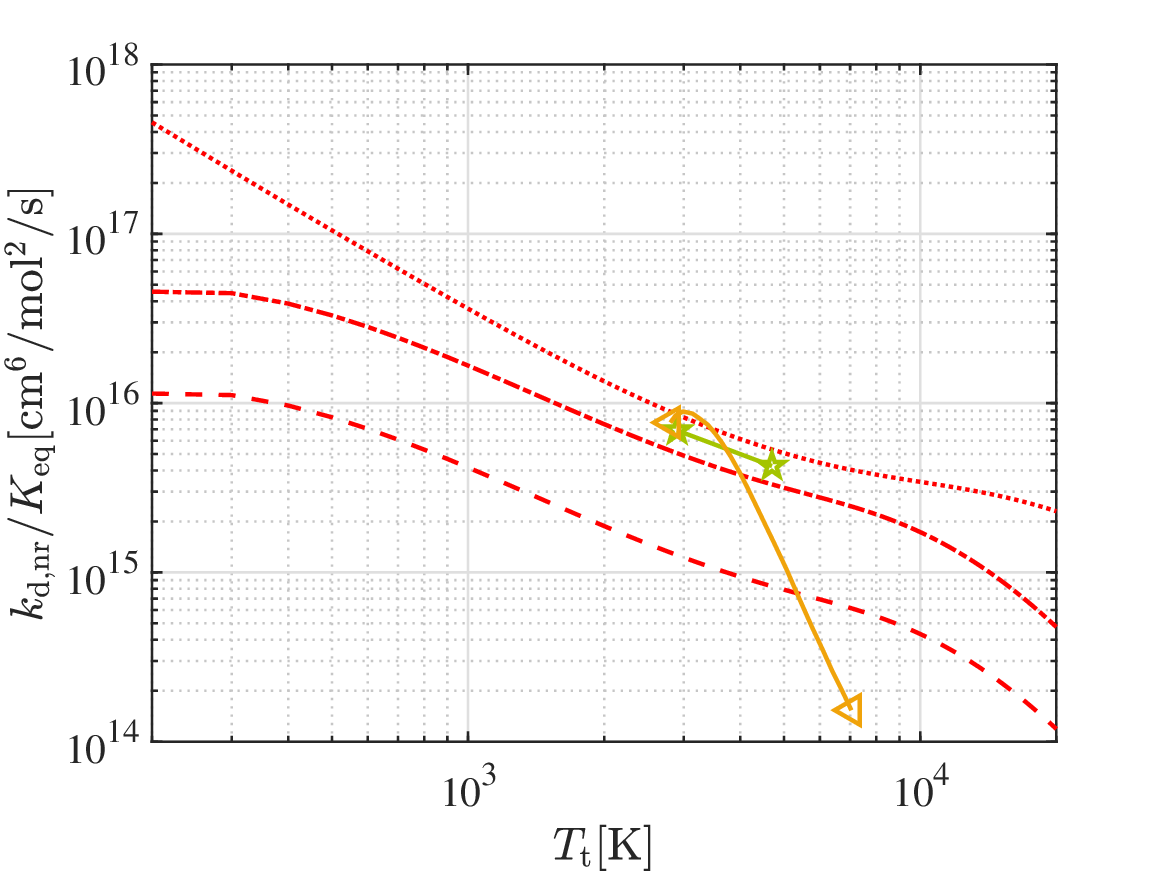}
        \caption{M = H}
    \end{subfigure}
    \hfill
    \begin{subfigure}[b]{0.32\textwidth}
        \centering
        \includegraphics[width=\textwidth,trim={1cm 0cm 1.3cm 0cm},clip]{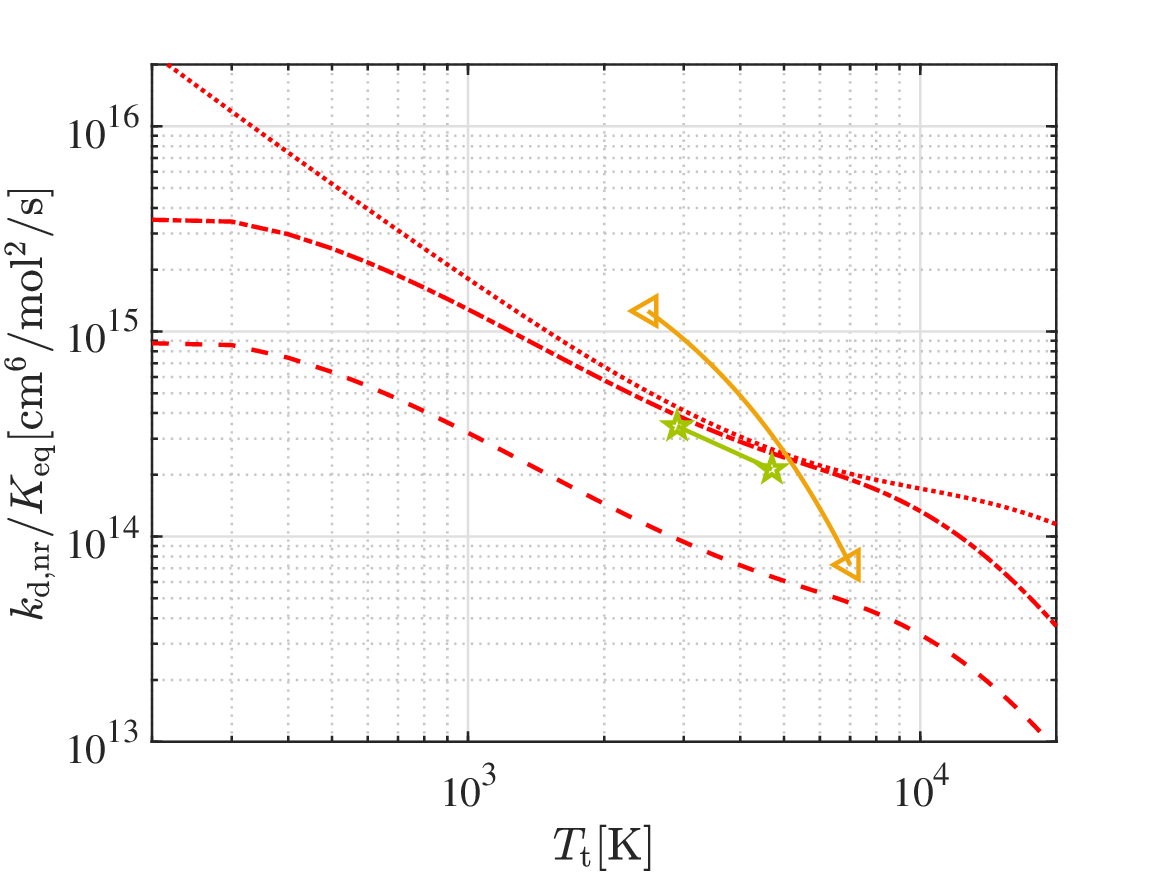}
        \caption{M = \{noble gas\}}
    \end{subfigure}
    \caption{Rate constant fits of Leibowitz~\cite{Leibowitz1973_1}, Leibowitz et al.~\cite{Leibowitz1973_2}, and Leibowitz and Kuo~\cite{Leibowitz1976} compared to those from the shock tube studies of Jacobs et al.~\cite{Jacobs1967} and Hurle et al.~\cite{Hurle1969}.}
    \label{fig:leib}
\end{figure*}

Figure~\ref{fig:leib} shows the rate constants (plotted as $k_{\rm d,nr}/K_{\rm eq}$) for all three formulations compared to the fits from the shock tube studies of Jacobs et al.~\cite{Jacobs1967} and Hurle et al.~\cite{Hurle1969}. While the rate constants of Leibowitz~\cite{Leibowitz1973_1} and Leibowitz and Kuo~\cite{Leibowitz1976} are similar to each other and to the rate constants of Jacobs et al.~\cite{Jacobs1967} for the temperature range between approximately 3,000 and 5,000 K, the rate constants of Leibowitz et al.~\cite{Leibowitz1973_2} are consistently lower for all three third-bodies. Additionally, despite citing the shock tube study by Hurle et al.~\cite{Hurle1969}, the rate constants of Leibowitz et al.~\cite{Leibowitz1976} do not follow the rate constants proposed by Hurle et al.~\cite{Hurle1969}.

Although Fig.~\ref{fig:leib} highlights that there are apparent differences between the three sets of rate constants proposed by Leibowitz~\cite{Leibowitz1973_1,Leibowitz1973_2,Leibowitz1976}, many of the studies in the ice and gas giant entry flow literature cite the three different formulations interchangeably. To the authors' knowledge, all of the studies that have implemented and used the rate constants attributed to Leibowitz~\cite{Palmer2014,Higdon2018,Liu2021,Hansson2021,Carroll2023_conv,Carroll2023,Coelho2023,Rataczak2024} have used the rate constants from the first formulation of Leibowitz~\cite{Leibowitz1973_1}, as they follow the modified Arrhenius form and thus can be readily implemented in most CFD codes. For this reason, comparisons are only made to this first set of rate constants from Leibowitz~\cite{Leibowitz1973_1} in section~\ref{sec:rates}.

\bibliography{main}

\begin{thebibliography}{129}%
\makeatletter
\providecommand \@ifxundefined [1]{%
 \@ifx{#1\undefined}
}%
\providecommand \@ifnum [1]{%
 \ifnum #1\expandafter \@firstoftwo
 \else \expandafter \@secondoftwo
 \fi
}%
\providecommand \@ifx [1]{%
 \ifx #1\expandafter \@firstoftwo
 \else \expandafter \@secondoftwo
 \fi
}%
\providecommand \natexlab [1]{#1}%
\providecommand \enquote  [1]{``#1''}%
\providecommand \bibnamefont  [1]{#1}%
\providecommand \bibfnamefont [1]{#1}%
\providecommand \citenamefont [1]{#1}%
\providecommand \href@noop [0]{\@secondoftwo}%
\providecommand \href [0]{\begingroup \@sanitize@url \@href}%
\providecommand \@href[1]{\@@startlink{#1}\@@href}%
\providecommand \@@href[1]{\endgroup#1\@@endlink}%
\providecommand \@sanitize@url [0]{\catcode `\\12\catcode `\$12\catcode `\&12\catcode `\#12\catcode `\^12\catcode `\_12\catcode `\%12\relax}%
\providecommand \@@startlink[1]{}%
\providecommand \@@endlink[0]{}%
\providecommand \url  [0]{\begingroup\@sanitize@url \@url }%
\providecommand \@url [1]{\endgroup\@href {#1}{\urlprefix }}%
\providecommand \urlprefix  [0]{URL }%
\providecommand \Eprint [0]{\href }%
\providecommand \doibase [0]{http://dx.doi.org/}%
\providecommand \selectlanguage [0]{\@gobble}%
\providecommand \bibinfo  [0]{\@secondoftwo}%
\providecommand \bibfield  [0]{\@secondoftwo}%
\providecommand \translation [1]{[#1]}%
\providecommand \BibitemOpen [0]{}%
\providecommand \bibitemStop [0]{}%
\providecommand \bibitemNoStop [0]{.\EOS\space}%
\providecommand \EOS [0]{\spacefactor3000\relax}%
\providecommand \BibitemShut  [1]{\csname bibitem#1\endcsname}%
\let\auto@bib@innerbib\@empty
\bibitem [{\citenamefont {Draine}, \citenamefont {Roberge},\ and\ \citenamefont {Dalgarno}(1983)}]{Draine1983}%
  \BibitemOpen
  \bibfield  {author} {\bibinfo {author} {\bibfnamefont {B.~T.}\ \bibnamefont {Draine}}, \bibinfo {author} {\bibfnamefont {W.~G.}\ \bibnamefont {Roberge}}, \ and\ \bibinfo {author} {\bibfnamefont {A.}~\bibnamefont {Dalgarno}},\ }\bibfield  {title} {\enquote {\bibinfo {title} {Magnetohydrodynamic shock waves in molecular clouds},}\ }\href {\doibase 10.1086/160617} {\bibfield  {journal} {\bibinfo  {journal} {The Astrophysical Journal}\ }\textbf {\bibinfo {volume} {264}},\ \bibinfo {pages} {485} (\bibinfo {year} {1983})}\BibitemShut {NoStop}%
\bibitem [{\citenamefont {Hollenbach}\ and\ \citenamefont {McKee}(1989)}]{Hollenbach1989}%
  \BibitemOpen
  \bibfield  {author} {\bibinfo {author} {\bibfnamefont {D.}~\bibnamefont {Hollenbach}}\ and\ \bibinfo {author} {\bibfnamefont {C.~F.}\ \bibnamefont {McKee}},\ }\bibfield  {title} {\enquote {\bibinfo {title} {Molecule formation and infrared emission in fast interstellar shocks. {III} - results for {J} shocks in molecular clouds},}\ }\href {\doibase 10.1086/167595} {\bibfield  {journal} {\bibinfo  {journal} {The Astrophysical Journal}\ }\textbf {\bibinfo {volume} {342}},\ \bibinfo {pages} {306} (\bibinfo {year} {1989})}\BibitemShut {NoStop}%
\bibitem [{\citenamefont {Chang}\ and\ \citenamefont {Martin}(1991)}]{Chang1991}%
  \BibitemOpen
  \bibfield  {author} {\bibinfo {author} {\bibfnamefont {C.~A.}\ \bibnamefont {Chang}}\ and\ \bibinfo {author} {\bibfnamefont {P.~G.}\ \bibnamefont {Martin}},\ }\bibfield  {title} {\enquote {\bibinfo {title} {Partially dissociative jump shocks in molecular hydrogen},}\ }\href {\doibase 10.1086/170420} {\bibfield  {journal} {\bibinfo  {journal} {The Astrophysical Journal}\ }\textbf {\bibinfo {volume} {378}},\ \bibinfo {pages} {202} (\bibinfo {year} {1991})}\BibitemShut {NoStop}%
\bibitem [{\citenamefont {Bell}\ and\ \citenamefont {Cowan}(2018)}]{Bell2018}%
  \BibitemOpen
  \bibfield  {author} {\bibinfo {author} {\bibfnamefont {T.~J.}\ \bibnamefont {Bell}}\ and\ \bibinfo {author} {\bibfnamefont {N.~B.}\ \bibnamefont {Cowan}},\ }\bibfield  {title} {\enquote {\bibinfo {title} {Increased heat transport in ultra-hot {Jupiter} atmospheres through {$\rm H_2$} dissociation and recombination},}\ }\href {\doibase 10.3847/2041-8213/aabcc8} {\bibfield  {journal} {\bibinfo  {journal} {The Astrophysical Journal Letters}\ }\textbf {\bibinfo {volume} {857}},\ \bibinfo {pages} {L20} (\bibinfo {year} {2018})}\BibitemShut {NoStop}%
\bibitem [{\citenamefont {Kristensen}\ \emph {et~al.}(2023)\citenamefont {Kristensen}, \citenamefont {Godard}, \citenamefont {Guillard}, \citenamefont {Gusdorf},\ and\ \citenamefont {Pineau~des For\^ets}}]{Kristensen2023}%
  \BibitemOpen
  \bibfield  {author} {\bibinfo {author} {\bibfnamefont {L.~E.}\ \bibnamefont {Kristensen}}, \bibinfo {author} {\bibfnamefont {B.}~\bibnamefont {Godard}}, \bibinfo {author} {\bibfnamefont {P.}~\bibnamefont {Guillard}}, \bibinfo {author} {\bibfnamefont {A.}~\bibnamefont {Gusdorf}}, \ and\ \bibinfo {author} {\bibfnamefont {G.}~\bibnamefont {Pineau~des For\^ets}},\ }\bibfield  {title} {\enquote {\bibinfo {title} {Shock excitation of {$\rm H_2$} in the {James Webb} space telescope era},}\ }\href {\doibase 10.1051/0004-6361/202346254} {\bibfield  {journal} {\bibinfo  {journal} {Astronomy \& Astrophysics}\ }\textbf {\bibinfo {volume} {675}},\ \bibinfo {pages} {A86} (\bibinfo {year} {2023})}\BibitemShut {NoStop}%
\bibitem [{\citenamefont {Matveyev}\ and\ \citenamefont {Silakov}(1995)}]{Matveyev1995}%
  \BibitemOpen
  \bibfield  {author} {\bibinfo {author} {\bibfnamefont {A.~A.}\ \bibnamefont {Matveyev}}\ and\ \bibinfo {author} {\bibfnamefont {V.~P.}\ \bibnamefont {Silakov}},\ }\bibfield  {title} {\enquote {\bibinfo {title} {Kinetic processes in a highly-ionized non-equilibrium hydrogen plasma},}\ }\href {\doibase 10.1088/0963-0252/4/4/012} {\bibfield  {journal} {\bibinfo  {journal} {Plasma Sources Science and Technology}\ }\textbf {\bibinfo {volume} {4}},\ \bibinfo {pages} {606–617} (\bibinfo {year} {1995})}\BibitemShut {NoStop}%
\bibitem [{\citenamefont {Silakov}\ \emph {et~al.}(1996)\citenamefont {Silakov}, \citenamefont {Matveyev}, \citenamefont {Chebotarev},\ and\ \citenamefont {Otorbaev}}]{VPSilakov1996}%
  \BibitemOpen
  \bibfield  {author} {\bibinfo {author} {\bibfnamefont {V.~P.}\ \bibnamefont {Silakov}}, \bibinfo {author} {\bibfnamefont {A.~A.}\ \bibnamefont {Matveyev}}, \bibinfo {author} {\bibfnamefont {A.~V.}\ \bibnamefont {Chebotarev}}, \ and\ \bibinfo {author} {\bibfnamefont {D.~K.}\ \bibnamefont {Otorbaev}},\ }\bibfield  {title} {\enquote {\bibinfo {title} {Non-equilibrium properties of a flowing hydrogen cascaded arc plasma: Kinetic modelling},}\ }\href {\doibase 10.1088/0022-3727/29/8/008} {\bibfield  {journal} {\bibinfo  {journal} {Journal of Physics D: Applied Physics}\ }\textbf {\bibinfo {volume} {29}},\ \bibinfo {pages} {2111–2118} (\bibinfo {year} {1996})}\BibitemShut {NoStop}%
\bibitem [{\citenamefont {Capitelli}\ \emph {et~al.}(2002)\citenamefont {Capitelli}, \citenamefont {Celiberto}, \citenamefont {Esposito}, \citenamefont {Laricchiuta}, \citenamefont {Hassouni},\ and\ \citenamefont {Longo}}]{Capitelli2002}%
  \BibitemOpen
  \bibfield  {author} {\bibinfo {author} {\bibfnamefont {M.}~\bibnamefont {Capitelli}}, \bibinfo {author} {\bibfnamefont {R.}~\bibnamefont {Celiberto}}, \bibinfo {author} {\bibfnamefont {F.}~\bibnamefont {Esposito}}, \bibinfo {author} {\bibfnamefont {A.}~\bibnamefont {Laricchiuta}}, \bibinfo {author} {\bibfnamefont {K.}~\bibnamefont {Hassouni}}, \ and\ \bibinfo {author} {\bibfnamefont {S.}~\bibnamefont {Longo}},\ }\bibfield  {title} {\enquote {\bibinfo {title} {Elementary processes and kinetics of {$\rm H_2$} plasmas for different technological applications},}\ }\href {\doibase 10.1088/0963-0252/11/3a/302} {\bibfield  {journal} {\bibinfo  {journal} {Plasma Sources Science and Technology}\ }\textbf {\bibinfo {volume} {11}},\ \bibinfo {pages} {A7–A25} (\bibinfo {year} {2002})}\BibitemShut {NoStop}%
\bibitem [{\citenamefont {Shakhatov}\ and\ \citenamefont {Lebedev}(2018)}]{Shakhatov2018}%
  \BibitemOpen
  \bibfield  {author} {\bibinfo {author} {\bibfnamefont {V.~A.}\ \bibnamefont {Shakhatov}}\ and\ \bibinfo {author} {\bibfnamefont {Y.~A.}\ \bibnamefont {Lebedev}},\ }\bibfield  {title} {\enquote {\bibinfo {title} {Kinetics of populations of singlet and triplet states in non-equilibrium hydrogen plasma},}\ }\href {\doibase 10.1088/1361-6463/aab8ca} {\bibfield  {journal} {\bibinfo  {journal} {Journal of Physics D: Applied Physics}\ }\textbf {\bibinfo {volume} {51}},\ \bibinfo {pages} {213001} (\bibinfo {year} {2018})}\BibitemShut {NoStop}%
\bibitem [{\citenamefont {Smith}\ \emph {et~al.}(2024)\citenamefont {Smith}, \citenamefont {Diomede}, \citenamefont {Gibson}, \citenamefont {Doyle}, \citenamefont {Guerra}, \citenamefont {Kushner}, \citenamefont {Gans},\ and\ \citenamefont {Dedrick}}]{Smith2024}%
  \BibitemOpen
  \bibfield  {author} {\bibinfo {author} {\bibfnamefont {G.~J.}\ \bibnamefont {Smith}}, \bibinfo {author} {\bibfnamefont {P.}~\bibnamefont {Diomede}}, \bibinfo {author} {\bibfnamefont {A.~R.}\ \bibnamefont {Gibson}}, \bibinfo {author} {\bibfnamefont {S.~J.}\ \bibnamefont {Doyle}}, \bibinfo {author} {\bibfnamefont {V.}~\bibnamefont {Guerra}}, \bibinfo {author} {\bibfnamefont {M.~J.}\ \bibnamefont {Kushner}}, \bibinfo {author} {\bibfnamefont {T.}~\bibnamefont {Gans}}, \ and\ \bibinfo {author} {\bibfnamefont {J.~P.}\ \bibnamefont {Dedrick}},\ }\bibfield  {title} {\enquote {\bibinfo {title} {Low-pressure inductively coupled plasmas in hydrogen: Impact of gas heating on the spatial distribution of atomic hydrogen and vibrationally excited states},}\ }\href {\doibase 10.1088/1361-6595/ad1ece} {\bibfield  {journal} {\bibinfo  {journal} {Plasma Sources Science and Technology}\ }\textbf {\bibinfo {volume} {33}},\ \bibinfo {pages} {025002} (\bibinfo {year} {2024})}\BibitemShut {NoStop}%
\bibitem [{\citenamefont {Taylor}\ \emph {et~al.}(2013)\citenamefont {Taylor}, \citenamefont {Kessler}, \citenamefont {Gamezo},\ and\ \citenamefont {Oran}}]{Taylor2013}%
  \BibitemOpen
  \bibfield  {author} {\bibinfo {author} {\bibfnamefont {B.}~\bibnamefont {Taylor}}, \bibinfo {author} {\bibfnamefont {D.}~\bibnamefont {Kessler}}, \bibinfo {author} {\bibfnamefont {V.}~\bibnamefont {Gamezo}}, \ and\ \bibinfo {author} {\bibfnamefont {E.}~\bibnamefont {Oran}},\ }\bibfield  {title} {\enquote {\bibinfo {title} {Numerical simulations of hydrogen detonations with detailed chemical kinetics},}\ }\href {\doibase 10.1016/j.proci.2012.05.045} {\bibfield  {journal} {\bibinfo  {journal} {Proceedings of the Combustion Institute}\ }\textbf {\bibinfo {volume} {34}},\ \bibinfo {pages} {2009–2016} (\bibinfo {year} {2013})}\BibitemShut {NoStop}%
\bibitem [{\citenamefont {Voelkel}\ \emph {et~al.}(2016)\citenamefont {Voelkel}, \citenamefont {Masselot}, \citenamefont {Varghese},\ and\ \citenamefont {Raman}}]{Voelkel2016}%
  \BibitemOpen
  \bibfield  {author} {\bibinfo {author} {\bibfnamefont {S.}~\bibnamefont {Voelkel}}, \bibinfo {author} {\bibfnamefont {D.}~\bibnamefont {Masselot}}, \bibinfo {author} {\bibfnamefont {P.~L.}\ \bibnamefont {Varghese}}, \ and\ \bibinfo {author} {\bibfnamefont {V.}~\bibnamefont {Raman}},\ }\bibfield  {title} {\enquote {\bibinfo {title} {Analysis of hydrogen-air detonation waves with vibrational nonequilibrium},}\ }in\ \href {\doibase 10.1063/1.4967591} {\emph {\bibinfo {booktitle} {AIP Conference Proceedings}}}\ (\bibinfo {year} {2016})\BibitemShut {NoStop}%
\bibitem [{\citenamefont {Shi}\ \emph {et~al.}(2016)\citenamefont {Shi}, \citenamefont {Shen}, \citenamefont {Zhang}, \citenamefont {Zhang},\ and\ \citenamefont {Wen}}]{Shi2016}%
  \BibitemOpen
  \bibfield  {author} {\bibinfo {author} {\bibfnamefont {L.}~\bibnamefont {Shi}}, \bibinfo {author} {\bibfnamefont {H.}~\bibnamefont {Shen}}, \bibinfo {author} {\bibfnamefont {P.}~\bibnamefont {Zhang}}, \bibinfo {author} {\bibfnamefont {D.}~\bibnamefont {Zhang}}, \ and\ \bibinfo {author} {\bibfnamefont {C.}~\bibnamefont {Wen}},\ }\bibfield  {title} {\enquote {\bibinfo {title} {Assessment of vibrational non-equilibrium effect on detonation cell size},}\ }\href {\doibase 10.1080/00102202.2016.1260561} {\bibfield  {journal} {\bibinfo  {journal} {Combustion Science and Technology}\ }\textbf {\bibinfo {volume} {189}},\ \bibinfo {pages} {841–853} (\bibinfo {year} {2016})}\BibitemShut {NoStop}%
\bibitem [{\citenamefont {Vargas}\ \emph {et~al.}(2022)\citenamefont {Vargas}, \citenamefont {Mével}, \citenamefont {Lino~da Silva},\ and\ \citenamefont {Lacoste}}]{Vargas2022}%
  \BibitemOpen
  \bibfield  {author} {\bibinfo {author} {\bibfnamefont {J.}~\bibnamefont {Vargas}}, \bibinfo {author} {\bibfnamefont {R.}~\bibnamefont {Mével}}, \bibinfo {author} {\bibfnamefont {M.}~\bibnamefont {Lino~da Silva}}, \ and\ \bibinfo {author} {\bibfnamefont {D.~A.}\ \bibnamefont {Lacoste}},\ }\bibfield  {title} {\enquote {\bibinfo {title} {Development of a steady detonation reactor with state-to-state thermochemical modeling},}\ }\href {\doibase 10.1007/s00193-022-01105-2} {\bibfield  {journal} {\bibinfo  {journal} {Shock Waves}\ }\textbf {\bibinfo {volume} {32}},\ \bibinfo {pages} {679–689} (\bibinfo {year} {2022})}\BibitemShut {NoStop}%
\bibitem [{\citenamefont {Baumgart}, \citenamefont {Yao},\ and\ \citenamefont {Blanquart}(2025)}]{Baumgart2025}%
  \BibitemOpen
  \bibfield  {author} {\bibinfo {author} {\bibfnamefont {A.}~\bibnamefont {Baumgart}}, \bibinfo {author} {\bibfnamefont {M.~X.}\ \bibnamefont {Yao}}, \ and\ \bibinfo {author} {\bibfnamefont {G.}~\bibnamefont {Blanquart}},\ }\bibfield  {title} {\enquote {\bibinfo {title} {Tabulated chemistry approach for detonation simulations},}\ }\href {\doibase 10.1016/j.combustflame.2024.113878} {\bibfield  {journal} {\bibinfo  {journal} {Combustion and Flame}\ }\textbf {\bibinfo {volume} {272}},\ \bibinfo {pages} {113878} (\bibinfo {year} {2025})}\BibitemShut {NoStop}%
\bibitem [{\citenamefont {Park}(2011)}]{Park2011}%
  \BibitemOpen
  \bibfield  {author} {\bibinfo {author} {\bibfnamefont {C.}~\bibnamefont {Park}},\ }\bibfield  {title} {\enquote {\bibinfo {title} {Nonequilibrium chemistry and radiation for {Neptune} entry},}\ }\href {\doibase 10.2514/1.51810} {\bibfield  {journal} {\bibinfo  {journal} {Journal of Spacecraft and Rockets}\ }\textbf {\bibinfo {volume} {48}},\ \bibinfo {pages} {897–903} (\bibinfo {year} {2011})}\BibitemShut {NoStop}%
\bibitem [{\citenamefont {Park}(2012)}]{Park2012}%
  \BibitemOpen
  \bibfield  {author} {\bibinfo {author} {\bibfnamefont {C.}~\bibnamefont {Park}},\ }\bibfield  {title} {\enquote {\bibinfo {title} {Nonequilibrium ionization and radiation in hydrogen-helium mixtures},}\ }\href {\doibase 10.2514/1.t3689} {\bibfield  {journal} {\bibinfo  {journal} {Journal of Thermophysics and Heat Transfer}\ }\textbf {\bibinfo {volume} {26}},\ \bibinfo {pages} {231–243} (\bibinfo {year} {2012})}\BibitemShut {NoStop}%
\bibitem [{\citenamefont {Palmer}, \citenamefont {Prabhu},\ and\ \citenamefont {Cruden}(2014)}]{Palmer2014}%
  \BibitemOpen
  \bibfield  {author} {\bibinfo {author} {\bibfnamefont {G.}~\bibnamefont {Palmer}}, \bibinfo {author} {\bibfnamefont {D.}~\bibnamefont {Prabhu}}, \ and\ \bibinfo {author} {\bibfnamefont {B.~A.}\ \bibnamefont {Cruden}},\ }\bibfield  {title} {\enquote {\bibinfo {title} {Aeroheating uncertainties in {Uranus} and {Saturn} entries by the {Monte} {Carlo} method},}\ }\href {\doibase 10.2514/1.a32768} {\bibfield  {journal} {\bibinfo  {journal} {Journal of Spacecraft and Rockets}\ }\textbf {\bibinfo {volume} {51}},\ \bibinfo {pages} {801--814} (\bibinfo {year} {2014})}\BibitemShut {NoStop}%
\bibitem [{\citenamefont {Erb}, \citenamefont {West},\ and\ \citenamefont {Johnston}(2019)}]{Erb2019}%
  \BibitemOpen
  \bibfield  {author} {\bibinfo {author} {\bibfnamefont {A.~J.}\ \bibnamefont {Erb}}, \bibinfo {author} {\bibfnamefont {T.~K.}\ \bibnamefont {West}}, \ and\ \bibinfo {author} {\bibfnamefont {C.~O.}\ \bibnamefont {Johnston}},\ }\bibfield  {title} {\enquote {\bibinfo {title} {Investigation of {Galileo} probe entry heating with coupled radiation and ablation},}\ }in\ \href {\doibase 10.2514/6.2019-3360} {\emph {\bibinfo {booktitle} {AIAA Aviation 2019 Forum}}}\ (\bibinfo  {publisher} {American Institute of Aeronautics and Astronautics},\ \bibinfo {year} {2019})\BibitemShut {NoStop}%
\bibitem [{\citenamefont {Carroll}\ and\ \citenamefont {Brandis}(2023)}]{Carroll2023_conv}%
  \BibitemOpen
  \bibfield  {author} {\bibinfo {author} {\bibfnamefont {A.~T.}\ \bibnamefont {Carroll}}\ and\ \bibinfo {author} {\bibfnamefont {A.~M.}\ \bibnamefont {Brandis}},\ }\bibfield  {title} {\enquote {\bibinfo {title} {Stagnation point convective heating correlations for entry into {$\rm H_2$}/{He} atmospheres},}\ }in\ \href {\doibase 10.2514/6.2023-0208} {\emph {\bibinfo {booktitle} {AIAA Scitech 2023 Forum}}}\ (\bibinfo  {publisher} {American Institute of Aeronautics and Astronautics},\ \bibinfo {year} {2023})\BibitemShut {NoStop}%
\bibitem [{\citenamefont {Scoggins}, \citenamefont {Hinkle},\ and\ \citenamefont {Shellabarger}(2024)}]{Scoggins2024}%
  \BibitemOpen
  \bibfield  {author} {\bibinfo {author} {\bibfnamefont {J.~B.}\ \bibnamefont {Scoggins}}, \bibinfo {author} {\bibfnamefont {A.~D.}\ \bibnamefont {Hinkle}}, \ and\ \bibinfo {author} {\bibfnamefont {E.}~\bibnamefont {Shellabarger}},\ }\bibfield  {title} {\enquote {\bibinfo {title} {Aeroheating environment of aerocapture systems for {Uranus} orbiters},}\ }in\ \href {\doibase 10.2514/6.2024-0951} {\emph {\bibinfo {booktitle} {AIAA Scitech 2024 Forum}}}\ (\bibinfo  {publisher} {American Institute of Aeronautics and Astronautics},\ \bibinfo {year} {2024})\BibitemShut {NoStop}%
\bibitem [{\citenamefont {Steer}\ \emph {et~al.}(2024)\citenamefont {Steer}, \citenamefont {Collen}, \citenamefont {Glenn}, \citenamefont {Hambidge}, \citenamefont {Doherty}, \citenamefont {McGilvray}, \citenamefont {Sopek}, \citenamefont {Loehle},\ and\ \citenamefont {Walpot}}]{Steer2024}%
  \BibitemOpen
  \bibfield  {author} {\bibinfo {author} {\bibfnamefont {J.}~\bibnamefont {Steer}}, \bibinfo {author} {\bibfnamefont {P.}~\bibnamefont {Collen}}, \bibinfo {author} {\bibfnamefont {A.}~\bibnamefont {Glenn}}, \bibinfo {author} {\bibfnamefont {C.}~\bibnamefont {Hambidge}}, \bibinfo {author} {\bibfnamefont {L.~J.}\ \bibnamefont {Doherty}}, \bibinfo {author} {\bibfnamefont {M.}~\bibnamefont {McGilvray}}, \bibinfo {author} {\bibfnamefont {T.}~\bibnamefont {Sopek}}, \bibinfo {author} {\bibfnamefont {S.}~\bibnamefont {Loehle}}, \ and\ \bibinfo {author} {\bibfnamefont {L.}~\bibnamefont {Walpot}},\ }\bibfield  {title} {\enquote {\bibinfo {title} {Commissioning of upgrades to {T6} to study giant planet entry},}\ }\href {\doibase 10.2514/1.a35893} {\bibfield  {journal} {\bibinfo  {journal} {Journal of Spacecraft and Rockets}\ ,\ \bibinfo {pages} {1–18}} (\bibinfo {year} {2024})}\BibitemShut {NoStop}%
\bibitem [{\citenamefont {Steuer}\ \emph {et~al.}(2024)\citenamefont {Steuer}, \citenamefont {Donaldson}, \citenamefont {D\"{u}rnhofer}, \citenamefont {Leiser}, \citenamefont {Fasoulas},\ and\ \citenamefont {Loehle}}]{Steuer2024}%
  \BibitemOpen
  \bibfield  {author} {\bibinfo {author} {\bibfnamefont {D.~C.}\ \bibnamefont {Steuer}}, \bibinfo {author} {\bibfnamefont {N.}~\bibnamefont {Donaldson}}, \bibinfo {author} {\bibfnamefont {C.~A.}\ \bibnamefont {D\"{u}rnhofer}}, \bibinfo {author} {\bibfnamefont {D.}~\bibnamefont {Leiser}}, \bibinfo {author} {\bibfnamefont {S.}~\bibnamefont {Fasoulas}}, \ and\ \bibinfo {author} {\bibfnamefont {S.}~\bibnamefont {Loehle}},\ }\bibfield  {title} {\enquote {\bibinfo {title} {Investigation on the usability of arc-jet generators for the experimental simulation of giant planet entry},}\ }in\ \href {\doibase 10.2514/6.2024-3552} {\emph {\bibinfo {booktitle} {AIAA Aviation FORUM AND ASCEND 2024}}}\ (\bibinfo  {publisher} {American Institute of Aeronautics and Astronautics},\ \bibinfo {year} {2024})\BibitemShut {NoStop}%
\bibitem [{\citenamefont {Rataczak}, \citenamefont {Boyd},\ and\ \citenamefont {McMahon}(2024)}]{Rataczak2024}%
  \BibitemOpen
  \bibfield  {author} {\bibinfo {author} {\bibfnamefont {J.~A.}\ \bibnamefont {Rataczak}}, \bibinfo {author} {\bibfnamefont {I.~D.}\ \bibnamefont {Boyd}}, \ and\ \bibinfo {author} {\bibfnamefont {J.~W.}\ \bibnamefont {McMahon}},\ }\bibfield  {title} {\enquote {\bibinfo {title} {Parametric sensitivity analysis and uncertainty quantification of ice giant aerocapture aerothermodynamics},}\ }in\ \href {\doibase 10.2514/6.2024-4472} {\emph {\bibinfo {booktitle} {AIAA Aviation FORUM AND ASCEND 2024}}}\ (\bibinfo {year} {2024})\BibitemShut {NoStop}%
\bibitem [{\citenamefont {Marrone}\ and\ \citenamefont {Treanor}(1963)}]{Marrone1963}%
  \BibitemOpen
  \bibfield  {author} {\bibinfo {author} {\bibfnamefont {P.~V.}\ \bibnamefont {Marrone}}\ and\ \bibinfo {author} {\bibfnamefont {C.~E.}\ \bibnamefont {Treanor}},\ }\bibfield  {title} {\enquote {\bibinfo {title} {Chemical relaxation with preferential dissociation from excited vibrational levels},}\ }\href {\doibase 10.1063/1.1706888} {\bibfield  {journal} {\bibinfo  {journal} {The Physics of Fluids}\ }\textbf {\bibinfo {volume} {6}},\ \bibinfo {pages} {1215–1221} (\bibinfo {year} {1963})}\BibitemShut {NoStop}%
\bibitem [{\citenamefont {Macheret}\ \emph {et~al.}(1994)\citenamefont {Macheret}, \citenamefont {Fridman}, \citenamefont {Adamovich}, \citenamefont {Rich},\ and\ \citenamefont {Treanor}}]{Macheret1994}%
  \BibitemOpen
  \bibfield  {author} {\bibinfo {author} {\bibfnamefont {S.~O.}\ \bibnamefont {Macheret}}, \bibinfo {author} {\bibfnamefont {A.~A.}\ \bibnamefont {Fridman}}, \bibinfo {author} {\bibfnamefont {I.~V.}\ \bibnamefont {Adamovich}}, \bibinfo {author} {\bibfnamefont {J.~W.}\ \bibnamefont {Rich}}, \ and\ \bibinfo {author} {\bibfnamefont {C.~E.}\ \bibnamefont {Treanor}},\ }\bibfield  {title} {\enquote {\bibinfo {title} {Mechanisms of nonequilibrium dissociation of diatomic molecules},}\ }in\ \href {\doibase 10.2514/6.1994-1984} {\emph {\bibinfo {booktitle} {6th Joint Thermophysics and Heat Transfer Conference}}}\ (\bibinfo  {publisher} {American Institute of Aeronautics and Astronautics},\ \bibinfo {year} {1994})\BibitemShut {NoStop}%
\bibitem [{\citenamefont {Porter}(1974)}]{Porter1974}%
  \BibitemOpen
  \bibfield  {author} {\bibinfo {author} {\bibfnamefont {R.~N.}\ \bibnamefont {Porter}},\ }\bibfield  {title} {\enquote {\bibinfo {title} {Molecular trajectory calculations},}\ }\href {\doibase 10.1146/annurev.pc.25.100174.001533} {\bibfield  {journal} {\bibinfo  {journal} {Annual Review of Physical Chemistry}\ }\textbf {\bibinfo {volume} {25}},\ \bibinfo {pages} {317–355} (\bibinfo {year} {1974})}\BibitemShut {NoStop}%
\bibitem [{\citenamefont {Koura}(1997)}]{Koura1997}%
  \BibitemOpen
  \bibfield  {author} {\bibinfo {author} {\bibfnamefont {K.}~\bibnamefont {Koura}},\ }\bibfield  {title} {\enquote {\bibinfo {title} {{Monte} {Carlo} direct simulation of rotational relaxation of diatomic molecules using classical trajectory calculations: Nitrogen shock wave},}\ }\href {\doibase 10.1063/1.869462} {\bibfield  {journal} {\bibinfo  {journal} {Physics of Fluids}\ }\textbf {\bibinfo {volume} {9}},\ \bibinfo {pages} {3543–3549} (\bibinfo {year} {1997})}\BibitemShut {NoStop}%
\bibitem [{\citenamefont {Kim}, \citenamefont {Kwon},\ and\ \citenamefont {Park}(2009)}]{Kim2009}%
  \BibitemOpen
  \bibfield  {author} {\bibinfo {author} {\bibfnamefont {J.~G.}\ \bibnamefont {Kim}}, \bibinfo {author} {\bibfnamefont {O.~J.}\ \bibnamefont {Kwon}}, \ and\ \bibinfo {author} {\bibfnamefont {C.}~\bibnamefont {Park}},\ }\bibfield  {title} {\enquote {\bibinfo {title} {Master equation study and nonequilibrium chemical reactions for {H}+{$\rm H_2$} and {He}+{$\rm H_2$}},}\ }\href {\doibase 10.2514/1.41741} {\bibfield  {journal} {\bibinfo  {journal} {Journal of Thermophysics and Heat Transfer}\ }\textbf {\bibinfo {volume} {23}},\ \bibinfo {pages} {443--453} (\bibinfo {year} {2009})}\BibitemShut {NoStop}%
\bibitem [{\citenamefont {Kim}\ and\ \citenamefont {Boyd}(2012)}]{Kim2012}%
  \BibitemOpen
  \bibfield  {author} {\bibinfo {author} {\bibfnamefont {J.~G.}\ \bibnamefont {Kim}}\ and\ \bibinfo {author} {\bibfnamefont {I.~D.}\ \bibnamefont {Boyd}},\ }\bibfield  {title} {\enquote {\bibinfo {title} {State-resolved thermochemical nonequilibrium analysis of hydrogen mixture flows},}\ }\href {\doibase 10.1063/1.4747340} {\bibfield  {journal} {\bibinfo  {journal} {Physics of Fluids}\ }\textbf {\bibinfo {volume} {24}} (\bibinfo {year} {2012}),\ 10.1063/1.4747340}\BibitemShut {NoStop}%
\bibitem [{\citenamefont {Panesi}\ \emph {et~al.}(2013)\citenamefont {Panesi}, \citenamefont {Jaffe}, \citenamefont {Schwenke},\ and\ \citenamefont {Magin}}]{Panesi2013}%
  \BibitemOpen
  \bibfield  {author} {\bibinfo {author} {\bibfnamefont {M.}~\bibnamefont {Panesi}}, \bibinfo {author} {\bibfnamefont {R.~L.}\ \bibnamefont {Jaffe}}, \bibinfo {author} {\bibfnamefont {D.~W.}\ \bibnamefont {Schwenke}}, \ and\ \bibinfo {author} {\bibfnamefont {T.~E.}\ \bibnamefont {Magin}},\ }\bibfield  {title} {\enquote {\bibinfo {title} {Rovibrational internal energy transfer and dissociation of $\mathbf { \rm N_2(^1\Sigma _g^+)-\rm N(^4S_u)}$ system in hypersonic flows},}\ }\href {\doibase 10.1063/1.4774412} {\bibfield  {journal} {\bibinfo  {journal} {The Journal of Chemical Physics}\ }\textbf {\bibinfo {volume} {138}} (\bibinfo {year} {2013}),\ 10.1063/1.4774412}\BibitemShut {NoStop}%
\bibitem [{\citenamefont {Andrienko}\ and\ \citenamefont {Boyd}(2015)}]{Andrienko2015}%
  \BibitemOpen
  \bibfield  {author} {\bibinfo {author} {\bibfnamefont {D.}~\bibnamefont {Andrienko}}\ and\ \bibinfo {author} {\bibfnamefont {I.~D.}\ \bibnamefont {Boyd}},\ }\bibfield  {title} {\enquote {\bibinfo {title} {Investigation of oxygen vibrational relaxation by quasi-classical trajectory method},}\ }\href {\doibase 10.1016/j.chemphys.2015.07.023} {\bibfield  {journal} {\bibinfo  {journal} {Chemical Physics}\ }\textbf {\bibinfo {volume} {459}},\ \bibinfo {pages} {1–13} (\bibinfo {year} {2015})}\BibitemShut {NoStop}%
\bibitem [{\citenamefont {Norman}, \citenamefont {Valentini},\ and\ \citenamefont {Schwartzentruber}(2013)}]{Norman2013}%
  \BibitemOpen
  \bibfield  {author} {\bibinfo {author} {\bibfnamefont {P.}~\bibnamefont {Norman}}, \bibinfo {author} {\bibfnamefont {P.}~\bibnamefont {Valentini}}, \ and\ \bibinfo {author} {\bibfnamefont {T.}~\bibnamefont {Schwartzentruber}},\ }\bibfield  {title} {\enquote {\bibinfo {title} {{GPU}-accelerated classical trajectory calculation direct simulation {Monte} {Carlo} applied to shock waves},}\ }\href {\doibase 10.1016/j.jcp.2013.03.060} {\bibfield  {journal} {\bibinfo  {journal} {Journal of Computational Physics}\ }\textbf {\bibinfo {volume} {247}},\ \bibinfo {pages} {153–167} (\bibinfo {year} {2013})}\BibitemShut {NoStop}%
\bibitem [{\citenamefont {Valentini}\ \emph {et~al.}(2014)\citenamefont {Valentini}, \citenamefont {Norman}, \citenamefont {Zhang},\ and\ \citenamefont {Schwartzentruber}}]{Valentini2014}%
  \BibitemOpen
  \bibfield  {author} {\bibinfo {author} {\bibfnamefont {P.}~\bibnamefont {Valentini}}, \bibinfo {author} {\bibfnamefont {P.}~\bibnamefont {Norman}}, \bibinfo {author} {\bibfnamefont {C.}~\bibnamefont {Zhang}}, \ and\ \bibinfo {author} {\bibfnamefont {T.~E.}\ \bibnamefont {Schwartzentruber}},\ }\bibfield  {title} {\enquote {\bibinfo {title} {Rovibrational coupling in molecular nitrogen at high temperature: An atomic-level study},}\ }\href {\doibase 10.1063/1.4875279} {\bibfield  {journal} {\bibinfo  {journal} {Physics of Fluids}\ }\textbf {\bibinfo {volume} {26}} (\bibinfo {year} {2014}),\ 10.1063/1.4875279}\BibitemShut {NoStop}%
\bibitem [{\citenamefont {Schwartzentruber}, \citenamefont {Grover},\ and\ \citenamefont {Valentini}(2018)}]{Schwartzentruber2018}%
  \BibitemOpen
  \bibfield  {author} {\bibinfo {author} {\bibfnamefont {T.~E.}\ \bibnamefont {Schwartzentruber}}, \bibinfo {author} {\bibfnamefont {M.~S.}\ \bibnamefont {Grover}}, \ and\ \bibinfo {author} {\bibfnamefont {P.}~\bibnamefont {Valentini}},\ }\bibfield  {title} {\enquote {\bibinfo {title} {Direct molecular simulation of nonequilibrium dilute gases},}\ }\href {\doibase 10.2514/1.t5188} {\bibfield  {journal} {\bibinfo  {journal} {Journal of Thermophysics and Heat Transfer}\ }\textbf {\bibinfo {volume} {32}},\ \bibinfo {pages} {892–903} (\bibinfo {year} {2018})}\BibitemShut {NoStop}%
\bibitem [{\citenamefont {Park}(1988)}]{Park1988}%
  \BibitemOpen
  \bibfield  {author} {\bibinfo {author} {\bibfnamefont {C.}~\bibnamefont {Park}},\ }\bibfield  {title} {\enquote {\bibinfo {title} {Assessment of a two-temperature kinetic model for dissociating and weakly ionizing nitrogen},}\ }\href {\doibase 10.2514/3.55} {\bibfield  {journal} {\bibinfo  {journal} {Journal of Thermophysics and Heat Transfer}\ }\textbf {\bibinfo {volume} {2}},\ \bibinfo {pages} {8–16} (\bibinfo {year} {1988})}\BibitemShut {NoStop}%
\bibitem [{\citenamefont {Park}(1989{\natexlab{a}})}]{Park1989}%
  \BibitemOpen
  \bibfield  {author} {\bibinfo {author} {\bibfnamefont {C.}~\bibnamefont {Park}},\ }\bibfield  {title} {\enquote {\bibinfo {title} {Assessment of two-temperature kinetic model for ionizing air},}\ }\href {\doibase 10.2514/3.28771} {\bibfield  {journal} {\bibinfo  {journal} {Journal of Thermophysics and Heat Transfer}\ }\textbf {\bibinfo {volume} {3}},\ \bibinfo {pages} {233–244} (\bibinfo {year} {1989}{\natexlab{a}})}\BibitemShut {NoStop}%
\bibitem [{\citenamefont {Park}(1989{\natexlab{b}})}]{Park1989NonequilibriumHA}%
  \BibitemOpen
  \bibfield  {author} {\bibinfo {author} {\bibfnamefont {C.~H.}\ \bibnamefont {Park}},\ }\href@noop {} {\emph {\bibinfo {title} {Nonequilibrium Hypersonic Aerothermodynamics}}}\ (\bibinfo  {publisher} {Wiley},\ \bibinfo {year} {1989})\BibitemShut {NoStop}%
\bibitem [{\citenamefont {Chaudhry}\ and\ \citenamefont {Candler}(2019)}]{Chaudhry2019}%
  \BibitemOpen
  \bibfield  {author} {\bibinfo {author} {\bibfnamefont {R.~S.}\ \bibnamefont {Chaudhry}}\ and\ \bibinfo {author} {\bibfnamefont {G.~V.}\ \bibnamefont {Candler}},\ }\bibfield  {title} {\enquote {\bibinfo {title} {Statistical analyses of quasiclassical trajectory data for air dissociation},}\ }in\ \href {\doibase 10.2514/6.2019-0789} {\emph {\bibinfo {booktitle} {AIAA Scitech 2019 Forum}}}\ (\bibinfo  {publisher} {American Institute of Aeronautics and Astronautics},\ \bibinfo {year} {2019})\BibitemShut {NoStop}%
\bibitem [{\citenamefont {Chaudhry}\ \emph {et~al.}(2020)\citenamefont {Chaudhry}, \citenamefont {Boyd}, \citenamefont {Torres}, \citenamefont {Schwartzentruber},\ and\ \citenamefont {Candler}}]{Chaudhry2020}%
  \BibitemOpen
  \bibfield  {author} {\bibinfo {author} {\bibfnamefont {R.~S.}\ \bibnamefont {Chaudhry}}, \bibinfo {author} {\bibfnamefont {I.~D.}\ \bibnamefont {Boyd}}, \bibinfo {author} {\bibfnamefont {E.}~\bibnamefont {Torres}}, \bibinfo {author} {\bibfnamefont {T.~E.}\ \bibnamefont {Schwartzentruber}}, \ and\ \bibinfo {author} {\bibfnamefont {G.~V.}\ \bibnamefont {Candler}},\ }\bibfield  {title} {\enquote {\bibinfo {title} {Implementation of a chemical kinetics model for hypersonic flows in air for high-performance {CFD}},}\ }in\ \href {\doibase 10.2514/6.2020-2191} {\emph {\bibinfo {booktitle} {AIAA Scitech 2020 Forum}}}\ (\bibinfo  {publisher} {American Institute of Aeronautics and Astronautics},\ \bibinfo {year} {2020})\BibitemShut {NoStop}%
\bibitem [{\citenamefont {Singh}\ and\ \citenamefont {Schwartzentruber}(2020{\natexlab{a}})}]{Singh2020_1}%
  \BibitemOpen
  \bibfield  {author} {\bibinfo {author} {\bibfnamefont {N.}~\bibnamefont {Singh}}\ and\ \bibinfo {author} {\bibfnamefont {T.}~\bibnamefont {Schwartzentruber}},\ }\bibfield  {title} {\enquote {\bibinfo {title} {Consistent kinetic–continuum dissociation model {I.} kinetic formulation},}\ }\href {\doibase 10.1063/1.5142752} {\bibfield  {journal} {\bibinfo  {journal} {The Journal of Chemical Physics}\ }\textbf {\bibinfo {volume} {152}} (\bibinfo {year} {2020}{\natexlab{a}}),\ 10.1063/1.5142752}\BibitemShut {NoStop}%
\bibitem [{\citenamefont {Singh}\ and\ \citenamefont {Schwartzentruber}(2020{\natexlab{b}})}]{Singh2020_2}%
  \BibitemOpen
  \bibfield  {author} {\bibinfo {author} {\bibfnamefont {N.}~\bibnamefont {Singh}}\ and\ \bibinfo {author} {\bibfnamefont {T.}~\bibnamefont {Schwartzentruber}},\ }\bibfield  {title} {\enquote {\bibinfo {title} {Consistent kinetic-continuum dissociation model. {II.} continuum formulation and verification},}\ }\href {\doibase 10.1063/1.5142754} {\bibfield  {journal} {\bibinfo  {journal} {The Journal of Chemical Physics}\ }\textbf {\bibinfo {volume} {152}} (\bibinfo {year} {2020}{\natexlab{b}}),\ 10.1063/1.5142754}\BibitemShut {NoStop}%
\bibitem [{\citenamefont {Magin}\ \emph {et~al.}(2012)\citenamefont {Magin}, \citenamefont {Panesi}, \citenamefont {Bourdon}, \citenamefont {Jaffe},\ and\ \citenamefont {Schwenke}}]{Magin2012}%
  \BibitemOpen
  \bibfield  {author} {\bibinfo {author} {\bibfnamefont {T.~E.}\ \bibnamefont {Magin}}, \bibinfo {author} {\bibfnamefont {M.}~\bibnamefont {Panesi}}, \bibinfo {author} {\bibfnamefont {A.}~\bibnamefont {Bourdon}}, \bibinfo {author} {\bibfnamefont {R.~L.}\ \bibnamefont {Jaffe}}, \ and\ \bibinfo {author} {\bibfnamefont {D.~W.}\ \bibnamefont {Schwenke}},\ }\bibfield  {title} {\enquote {\bibinfo {title} {Coarse-grain model for internal energy excitation and dissociation of molecular nitrogen},}\ }\href {\doibase 10.1016/j.chemphys.2011.10.009} {\bibfield  {journal} {\bibinfo  {journal} {Chemical Physics}\ }\textbf {\bibinfo {volume} {398}},\ \bibinfo {pages} {90–95} (\bibinfo {year} {2012})}\BibitemShut {NoStop}%
\bibitem [{\citenamefont {Sahai}\ \emph {et~al.}(2017)\citenamefont {Sahai}, \citenamefont {Lopez}, \citenamefont {Johnston},\ and\ \citenamefont {Panesi}}]{Sahai2017}%
  \BibitemOpen
  \bibfield  {author} {\bibinfo {author} {\bibfnamefont {A.}~\bibnamefont {Sahai}}, \bibinfo {author} {\bibfnamefont {B.}~\bibnamefont {Lopez}}, \bibinfo {author} {\bibfnamefont {C.~O.}\ \bibnamefont {Johnston}}, \ and\ \bibinfo {author} {\bibfnamefont {M.}~\bibnamefont {Panesi}},\ }\bibfield  {title} {\enquote {\bibinfo {title} {Adaptive coarse graining method for energy transfer and dissociation kinetics of polyatomic species},}\ }\href {\doibase 10.1063/1.4996654} {\bibfield  {journal} {\bibinfo  {journal} {The Journal of Chemical Physics}\ }\textbf {\bibinfo {volume} {147}} (\bibinfo {year} {2017}),\ 10.1063/1.4996654}\BibitemShut {NoStop}%
\bibitem [{\citenamefont {Venturi}\ \emph {et~al.}(2020)\citenamefont {Venturi}, \citenamefont {Sharma}, \citenamefont {Lopez},\ and\ \citenamefont {Panesi}}]{Venturi2020}%
  \BibitemOpen
  \bibfield  {author} {\bibinfo {author} {\bibfnamefont {S.}~\bibnamefont {Venturi}}, \bibinfo {author} {\bibfnamefont {M.~P.}\ \bibnamefont {Sharma}}, \bibinfo {author} {\bibfnamefont {B.}~\bibnamefont {Lopez}}, \ and\ \bibinfo {author} {\bibfnamefont {M.}~\bibnamefont {Panesi}},\ }\bibfield  {title} {\enquote {\bibinfo {title} {Data-inspired and physics-driven model reduction for dissociation: Application to the {$\rm O_2$} + {O} system},}\ }\href {\doibase 10.1021/acs.jpca.0c04516} {\bibfield  {journal} {\bibinfo  {journal} {The Journal of Physical Chemistry A}\ }\textbf {\bibinfo {volume} {124}},\ \bibinfo {pages} {8359–8372} (\bibinfo {year} {2020})}\BibitemShut {NoStop}%
\bibitem [{\citenamefont {Liu}\ \emph {et~al.}(2015)\citenamefont {Liu}, \citenamefont {Panesi}, \citenamefont {Sahai},\ and\ \citenamefont {Vinokur}}]{Liu2015}%
  \BibitemOpen
  \bibfield  {author} {\bibinfo {author} {\bibfnamefont {Y.}~\bibnamefont {Liu}}, \bibinfo {author} {\bibfnamefont {M.}~\bibnamefont {Panesi}}, \bibinfo {author} {\bibfnamefont {A.}~\bibnamefont {Sahai}}, \ and\ \bibinfo {author} {\bibfnamefont {M.}~\bibnamefont {Vinokur}},\ }\bibfield  {title} {\enquote {\bibinfo {title} {General multi-group macroscopic modeling for thermo-chemical non-equilibrium gas mixtures},}\ }\href {\doibase 10.1063/1.4915926} {\bibfield  {journal} {\bibinfo  {journal} {The Journal of Chemical Physics}\ }\textbf {\bibinfo {volume} {142}} (\bibinfo {year} {2015}),\ 10.1063/1.4915926}\BibitemShut {NoStop}%
\bibitem [{\citenamefont {Park}(1993)}]{Park1993}%
  \BibitemOpen
  \bibfield  {author} {\bibinfo {author} {\bibfnamefont {C.}~\bibnamefont {Park}},\ }\bibfield  {title} {\enquote {\bibinfo {title} {Review of chemical-kinetic problems of future {NASA} missions. {I} - {Earth} entries},}\ }\href {\doibase 10.2514/3.431} {\bibfield  {journal} {\bibinfo  {journal} {Journal of Thermophysics and Heat Transfer}\ }\textbf {\bibinfo {volume} {7}},\ \bibinfo {pages} {385--398} (\bibinfo {year} {1993})}\BibitemShut {NoStop}%
\bibitem [{\citenamefont {Park}\ \emph {et~al.}(1994)\citenamefont {Park}, \citenamefont {Howe}, \citenamefont {Jaffe},\ and\ \citenamefont {Candler}}]{Park1994}%
  \BibitemOpen
  \bibfield  {author} {\bibinfo {author} {\bibfnamefont {C.}~\bibnamefont {Park}}, \bibinfo {author} {\bibfnamefont {J.~T.}\ \bibnamefont {Howe}}, \bibinfo {author} {\bibfnamefont {R.~L.}\ \bibnamefont {Jaffe}}, \ and\ \bibinfo {author} {\bibfnamefont {G.~V.}\ \bibnamefont {Candler}},\ }\bibfield  {title} {\enquote {\bibinfo {title} {Review of chemical-kinetic problems of future {NASA} missions. {II} - {Mars} entries},}\ }\href {\doibase 10.2514/3.496} {\bibfield  {journal} {\bibinfo  {journal} {Journal of Thermophysics and Heat Transfer}\ }\textbf {\bibinfo {volume} {8}},\ \bibinfo {pages} {9--23} (\bibinfo {year} {1994})}\BibitemShut {NoStop}%
\bibitem [{\citenamefont {Park}, \citenamefont {Jaffe},\ and\ \citenamefont {Partridge}(2001)}]{Park2001}%
  \BibitemOpen
  \bibfield  {author} {\bibinfo {author} {\bibfnamefont {C.}~\bibnamefont {Park}}, \bibinfo {author} {\bibfnamefont {R.~L.}\ \bibnamefont {Jaffe}}, \ and\ \bibinfo {author} {\bibfnamefont {H.}~\bibnamefont {Partridge}},\ }\bibfield  {title} {\enquote {\bibinfo {title} {Chemical-kinetic parameters of hyperbolic {Earth} entry},}\ }\href {\doibase 10.2514/2.6582} {\bibfield  {journal} {\bibinfo  {journal} {Journal of Thermophysics and Heat Transfer}\ }\textbf {\bibinfo {volume} {15}},\ \bibinfo {pages} {76–90} (\bibinfo {year} {2001})}\BibitemShut {NoStop}%
\bibitem [{\citenamefont {Sarma}(2000)}]{Sarma2000}%
  \BibitemOpen
  \bibfield  {author} {\bibinfo {author} {\bibfnamefont {G.}~\bibnamefont {Sarma}},\ }\bibfield  {title} {\enquote {\bibinfo {title} {Physico–chemical modelling in hypersonic flow simulation},}\ }\href {\doibase 10.1016/s0376-0421(00)00004-x} {\bibfield  {journal} {\bibinfo  {journal} {Progress in Aerospace Sciences}\ }\textbf {\bibinfo {volume} {36}},\ \bibinfo {pages} {281–349} (\bibinfo {year} {2000})}\BibitemShut {NoStop}%
\bibitem [{\citenamefont {Pietanza}\ \emph {et~al.}(2021)\citenamefont {Pietanza}, \citenamefont {Guaitella}, \citenamefont {Aquilanti}, \citenamefont {Armenise}, \citenamefont {Bogaerts}, \citenamefont {Capitelli}, \citenamefont {Colonna}, \citenamefont {Guerra}, \citenamefont {Engeln}, \citenamefont {Kustova}, \citenamefont {Lombardi}, \citenamefont {Palazzetti},\ and\ \citenamefont {Silva}}]{Pietanza2021}%
  \BibitemOpen
  \bibfield  {author} {\bibinfo {author} {\bibfnamefont {L.~D.}\ \bibnamefont {Pietanza}}, \bibinfo {author} {\bibfnamefont {O.}~\bibnamefont {Guaitella}}, \bibinfo {author} {\bibfnamefont {V.}~\bibnamefont {Aquilanti}}, \bibinfo {author} {\bibfnamefont {I.}~\bibnamefont {Armenise}}, \bibinfo {author} {\bibfnamefont {A.}~\bibnamefont {Bogaerts}}, \bibinfo {author} {\bibfnamefont {M.}~\bibnamefont {Capitelli}}, \bibinfo {author} {\bibfnamefont {G.}~\bibnamefont {Colonna}}, \bibinfo {author} {\bibfnamefont {V.}~\bibnamefont {Guerra}}, \bibinfo {author} {\bibfnamefont {R.}~\bibnamefont {Engeln}}, \bibinfo {author} {\bibfnamefont {E.}~\bibnamefont {Kustova}}, \bibinfo {author} {\bibfnamefont {A.}~\bibnamefont {Lombardi}}, \bibinfo {author} {\bibfnamefont {F.}~\bibnamefont {Palazzetti}}, \ and\ \bibinfo {author} {\bibfnamefont {T.}~\bibnamefont {Silva}},\ }\bibfield  {title} {\enquote {\bibinfo {title} {Advances in non-equilibrium {$\rm CO_2$} plasma kinetics: A theoretical and experimental review},}\ }\href
  {\doibase 10.1140/epjd/s10053-021-00226-0} {\bibfield  {journal} {\bibinfo  {journal} {The European Physical Journal D}\ }\textbf {\bibinfo {volume} {75}} (\bibinfo {year} {2021}),\ 10.1140/epjd/s10053-021-00226-0}\BibitemShut {NoStop}%
\bibitem [{\citenamefont {{National Academies of Sciences, Engineering, and Medicine}}(2022)}]{NAP26522}%
  \BibitemOpen
  \bibfield  {author} {\bibinfo {author} {\bibnamefont {{National Academies of Sciences, Engineering, and Medicine}}},\ }\href {\doibase 10.17226/26522} {\emph {\bibinfo {title} {Origins, Worlds, and Life: A Decadal Strategy for Planetary Science and Astrobiology 2023-2032}}}\ (\bibinfo  {publisher} {The National Academies Press},\ \bibinfo {address} {Washington, DC},\ \bibinfo {year} {2022})\BibitemShut {NoStop}%
\bibitem [{\citenamefont {Higdon}\ \emph {et~al.}(2018)\citenamefont {Higdon}, \citenamefont {Cruden}, \citenamefont {Brandis}, \citenamefont {Liechty}, \citenamefont {Goldstein},\ and\ \citenamefont {Varghese}}]{Higdon2018}%
  \BibitemOpen
  \bibfield  {author} {\bibinfo {author} {\bibfnamefont {K.~J.}\ \bibnamefont {Higdon}}, \bibinfo {author} {\bibfnamefont {B.~A.}\ \bibnamefont {Cruden}}, \bibinfo {author} {\bibfnamefont {A.~M.}\ \bibnamefont {Brandis}}, \bibinfo {author} {\bibfnamefont {D.~S.}\ \bibnamefont {Liechty}}, \bibinfo {author} {\bibfnamefont {D.~B.}\ \bibnamefont {Goldstein}}, \ and\ \bibinfo {author} {\bibfnamefont {P.~L.}\ \bibnamefont {Varghese}},\ }\bibfield  {title} {\enquote {\bibinfo {title} {{Direct Simulation Monte Carlo} shock simulation of {Saturn} entry probe conditions},}\ }\href {\doibase 10.2514/1.t5275} {\bibfield  {journal} {\bibinfo  {journal} {Journal of Thermophysics and Heat Transfer}\ }\textbf {\bibinfo {volume} {32}},\ \bibinfo {pages} {680--690} (\bibinfo {year} {2018})}\BibitemShut {NoStop}%
\bibitem [{\citenamefont {Liu}\ \emph {et~al.}(2021)\citenamefont {Liu}, \citenamefont {James}, \citenamefont {Morgan}, \citenamefont {Jacobs}, \citenamefont {Gollan},\ and\ \citenamefont {McIntyre}}]{Liu2021}%
  \BibitemOpen
  \bibfield  {author} {\bibinfo {author} {\bibfnamefont {Y.}~\bibnamefont {Liu}}, \bibinfo {author} {\bibfnamefont {C.~M.}\ \bibnamefont {James}}, \bibinfo {author} {\bibfnamefont {R.~G.}\ \bibnamefont {Morgan}}, \bibinfo {author} {\bibfnamefont {P.}~\bibnamefont {Jacobs}}, \bibinfo {author} {\bibfnamefont {R.}~\bibnamefont {Gollan}}, \ and\ \bibinfo {author} {\bibfnamefont {T.}~\bibnamefont {McIntyre}},\ }\bibfield  {title} {\enquote {\bibinfo {title} {Electron number density measurements in a {Saturn} entry condition},}\ }in\ \href {\doibase 10.2514/6.2021-0209} {\emph {\bibinfo {booktitle} {{AIAA} Scitech 2021 Forum}}}\ (\bibinfo  {publisher} {American Institute of Aeronautics and Astronautics},\ \bibinfo {year} {2021})\BibitemShut {NoStop}%
\bibitem [{\citenamefont {Hansson}\ \emph {et~al.}(2021)\citenamefont {Hansson}, \citenamefont {Carroll}, \citenamefont {Poovathingal},\ and\ \citenamefont {Boyd}}]{Hansson2021}%
  \BibitemOpen
  \bibfield  {author} {\bibinfo {author} {\bibfnamefont {K.}~\bibnamefont {Hansson}}, \bibinfo {author} {\bibfnamefont {A.~T.}\ \bibnamefont {Carroll}}, \bibinfo {author} {\bibfnamefont {S.~J.}\ \bibnamefont {Poovathingal}}, \ and\ \bibinfo {author} {\bibfnamefont {I.~D.}\ \bibnamefont {Boyd}},\ }\bibfield  {title} {\enquote {\bibinfo {title} {Analysis of chemical kinetic parameters for hydrogen atmospheres},}\ }in\ \href {\doibase 10.2514/6.2021-0706} {\emph {\bibinfo {booktitle} {{AIAA} Scitech 2021 Forum}}}\ (\bibinfo  {publisher} {American Institute of Aeronautics and Astronautics},\ \bibinfo {year} {2021})\BibitemShut {NoStop}%
\bibitem [{\citenamefont {Carroll}\ \emph {et~al.}(2023)\citenamefont {Carroll}, \citenamefont {Blanquart}, \citenamefont {Brandis},\ and\ \citenamefont {Cruden}}]{Carroll2023}%
  \BibitemOpen
  \bibfield  {author} {\bibinfo {author} {\bibfnamefont {A.~T.}\ \bibnamefont {Carroll}}, \bibinfo {author} {\bibfnamefont {G.}~\bibnamefont {Blanquart}}, \bibinfo {author} {\bibfnamefont {A.~M.}\ \bibnamefont {Brandis}}, \ and\ \bibinfo {author} {\bibfnamefont {B.~A.}\ \bibnamefont {Cruden}},\ }\bibfield  {title} {\enquote {\bibinfo {title} {Kinetic and transport modeling for entry flows in hydrogen-helium atmospheres},}\ }in\ \href {\doibase 10.2514/6.2023-3728} {\emph {\bibinfo {booktitle} {AIAA Aviation 2023 Forum}}}\ (\bibinfo  {publisher} {American Institute of Aeronautics and Astronautics},\ \bibinfo {year} {2023})\BibitemShut {NoStop}%
\bibitem [{\citenamefont {Coelho}\ and\ \citenamefont {da~Silva}(2023)}]{Coelho2023}%
  \BibitemOpen
  \bibfield  {author} {\bibinfo {author} {\bibfnamefont {J.}~\bibnamefont {Coelho}}\ and\ \bibinfo {author} {\bibfnamefont {M.~L.}\ \bibnamefont {da~Silva}},\ }\bibfield  {title} {\enquote {\bibinfo {title} {Aerothermodynamic analysis of {Neptune} ballistic entry and aerocapture flows},}\ }\href {\doibase 10.1016/j.asr.2022.12.024} {\bibfield  {journal} {\bibinfo  {journal} {Advances in Space Research}\ }\textbf {\bibinfo {volume} {71}},\ \bibinfo {pages} {3408--3432} (\bibinfo {year} {2023})}\BibitemShut {NoStop}%
\bibitem [{\citenamefont {Leibowitz}(1973)}]{Leibowitz1973_1}%
  \BibitemOpen
  \bibfield  {author} {\bibinfo {author} {\bibfnamefont {L.~P.}\ \bibnamefont {Leibowitz}},\ }\bibfield  {title} {\enquote {\bibinfo {title} {Measurements of the structure of an ionizing shock wave in a hydrogen-helium mixture},}\ }\href {\doibase 10.1063/1.1694174} {\bibfield  {journal} {\bibinfo  {journal} {The Physics of Fluids}\ }\textbf {\bibinfo {volume} {16}},\ \bibinfo {pages} {59--68} (\bibinfo {year} {1973})}\BibitemShut {NoStop}%
\bibitem [{\citenamefont {Jacobs}, \citenamefont {Giedt},\ and\ \citenamefont {Cohen}(1967)}]{Jacobs1967}%
  \BibitemOpen
  \bibfield  {author} {\bibinfo {author} {\bibfnamefont {T.~A.}\ \bibnamefont {Jacobs}}, \bibinfo {author} {\bibfnamefont {R.~R.}\ \bibnamefont {Giedt}}, \ and\ \bibinfo {author} {\bibfnamefont {N.}~\bibnamefont {Cohen}},\ }\bibfield  {title} {\enquote {\bibinfo {title} {Kinetics of hydrogen halides in shock waves. {II}. a new measurement of the hydrogen dissociation rate},}\ }\href {\doibase 10.1063/1.1711890} {\bibfield  {journal} {\bibinfo  {journal} {The Journal of Chemical Physics}\ }\textbf {\bibinfo {volume} {47}},\ \bibinfo {pages} {54--57} (\bibinfo {year} {1967})}\BibitemShut {NoStop}%
\bibitem [{\citenamefont {Cohen}\ and\ \citenamefont {Westberg}(1983)}]{Cohen1983}%
  \BibitemOpen
  \bibfield  {author} {\bibinfo {author} {\bibfnamefont {N.}~\bibnamefont {Cohen}}\ and\ \bibinfo {author} {\bibfnamefont {K.~R.}\ \bibnamefont {Westberg}},\ }\bibfield  {title} {\enquote {\bibinfo {title} {Chemical kinetic data sheets for high-temperature chemical reactions},}\ }\href {\doibase 10.1063/1.555692} {\bibfield  {journal} {\bibinfo  {journal} {Journal of Physical and Chemical Reference Data}\ }\textbf {\bibinfo {volume} {12}},\ \bibinfo {pages} {531–590} (\bibinfo {year} {1983})}\BibitemShut {NoStop}%
\bibitem [{\citenamefont {Baulch}(1972)}]{Baulch1972}%
  \BibitemOpen
  \bibfield  {author} {\bibinfo {author} {\bibfnamefont {D.}~\bibnamefont {Baulch}},\ }\href@noop {} {\emph {\bibinfo {title} {Evaluated Kinetic Data for High Temperature Reactions}}}\ (\bibinfo  {publisher} {Butterworths},\ \bibinfo {year} {1972})\BibitemShut {NoStop}%
\bibitem [{\citenamefont {Baulch}\ \emph {et~al.}(2005)\citenamefont {Baulch}, \citenamefont {Bowman}, \citenamefont {Cobos}, \citenamefont {Cox}, \citenamefont {Just}, \citenamefont {Kerr}, \citenamefont {Pilling}, \citenamefont {Stocker}, \citenamefont {Troe}, \citenamefont {Tsang}, \citenamefont {Walker},\ and\ \citenamefont {Warnatz}}]{Baulch2005}%
  \BibitemOpen
  \bibfield  {author} {\bibinfo {author} {\bibfnamefont {D.~L.}\ \bibnamefont {Baulch}}, \bibinfo {author} {\bibfnamefont {C.~T.}\ \bibnamefont {Bowman}}, \bibinfo {author} {\bibfnamefont {C.~J.}\ \bibnamefont {Cobos}}, \bibinfo {author} {\bibfnamefont {R.~A.}\ \bibnamefont {Cox}}, \bibinfo {author} {\bibfnamefont {T.}~\bibnamefont {Just}}, \bibinfo {author} {\bibfnamefont {J.~A.}\ \bibnamefont {Kerr}}, \bibinfo {author} {\bibfnamefont {M.~J.}\ \bibnamefont {Pilling}}, \bibinfo {author} {\bibfnamefont {D.}~\bibnamefont {Stocker}}, \bibinfo {author} {\bibfnamefont {J.}~\bibnamefont {Troe}}, \bibinfo {author} {\bibfnamefont {W.}~\bibnamefont {Tsang}}, \bibinfo {author} {\bibfnamefont {R.~W.}\ \bibnamefont {Walker}}, \ and\ \bibinfo {author} {\bibfnamefont {J.}~\bibnamefont {Warnatz}},\ }\bibfield  {title} {\enquote {\bibinfo {title} {Evaluated kinetic data for combustion modeling: Supplement {II}},}\ }\href {\doibase 10.1063/1.1748524} {\bibfield  {journal} {\bibinfo  {journal} {Journal of Physical and Chemical
  Reference Data}\ }\textbf {\bibinfo {volume} {34}},\ \bibinfo {pages} {757--1397} (\bibinfo {year} {2005})}\BibitemShut {NoStop}%
\bibitem [{\citenamefont {Colonna}, \citenamefont {D’Angola},\ and\ \citenamefont {Capitelli}(2012)}]{Colonna2012}%
  \BibitemOpen
  \bibfield  {author} {\bibinfo {author} {\bibfnamefont {G.}~\bibnamefont {Colonna}}, \bibinfo {author} {\bibfnamefont {A.}~\bibnamefont {D’Angola}}, \ and\ \bibinfo {author} {\bibfnamefont {M.}~\bibnamefont {Capitelli}},\ }\bibfield  {title} {\enquote {\bibinfo {title} {Statistical thermodynamic description of {$\rm H_2$} molecules in normal ortho/para mixture},}\ }\href {\doibase 10.1016/j.ijhydene.2012.03.103} {\bibfield  {journal} {\bibinfo  {journal} {International Journal of Hydrogen Energy}\ }\textbf {\bibinfo {volume} {37}},\ \bibinfo {pages} {9656–9668} (\bibinfo {year} {2012})}\BibitemShut {NoStop}%
\bibitem [{\citenamefont {Popovas}\ and\ \citenamefont {J{\o}rgensen}(2016)}]{Popovas2016}%
  \BibitemOpen
  \bibfield  {author} {\bibinfo {author} {\bibfnamefont {A.}~\bibnamefont {Popovas}}\ and\ \bibinfo {author} {\bibfnamefont {U.~G.}\ \bibnamefont {J{\o}rgensen}},\ }\bibfield  {title} {\enquote {\bibinfo {title} {Partition functions - {I.} improved partition functions and thermodynamic quantities for normal, equilibrium, and ortho and para molecular hydrogen},}\ }\href {\doibase 10.1051/0004-6361/201527209} {\bibfield  {journal} {\bibinfo  {journal} {Astronomy \& Astrophysics}\ }\textbf {\bibinfo {volume} {595}},\ \bibinfo {pages} {A130} (\bibinfo {year} {2016})}\BibitemShut {NoStop}%
\bibitem [{\citenamefont {Kim}\ and\ \citenamefont {Boyd}(2013)}]{Kim2013}%
  \BibitemOpen
  \bibfield  {author} {\bibinfo {author} {\bibfnamefont {J.~G.}\ \bibnamefont {Kim}}\ and\ \bibinfo {author} {\bibfnamefont {I.~D.}\ \bibnamefont {Boyd}},\ }\bibfield  {title} {\enquote {\bibinfo {title} {State-resolved master equation analysis of thermochemical nonequilibrium of nitrogen},}\ }\href {\doibase 10.1016/j.chemphys.2013.01.027} {\bibfield  {journal} {\bibinfo  {journal} {Chemical Physics}\ }\textbf {\bibinfo {volume} {415}},\ \bibinfo {pages} {237–246} (\bibinfo {year} {2013})}\BibitemShut {NoStop}%
\bibitem [{\citenamefont {Kim}, \citenamefont {Kwon},\ and\ \citenamefont {Park}(2010)}]{Kim2010}%
  \BibitemOpen
  \bibfield  {author} {\bibinfo {author} {\bibfnamefont {J.~G.}\ \bibnamefont {Kim}}, \bibinfo {author} {\bibfnamefont {O.~J.}\ \bibnamefont {Kwon}}, \ and\ \bibinfo {author} {\bibfnamefont {C.}~\bibnamefont {Park}},\ }\bibfield  {title} {\enquote {\bibinfo {title} {Master equation study and nonequilibrium chemical reactions for hydrogen molecule},}\ }\href {\doibase 10.2514/1.45283} {\bibfield  {journal} {\bibinfo  {journal} {Journal of Thermophysics and Heat Transfer}\ }\textbf {\bibinfo {volume} {24}},\ \bibinfo {pages} {281–290} (\bibinfo {year} {2010})}\BibitemShut {NoStop}%
\bibitem [{\citenamefont {Haug}, \citenamefont {Truhlar},\ and\ \citenamefont {Blais}(1987)}]{Haug1987}%
  \BibitemOpen
  \bibfield  {author} {\bibinfo {author} {\bibfnamefont {K.}~\bibnamefont {Haug}}, \bibinfo {author} {\bibfnamefont {D.~G.}\ \bibnamefont {Truhlar}}, \ and\ \bibinfo {author} {\bibfnamefont {N.~C.}\ \bibnamefont {Blais}},\ }\bibfield  {title} {\enquote {\bibinfo {title} {{Monte} {Carlo} trajectory and master equation simulation of the nonequilibrium dissociation rate coefficient for {Ar}+{$\rm H_2$}$\rightarrow${Ar}+2{H} at 4500 {K}},}\ }\href {\doibase 10.1063/1.452073} {\bibfield  {journal} {\bibinfo  {journal} {The Journal of Chemical Physics}\ }\textbf {\bibinfo {volume} {86}},\ \bibinfo {pages} {2697–2716} (\bibinfo {year} {1987})}\BibitemShut {NoStop}%
\bibitem [{\citenamefont {Kim}(2015)}]{Kim2015}%
  \BibitemOpen
  \bibfield  {author} {\bibinfo {author} {\bibfnamefont {J.~G.}\ \bibnamefont {Kim}},\ }\bibfield  {title} {\enquote {\bibinfo {title} {Rovibrational energy transitions and coupled chemical reaction modeling of {H}+{$\rm H_2$} and {He}+{$\rm H_2$} in {DSMC}},}\ }\href {\doibase 10.5139/ijass.2015.16.3.347} {\bibfield  {journal} {\bibinfo  {journal} {International Journal of Aeronautical and Space Sciences}\ }\textbf {\bibinfo {volume} {16}},\ \bibinfo {pages} {347–359} (\bibinfo {year} {2015})}\BibitemShut {NoStop}%
\bibitem [{Note1()}]{Note1}%
  \BibitemOpen
  \bibinfo {note} {Number densities, rovibrational temperatures, and rate constants for these studies were only reported as plots. Therefore, these values have all been obtained via digitizations.}\BibitemShut {Stop}%
\bibitem [{Note2()}]{Note2}%
  \BibitemOpen
  \bibinfo {note} {For the case of M=$\protect \rm H_2$, a modified equilibrium constant, $K_{\protect \rm eq,mod}=2K_{\protect \rm eq}$, was required to reproduce the equilibrium number densities plotted by Kim and Boyd~\cite {Kim2012}. Additionally, for the case of M=H, $k_{\protect \rm d}$ values were extracted from the number density profiles assuming $n_{\protect \rm M}$ was fixed and equal to $n_{\protect \rm H,0}$.}\BibitemShut {Stop}%
\bibitem [{\citenamefont {Vargas}, \citenamefont {Monge-Palacios},\ and\ \citenamefont {Lacoste}(2024)}]{Vargas2023}%
  \BibitemOpen
  \bibfield  {author} {\bibinfo {author} {\bibfnamefont {J.}~\bibnamefont {Vargas}}, \bibinfo {author} {\bibfnamefont {M.}~\bibnamefont {Monge-Palacios}}, \ and\ \bibinfo {author} {\bibfnamefont {D.~A.}\ \bibnamefont {Lacoste}},\ }\bibfield  {title} {\enquote {\bibinfo {title} {State-specific dissociation and inelastic rate constants for collisions of {$\rm H_2$} with {H} and {He}},}\ }\href {\doibase 10.2514/1.T6878} {\bibfield  {journal} {\bibinfo  {journal} {Journal of Thermophysics and Heat Transfer}\ }\textbf {\bibinfo {volume} {38}},\ \bibinfo {pages} {210--221} (\bibinfo {year} {2024})}\BibitemShut {NoStop}%
\bibitem [{\citenamefont {Munafò}, \citenamefont {Panesi},\ and\ \citenamefont {Magin}(2014)}]{Munafo2014}%
  \BibitemOpen
  \bibfield  {author} {\bibinfo {author} {\bibfnamefont {A.}~\bibnamefont {Munafò}}, \bibinfo {author} {\bibfnamefont {M.}~\bibnamefont {Panesi}}, \ and\ \bibinfo {author} {\bibfnamefont {T.~E.}\ \bibnamefont {Magin}},\ }\bibfield  {title} {\enquote {\bibinfo {title} {Boltzmann rovibrational collisional coarse-grained model for internal energy excitation and dissociation in hypersonic flows},}\ }\href {\doibase 10.1103/physreve.89.023001} {\bibfield  {journal} {\bibinfo  {journal} {Physical Review E}\ }\textbf {\bibinfo {volume} {89}} (\bibinfo {year} {2014}),\ 10.1103/physreve.89.023001}\BibitemShut {NoStop}%
\bibitem [{\citenamefont {Sahai}\ \emph {et~al.}(2019)\citenamefont {Sahai}, \citenamefont {Lopez}, \citenamefont {Johnston},\ and\ \citenamefont {Panesi}}]{Sahai2019}%
  \BibitemOpen
  \bibfield  {author} {\bibinfo {author} {\bibfnamefont {A.}~\bibnamefont {Sahai}}, \bibinfo {author} {\bibfnamefont {B.}~\bibnamefont {Lopez}}, \bibinfo {author} {\bibfnamefont {C.~O.}\ \bibnamefont {Johnston}}, \ and\ \bibinfo {author} {\bibfnamefont {M.}~\bibnamefont {Panesi}},\ }\bibfield  {title} {\enquote {\bibinfo {title} {Reduced-order modeling of non-equilibrium kinetics and radiation for {$\rm CO_2$} axisymmetric wake flows},}\ }in\ \href {\doibase 10.1063/1.5119622} {\emph {\bibinfo {booktitle} {31ST INTERNATIONAL SYMPOSIUM ON RAREFIED GAS DYNAMICS: RGD31}}},\ Vol.\ \bibinfo {volume} {2132}\ (\bibinfo  {publisher} {AIP Publishing},\ \bibinfo {year} {2019})\ p.\ \bibinfo {pages} {130002}\BibitemShut {NoStop}%
\bibitem [{\citenamefont {Gardiner}\ and\ \citenamefont {Kistiakowsky}(1961)}]{Gardiner1961}%
  \BibitemOpen
  \bibfield  {author} {\bibinfo {author} {\bibfnamefont {W.~C.}\ \bibnamefont {Gardiner}}\ and\ \bibinfo {author} {\bibfnamefont {G.~B.}\ \bibnamefont {Kistiakowsky}},\ }\bibfield  {title} {\enquote {\bibinfo {title} {Thermal dissociation rate of hydrogen},}\ }\href {\doibase 10.1063/1.1732141} {\bibfield  {journal} {\bibinfo  {journal} {The Journal of Chemical Physics}\ }\textbf {\bibinfo {volume} {35}},\ \bibinfo {pages} {1765–1770} (\bibinfo {year} {1961})}\BibitemShut {NoStop}%
\bibitem [{\citenamefont {Sutton}(1962)}]{Sutton1962}%
  \BibitemOpen
  \bibfield  {author} {\bibinfo {author} {\bibfnamefont {E.~A.}\ \bibnamefont {Sutton}},\ }\bibfield  {title} {\enquote {\bibinfo {title} {Measurement of the dissociation rates of hydrogen and deuterium},}\ }\href {\doibase 10.1063/1.1732403} {\bibfield  {journal} {\bibinfo  {journal} {The Journal of Chemical Physics}\ }\textbf {\bibinfo {volume} {36}},\ \bibinfo {pages} {2923--2931} (\bibinfo {year} {1962})}\BibitemShut {NoStop}%
\bibitem [{\citenamefont {Rink}(1962)}]{Rink1962}%
  \BibitemOpen
  \bibfield  {author} {\bibinfo {author} {\bibfnamefont {J.~P.}\ \bibnamefont {Rink}},\ }\bibfield  {title} {\enquote {\bibinfo {title} {Shock tube determination of dissociation rates of hydrogen},}\ }\href {\doibase 10.1063/1.1732309} {\bibfield  {journal} {\bibinfo  {journal} {The Journal of Chemical Physics}\ }\textbf {\bibinfo {volume} {36}},\ \bibinfo {pages} {262--265} (\bibinfo {year} {1962})}\BibitemShut {NoStop}%
\bibitem [{\citenamefont {Patch}(1962)}]{Patch1962}%
  \BibitemOpen
  \bibfield  {author} {\bibinfo {author} {\bibfnamefont {R.~W.}\ \bibnamefont {Patch}},\ }\bibfield  {title} {\enquote {\bibinfo {title} {Shock-tube measurement of dissociation rates of hydrogen},}\ }\href {\doibase 10.1063/1.1701291} {\bibfield  {journal} {\bibinfo  {journal} {The Journal of Chemical Physics}\ }\textbf {\bibinfo {volume} {36}},\ \bibinfo {pages} {1919--1924} (\bibinfo {year} {1962})}\BibitemShut {NoStop}%
\bibitem [{\citenamefont {Myerson}\ and\ \citenamefont {Watt}(1968)}]{Myerson1968}%
  \BibitemOpen
  \bibfield  {author} {\bibinfo {author} {\bibfnamefont {A.~L.}\ \bibnamefont {Myerson}}\ and\ \bibinfo {author} {\bibfnamefont {W.~S.}\ \bibnamefont {Watt}},\ }\bibfield  {title} {\enquote {\bibinfo {title} {Atom-formation rates behind shock waves in hydrogen and the effect of added oxygen},}\ }\href {\doibase 10.1063/1.1669839} {\bibfield  {journal} {\bibinfo  {journal} {The Journal of Chemical Physics}\ }\textbf {\bibinfo {volume} {49}},\ \bibinfo {pages} {425–433} (\bibinfo {year} {1968})}\BibitemShut {NoStop}%
\bibitem [{\citenamefont {Hurle}, \citenamefont {Jones},\ and\ \citenamefont {Rosenfeld}(1969)}]{Hurle1969}%
  \BibitemOpen
  \bibfield  {author} {\bibinfo {author} {\bibfnamefont {I.~R.}\ \bibnamefont {Hurle}}, \bibinfo {author} {\bibfnamefont {A.}~\bibnamefont {Jones}}, \ and\ \bibinfo {author} {\bibfnamefont {J.~L.~J.}\ \bibnamefont {Rosenfeld}},\ }\bibfield  {title} {\enquote {\bibinfo {title} {Shock-wave observations of rate constants for atomic hydrogen recombination from 2500 to 7000 {K}: Collisional stabilization by exchange of hydrogen atoms},}\ }\href {\doibase 10.1098/rspa.1969.0074} {\bibfield  {journal} {\bibinfo  {journal} {Proceedings of the Royal Society of London. A. Mathematical and Physical Sciences}\ }\textbf {\bibinfo {volume} {310}},\ \bibinfo {pages} {253--276} (\bibinfo {year} {1969})}\BibitemShut {NoStop}%
\bibitem [{\citenamefont {Breshears}\ and\ \citenamefont {Bird}(1972)}]{Breshears1972}%
  \BibitemOpen
  \bibfield  {author} {\bibinfo {author} {\bibfnamefont {W.}~\bibnamefont {Breshears}}\ and\ \bibinfo {author} {\bibfnamefont {P.}~\bibnamefont {Bird}},\ }\bibfield  {title} {\enquote {\bibinfo {title} {Precise measurements of diatomic dissociation rates in shock waves},}\ }\href {\doibase 10.2172/4653639} {\bibfield  {journal} {\bibinfo  {journal} {Symposium (International) on Combustion}\ } (\bibinfo {year} {1972}),\ 10.2172/4653639}\BibitemShut {NoStop}%
\bibitem [{\citenamefont {Gordon}\ and\ \citenamefont {McBride}(1996)}]{Gordon1996}%
  \BibitemOpen
  \bibfield  {author} {\bibinfo {author} {\bibfnamefont {S.}~\bibnamefont {Gordon}}\ and\ \bibinfo {author} {\bibfnamefont {B.~J.}\ \bibnamefont {McBride}},\ }\href@noop {} {\enquote {\bibinfo {title} {Computer program for calculation of complex chemical equilibrium compositions and applications},}\ }\bibinfo {type} {Tech. Rep.}\ (\bibinfo  {institution} {NASA Reference Publication 1311},\ \bibinfo {year} {1996})\BibitemShut {NoStop}%
\bibitem [{\citenamefont {Lynch}, \citenamefont {Schwab},\ and\ \citenamefont {Michael}(1976)}]{Lynch1976}%
  \BibitemOpen
  \bibfield  {author} {\bibinfo {author} {\bibfnamefont {K.~P.}\ \bibnamefont {Lynch}}, \bibinfo {author} {\bibfnamefont {T.~C.}\ \bibnamefont {Schwab}}, \ and\ \bibinfo {author} {\bibfnamefont {J.~V.}\ \bibnamefont {Michael}},\ }\bibfield  {title} {\enquote {\bibinfo {title} {Lyman-$\alpha$ absorption photometry at high pressure and atom density kinetic results for {H} recombination},}\ }\href {\doibase 10.1002/kin.550080503} {\bibfield  {journal} {\bibinfo  {journal} {International Journal of Chemical Kinetics}\ }\textbf {\bibinfo {volume} {8}},\ \bibinfo {pages} {651--671} (\bibinfo {year} {1976})}\BibitemShut {NoStop}%
\bibitem [{\citenamefont {Larkin}(1968)}]{Larkin1968}%
  \BibitemOpen
  \bibfield  {author} {\bibinfo {author} {\bibfnamefont {F.~S.}\ \bibnamefont {Larkin}},\ }\bibfield  {title} {\enquote {\bibinfo {title} {Homogeneous rate of recombination of hydrogen atoms},}\ }\href {\doibase 10.1139/v68-164} {\bibfield  {journal} {\bibinfo  {journal} {Canadian Journal of Chemistry}\ }\textbf {\bibinfo {volume} {46}},\ \bibinfo {pages} {1005–1015} (\bibinfo {year} {1968})}\BibitemShut {NoStop}%
\bibitem [{\citenamefont {Bennett}\ and\ \citenamefont {Blackmore}(1968)}]{Bennett1968}%
  \BibitemOpen
  \bibfield  {author} {\bibinfo {author} {\bibfnamefont {J.~E.}\ \bibnamefont {Bennett}}\ and\ \bibinfo {author} {\bibfnamefont {D.~R.}\ \bibnamefont {Blackmore}},\ }\bibfield  {title} {\enquote {\bibinfo {title} {The measurement of the rate of recombination of hydrogen atoms at room temperature by means of e. s. r. spectroscopy},}\ }\href {\doibase 10.1098/rspa.1968.0132} {\bibfield  {journal} {\bibinfo  {journal} {Proceedings of the Royal Society of London. Series A. Mathematical and Physical Sciences}\ }\textbf {\bibinfo {volume} {305}},\ \bibinfo {pages} {553--574} (\bibinfo {year} {1968})}\BibitemShut {NoStop}%
\bibitem [{\citenamefont {Bennett}\ and\ \citenamefont {Blackmore}(1970)}]{Bennett1970}%
  \BibitemOpen
  \bibfield  {author} {\bibinfo {author} {\bibfnamefont {J.~E.}\ \bibnamefont {Bennett}}\ and\ \bibinfo {author} {\bibfnamefont {D.~R.}\ \bibnamefont {Blackmore}},\ }\bibfield  {title} {\enquote {\bibinfo {title} {Gas-phase recombination rates of hydrogen and deuterium atoms},}\ }\href {\doibase 10.1063/1.1673957} {\bibfield  {journal} {\bibinfo  {journal} {The Journal of Chemical Physics}\ }\textbf {\bibinfo {volume} {53}},\ \bibinfo {pages} {4400--4401} (\bibinfo {year} {1970})}\BibitemShut {NoStop}%
\bibitem [{\citenamefont {Bennett}\ and\ \citenamefont {Blackmore}(1971)}]{Bennett1971}%
  \BibitemOpen
  \bibfield  {author} {\bibinfo {author} {\bibfnamefont {J.}~\bibnamefont {Bennett}}\ and\ \bibinfo {author} {\bibfnamefont {D.}~\bibnamefont {Blackmore}},\ }\bibfield  {title} {\enquote {\bibinfo {title} {Rates of gas-phase hydrogen-atom recombination at room temperature in the presence of added gases},}\ }\href {\doibase 10.1016/s0082-0784(71)80009-7} {\bibfield  {journal} {\bibinfo  {journal} {Symposium (International) on Combustion}\ }\textbf {\bibinfo {volume} {13}},\ \bibinfo {pages} {51–59} (\bibinfo {year} {1971})}\BibitemShut {NoStop}%
\bibitem [{\citenamefont {Trainor}, \citenamefont {Ham},\ and\ \citenamefont {Kaufman}(1973)}]{Trainor1973}%
  \BibitemOpen
  \bibfield  {author} {\bibinfo {author} {\bibfnamefont {D.~W.}\ \bibnamefont {Trainor}}, \bibinfo {author} {\bibfnamefont {D.~O.}\ \bibnamefont {Ham}}, \ and\ \bibinfo {author} {\bibfnamefont {F.}~\bibnamefont {Kaufman}},\ }\bibfield  {title} {\enquote {\bibinfo {title} {Gas phase recombination of hydrogen and deuterium atoms},}\ }\href {\doibase 10.1063/1.1679024} {\bibfield  {journal} {\bibinfo  {journal} {The Journal of Chemical Physics}\ }\textbf {\bibinfo {volume} {58}},\ \bibinfo {pages} {4599--4609} (\bibinfo {year} {1973})}\BibitemShut {NoStop}%
\bibitem [{\citenamefont {Mitchell}\ and\ \citenamefont {LeRoy}(1977)}]{Mitchell1977}%
  \BibitemOpen
  \bibfield  {author} {\bibinfo {author} {\bibfnamefont {D.~N.}\ \bibnamefont {Mitchell}}\ and\ \bibinfo {author} {\bibfnamefont {D.~J.}\ \bibnamefont {LeRoy}},\ }\bibfield  {title} {\enquote {\bibinfo {title} {An experimental test of the orbiting resonance theory of hydrogen atom recombination at room temperature},}\ }\href {\doibase 10.1063/1.434981} {\bibfield  {journal} {\bibinfo  {journal} {The Journal of Chemical Physics}\ }\textbf {\bibinfo {volume} {67}},\ \bibinfo {pages} {1042--1050} (\bibinfo {year} {1977})}\BibitemShut {NoStop}%
\bibitem [{\citenamefont {Schwenke}(1990)}]{Schwenke1990}%
  \BibitemOpen
  \bibfield  {author} {\bibinfo {author} {\bibfnamefont {D.~W.}\ \bibnamefont {Schwenke}},\ }\bibfield  {title} {\enquote {\bibinfo {title} {A theoretical prediction of hydrogen molecule dissociation-recombination rates including an accurate treatment of internal state nonequilibrium effects},}\ }\href {\doibase 10.1063/1.458213} {\bibfield  {journal} {\bibinfo  {journal} {The Journal of Chemical Physics}\ }\textbf {\bibinfo {volume} {92}},\ \bibinfo {pages} {7267–7282} (\bibinfo {year} {1990})}\BibitemShut {NoStop}%
\bibitem [{\citenamefont {Schwenke}(1988)}]{Schwenke1988}%
  \BibitemOpen
  \bibfield  {author} {\bibinfo {author} {\bibfnamefont {D.~W.}\ \bibnamefont {Schwenke}},\ }\bibfield  {title} {\enquote {\bibinfo {title} {A new method for the direct calculation of resonance parameters with application to the quasibound states of the {$\rm H_2(X^1\Sigma_g^+)$} system},}\ }\href {\doibase 10.1007/bf01025840} {\bibfield  {journal} {\bibinfo  {journal} {Theoretica Chimica Acta}\ }\textbf {\bibinfo {volume} {74}},\ \bibinfo {pages} {381--402} (\bibinfo {year} {1988})}\BibitemShut {NoStop}%
\bibitem [{\citenamefont {Martin}, \citenamefont {Schwarz},\ and\ \citenamefont {Mandy}(1996)}]{Martin1996}%
  \BibitemOpen
  \bibfield  {author} {\bibinfo {author} {\bibfnamefont {P.~G.}\ \bibnamefont {Martin}}, \bibinfo {author} {\bibfnamefont {D.~H.}\ \bibnamefont {Schwarz}}, \ and\ \bibinfo {author} {\bibfnamefont {M.~E.}\ \bibnamefont {Mandy}},\ }\bibfield  {title} {\enquote {\bibinfo {title} {Master equation studies of the collisional excitation and dissociation of {$\rm H_2$} molecules by {H} atoms},}\ }\href {\doibase 10.1086/177053} {\bibfield  {journal} {\bibinfo  {journal} {The Astrophysical Journal}\ }\textbf {\bibinfo {volume} {461}},\ \bibinfo {pages} {265} (\bibinfo {year} {1996})}\BibitemShut {NoStop}%
\bibitem [{\citenamefont {Liu}(1973)}]{Liu1973}%
  \BibitemOpen
  \bibfield  {author} {\bibinfo {author} {\bibfnamefont {B.}~\bibnamefont {Liu}},\ }\bibfield  {title} {\enquote {\bibinfo {title} {Ab initio potential energy surface for linear {{$\rm H_3$}}},}\ }\href {\doibase 10.1063/1.1679454} {\bibfield  {journal} {\bibinfo  {journal} {The Journal of Chemical Physics}\ }\textbf {\bibinfo {volume} {58}},\ \bibinfo {pages} {1925–1937} (\bibinfo {year} {1973})}\BibitemShut {NoStop}%
\bibitem [{\citenamefont {Siegbahn}\ and\ \citenamefont {Liu}(1978)}]{Siegbahn1978}%
  \BibitemOpen
  \bibfield  {author} {\bibinfo {author} {\bibfnamefont {P.}~\bibnamefont {Siegbahn}}\ and\ \bibinfo {author} {\bibfnamefont {B.}~\bibnamefont {Liu}},\ }\bibfield  {title} {\enquote {\bibinfo {title} {An accurate three-dimensional potential energy surface for {{$\rm H_3$}}},}\ }\href {\doibase 10.1063/1.436018} {\bibfield  {journal} {\bibinfo  {journal} {The Journal of Chemical Physics}\ }\textbf {\bibinfo {volume} {68}},\ \bibinfo {pages} {2457–2465} (\bibinfo {year} {1978})}\BibitemShut {NoStop}%
\bibitem [{\citenamefont {Truhlar}\ and\ \citenamefont {Horowitz}(1978)}]{Truhlar1978}%
  \BibitemOpen
  \bibfield  {author} {\bibinfo {author} {\bibfnamefont {D.~G.}\ \bibnamefont {Truhlar}}\ and\ \bibinfo {author} {\bibfnamefont {C.~J.}\ \bibnamefont {Horowitz}},\ }\bibfield  {title} {\enquote {\bibinfo {title} {Functional representation of {Liu} and {Siegbahn’s} accurate ab-initio potential energy calculations for {H}+{$\rm H_2$}},}\ }\href {\doibase 10.1063/1.436019} {\bibfield  {journal} {\bibinfo  {journal} {The Journal of Chemical Physics}\ }\textbf {\bibinfo {volume} {68}},\ \bibinfo {pages} {2466–2476} (\bibinfo {year} {1978})}\BibitemShut {NoStop}%
\bibitem [{\citenamefont {Truhlar}\ and\ \citenamefont {Horowitz}(1979)}]{Truhlar1979}%
  \BibitemOpen
  \bibfield  {author} {\bibinfo {author} {\bibfnamefont {D.~G.}\ \bibnamefont {Truhlar}}\ and\ \bibinfo {author} {\bibfnamefont {C.~J.}\ \bibnamefont {Horowitz}},\ }\bibfield  {title} {\enquote {\bibinfo {title} {Erratum: Functional representation of {Liu and Siegbahn’s} accurate ab-initio potential energy calculations for {H}+{$\rm H_2$}},}\ }\href {\doibase 10.1063/1.438835} {\bibfield  {journal} {\bibinfo  {journal} {The Journal of Chemical Physics}\ }\textbf {\bibinfo {volume} {71}},\ \bibinfo {pages} {1514–1514} (\bibinfo {year} {1979})}\BibitemShut {NoStop}%
\bibitem [{\citenamefont {Furudate}, \citenamefont {Fujita},\ and\ \citenamefont {Abe}(2006)}]{Furudate2006}%
  \BibitemOpen
  \bibfield  {author} {\bibinfo {author} {\bibfnamefont {M.}~\bibnamefont {Furudate}}, \bibinfo {author} {\bibfnamefont {K.}~\bibnamefont {Fujita}}, \ and\ \bibinfo {author} {\bibfnamefont {T.}~\bibnamefont {Abe}},\ }\bibfield  {title} {\enquote {\bibinfo {title} {Coupled rotational-vibrational relaxation of molecular hydrogen at high temperatures},}\ }\href {\doibase 10.2514/1.16323} {\bibfield  {journal} {\bibinfo  {journal} {Journal of Thermophysics and Heat Transfer}\ }\textbf {\bibinfo {volume} {20}},\ \bibinfo {pages} {457–464} (\bibinfo {year} {2006})}\BibitemShut {NoStop}%
\bibitem [{\citenamefont {Boothroyd}\ \emph {et~al.}(1996)\citenamefont {Boothroyd}, \citenamefont {Keogh}, \citenamefont {Martin},\ and\ \citenamefont {Peterson}}]{Boothroyd1996}%
  \BibitemOpen
  \bibfield  {author} {\bibinfo {author} {\bibfnamefont {A.~I.}\ \bibnamefont {Boothroyd}}, \bibinfo {author} {\bibfnamefont {W.~J.}\ \bibnamefont {Keogh}}, \bibinfo {author} {\bibfnamefont {P.~G.}\ \bibnamefont {Martin}}, \ and\ \bibinfo {author} {\bibfnamefont {M.~R.}\ \bibnamefont {Peterson}},\ }\bibfield  {title} {\enquote {\bibinfo {title} {A refined {{$\rm H_3$}} potential energy surface},}\ }\href {\doibase 10.1063/1.471430} {\bibfield  {journal} {\bibinfo  {journal} {The Journal of Chemical Physics}\ }\textbf {\bibinfo {volume} {104}},\ \bibinfo {pages} {7139–7152} (\bibinfo {year} {1996})}\BibitemShut {NoStop}%
\bibitem [{\citenamefont {Boothroyd}, \citenamefont {Martin},\ and\ \citenamefont {Peterson}(2003)}]{Boothroyd2003}%
  \BibitemOpen
  \bibfield  {author} {\bibinfo {author} {\bibfnamefont {A.~I.}\ \bibnamefont {Boothroyd}}, \bibinfo {author} {\bibfnamefont {P.~G.}\ \bibnamefont {Martin}}, \ and\ \bibinfo {author} {\bibfnamefont {M.~R.}\ \bibnamefont {Peterson}},\ }\bibfield  {title} {\enquote {\bibinfo {title} {Accurate analytic {He}–{$\rm H_2$} potential energy surface from a greatly expanded set of ab initio energies},}\ }\href {\doibase 10.1063/1.1589734} {\bibfield  {journal} {\bibinfo  {journal} {The Journal of Chemical Physics}\ }\textbf {\bibinfo {volume} {119}},\ \bibinfo {pages} {3187–3207} (\bibinfo {year} {2003})}\BibitemShut {NoStop}%
\bibitem [{\citenamefont {Sharma}(1994)}]{Sharma1994}%
  \BibitemOpen
  \bibfield  {author} {\bibinfo {author} {\bibfnamefont {S.~P.}\ \bibnamefont {Sharma}},\ }\bibfield  {title} {\enquote {\bibinfo {title} {Rotational relaxation of molecular hydrogen at moderate temperatures},}\ }\href {\doibase 10.2514/3.498} {\bibfield  {journal} {\bibinfo  {journal} {Journal of Thermophysics and Heat Transfer}\ }\textbf {\bibinfo {volume} {8}},\ \bibinfo {pages} {35–39} (\bibinfo {year} {1994})}\BibitemShut {NoStop}%
\bibitem [{\citenamefont {Shui}, \citenamefont {Appleton},\ and\ \citenamefont {Keck}(1971)}]{Shui1971_Ar}%
  \BibitemOpen
  \bibfield  {author} {\bibinfo {author} {\bibfnamefont {V.~H.}\ \bibnamefont {Shui}}, \bibinfo {author} {\bibfnamefont {J.~P.}\ \bibnamefont {Appleton}}, \ and\ \bibinfo {author} {\bibfnamefont {J.~C.}\ \bibnamefont {Keck}},\ }\bibfield  {title} {\enquote {\bibinfo {title} {The three-body recombination and dissociation of diatomic molecules: A comparison between theory and experiment},}\ }\href {\doibase 10.1016/s0082-0784(71)80007-3} {\bibfield  {journal} {\bibinfo  {journal} {Symposium (International) on Combustion}\ }\textbf {\bibinfo {volume} {13}},\ \bibinfo {pages} {21–35} (\bibinfo {year} {1971})}\BibitemShut {NoStop}%
\bibitem [{\citenamefont {Shui}\ and\ \citenamefont {Appleton}(1971)}]{Shui1971}%
  \BibitemOpen
  \bibfield  {author} {\bibinfo {author} {\bibfnamefont {V.~H.}\ \bibnamefont {Shui}}\ and\ \bibinfo {author} {\bibfnamefont {J.~P.}\ \bibnamefont {Appleton}},\ }\bibfield  {title} {\enquote {\bibinfo {title} {Gas-phase recombination of hydrogen: A comparison between theory and experiment},}\ }\href {\doibase 10.1063/1.1676558} {\bibfield  {journal} {\bibinfo  {journal} {The Journal of Chemical Physics}\ }\textbf {\bibinfo {volume} {55}},\ \bibinfo {pages} {3126–3132} (\bibinfo {year} {1971})}\BibitemShut {NoStop}%
\bibitem [{\citenamefont {Shui}(1973)}]{Shui1973}%
  \BibitemOpen
  \bibfield  {author} {\bibinfo {author} {\bibfnamefont {V.~H.}\ \bibnamefont {Shui}},\ }\bibfield  {title} {\enquote {\bibinfo {title} {Thermal dissociation and recombination of hydrogen according to the reactions {$\rm H_2$}+{H} $\leftrightarrow$ {H}+{H}+{H}},}\ }\href {\doibase 10.1063/1.1679071} {\bibfield  {journal} {\bibinfo  {journal} {The Journal of Chemical Physics}\ }\textbf {\bibinfo {volume} {58}},\ \bibinfo {pages} {4868–4879} (\bibinfo {year} {1973})}\BibitemShut {NoStop}%
\bibitem [{\citenamefont {Roberge}\ and\ \citenamefont {Dalgarno}(1982)}]{Roberge1982}%
  \BibitemOpen
  \bibfield  {author} {\bibinfo {author} {\bibfnamefont {W.}~\bibnamefont {Roberge}}\ and\ \bibinfo {author} {\bibfnamefont {A.}~\bibnamefont {Dalgarno}},\ }\bibfield  {title} {\enquote {\bibinfo {title} {Collision-induced dissociation of {$\rm H_2$} and {CO} molecules},}\ }\href {\doibase 10.1086/159815} {\bibfield  {journal} {\bibinfo  {journal} {The Astrophysical Journal}\ }\textbf {\bibinfo {volume} {255}},\ \bibinfo {pages} {176} (\bibinfo {year} {1982})}\BibitemShut {NoStop}%
\bibitem [{\citenamefont {Esposito}, \citenamefont {Gorse},\ and\ \citenamefont {Capitelli}(1999)}]{Esposito1999}%
  \BibitemOpen
  \bibfield  {author} {\bibinfo {author} {\bibfnamefont {F.}~\bibnamefont {Esposito}}, \bibinfo {author} {\bibfnamefont {C.}~\bibnamefont {Gorse}}, \ and\ \bibinfo {author} {\bibfnamefont {M.}~\bibnamefont {Capitelli}},\ }\bibfield  {title} {\enquote {\bibinfo {title} {Quasi-classical dynamics calculations and state-selected rate coefficients for {H}+{$\rm H_2$}(v, j)$\rightarrow$3{H} processes: Application to the global dissociation rate under thermal conditions},}\ }\href {\doibase 10.1016/s0009-2614(99)00241-9} {\bibfield  {journal} {\bibinfo  {journal} {Chemical Physics Letters}\ }\textbf {\bibinfo {volume} {303}},\ \bibinfo {pages} {636--640} (\bibinfo {year} {1999})}\BibitemShut {NoStop}%
\bibitem [{\citenamefont {Dove}\ and\ \citenamefont {Jones}(1972)}]{Dove1972}%
  \BibitemOpen
  \bibfield  {author} {\bibinfo {author} {\bibfnamefont {J.}~\bibnamefont {Dove}}\ and\ \bibinfo {author} {\bibfnamefont {D.}~\bibnamefont {Jones}},\ }\bibfield  {title} {\enquote {\bibinfo {title} {Effect of rotation on the computed thermal dissociation rate of {$\rm H_2$}},}\ }\href {\doibase 10.1016/0009-2614(72)80344-0} {\bibfield  {journal} {\bibinfo  {journal} {Chemical Physics Letters}\ }\textbf {\bibinfo {volume} {17}},\ \bibinfo {pages} {134–136} (\bibinfo {year} {1972})}\BibitemShut {NoStop}%
\bibitem [{\citenamefont {Lepp}\ and\ \citenamefont {Shull}(1983)}]{Lepp1983}%
  \BibitemOpen
  \bibfield  {author} {\bibinfo {author} {\bibfnamefont {S.}~\bibnamefont {Lepp}}\ and\ \bibinfo {author} {\bibfnamefont {J.~M.}\ \bibnamefont {Shull}},\ }\bibfield  {title} {\enquote {\bibinfo {title} {The kinetic theory of {$\rm H_2$} dissociation},}\ }\href {\doibase 10.1086/161149} {\bibfield  {journal} {\bibinfo  {journal} {The Astrophysical Journal}\ }\textbf {\bibinfo {volume} {270}},\ \bibinfo {pages} {578} (\bibinfo {year} {1983})}\BibitemShut {NoStop}%
\bibitem [{\citenamefont {Blais}\ and\ \citenamefont {Truhlar}(1979)}]{Blais1979}%
  \BibitemOpen
  \bibfield  {author} {\bibinfo {author} {\bibfnamefont {N.~C.}\ \bibnamefont {Blais}}\ and\ \bibinfo {author} {\bibfnamefont {D.~G.}\ \bibnamefont {Truhlar}},\ }\bibfield  {title} {\enquote {\bibinfo {title} {{Monte} {Carlo} trajectory study of {Ar}+{$\rm H_2$}: Vibrational selectivity of dissociative collisions at 4500 {K} and the characteristics of dissociation under equilibrium conditions},}\ }\href {\doibase 10.1063/1.437835} {\bibfield  {journal} {\bibinfo  {journal} {The Journal of Chemical Physics}\ }\textbf {\bibinfo {volume} {70}},\ \bibinfo {pages} {2962–2978} (\bibinfo {year} {1979})}\BibitemShut {NoStop}%
\bibitem [{\citenamefont {Dove}\ and\ \citenamefont {Raynor}(1979)}]{Dove1979}%
  \BibitemOpen
  \bibfield  {author} {\bibinfo {author} {\bibfnamefont {J.~E.}\ \bibnamefont {Dove}}\ and\ \bibinfo {author} {\bibfnamefont {S.}~\bibnamefont {Raynor}},\ }\bibfield  {title} {\enquote {\bibinfo {title} {An ab initio calculation of the rate of vibrational relaxation and thermal dissociation of hydrogen by helium at high temperatures},}\ }\href {\doibase 10.1021/j100464a020} {\bibfield  {journal} {\bibinfo  {journal} {The Journal of Physical Chemistry}\ }\textbf {\bibinfo {volume} {83}},\ \bibinfo {pages} {127–133} (\bibinfo {year} {1979})}\BibitemShut {NoStop}%
\bibitem [{\citenamefont {Dove}\ \emph {et~al.}(1987)\citenamefont {Dove}, \citenamefont {Rusk}, \citenamefont {Cribb},\ and\ \citenamefont {Martin}}]{Dove1987}%
  \BibitemOpen
  \bibfield  {author} {\bibinfo {author} {\bibfnamefont {J.~E.}\ \bibnamefont {Dove}}, \bibinfo {author} {\bibfnamefont {A.~C.~M.}\ \bibnamefont {Rusk}}, \bibinfo {author} {\bibfnamefont {P.~H.}\ \bibnamefont {Cribb}}, \ and\ \bibinfo {author} {\bibfnamefont {P.~G.}\ \bibnamefont {Martin}},\ }\bibfield  {title} {\enquote {\bibinfo {title} {Excitation and dissociation of molecular hydrogen in shock waves at interstellar densities},}\ }\href {\doibase 10.1086/165375} {\bibfield  {journal} {\bibinfo  {journal} {The Astrophysical Journal}\ }\textbf {\bibinfo {volume} {318}},\ \bibinfo {pages} {379} (\bibinfo {year} {1987})}\BibitemShut {NoStop}%
\bibitem [{\citenamefont {Whitlock}, \citenamefont {Muckerman},\ and\ \citenamefont {Roberts}(1974)}]{Whitlock1974}%
  \BibitemOpen
  \bibfield  {author} {\bibinfo {author} {\bibfnamefont {P.~A.}\ \bibnamefont {Whitlock}}, \bibinfo {author} {\bibfnamefont {J.~T.}\ \bibnamefont {Muckerman}}, \ and\ \bibinfo {author} {\bibfnamefont {R.~E.}\ \bibnamefont {Roberts}},\ }\bibfield  {title} {\enquote {\bibinfo {title} {Classical mechanics of recombination via the resonance complex mechanism: {H} + {H} + {M} → {$\rm H_2$} + {M} for {M} = {H}, {$\rm H_2$}, {He}, and {Ar}},}\ }\href {\doibase 10.1063/1.1681586} {\bibfield  {journal} {\bibinfo  {journal} {The Journal of Chemical Physics}\ }\textbf {\bibinfo {volume} {60}},\ \bibinfo {pages} {3658–3673} (\bibinfo {year} {1974})}\BibitemShut {NoStop}%
\bibitem [{\citenamefont {McBride}, \citenamefont {Zehe},\ and\ \citenamefont {Gordon}(2002)}]{McBride2002}%
  \BibitemOpen
  \bibfield  {author} {\bibinfo {author} {\bibfnamefont {B.~J.}\ \bibnamefont {McBride}}, \bibinfo {author} {\bibfnamefont {M.~J.}\ \bibnamefont {Zehe}}, \ and\ \bibinfo {author} {\bibfnamefont {S.}~\bibnamefont {Gordon}},\ }\href@noop {} {\enquote {\bibinfo {title} {{NASA Glenn} coefficicients for calculating thermodynamic properties of individual species},}\ }\bibinfo {type} {Tech. Rep.}\ (\bibinfo  {institution} {NASA NTRS Report/ Patent Number: NASA/TP-2002-211556},\ \bibinfo {year} {2002})\BibitemShut {NoStop}%
\bibitem [{\citenamefont {Macdonald}(2024)}]{Macdonald2024}%
  \BibitemOpen
  \bibfield  {author} {\bibinfo {author} {\bibfnamefont {R.~L.}\ \bibnamefont {Macdonald}},\ }\bibfield  {title} {\enquote {\bibinfo {title} {State-to-state study of non-equilibrium recombination of oxygen and nitrogen molecules},}\ }\href {\doibase 10.1063/5.0195238} {\bibfield  {journal} {\bibinfo  {journal} {The Journal of Chemical Physics}\ }\textbf {\bibinfo {volume} {160}} (\bibinfo {year} {2024}),\ 10.1063/5.0195238}\BibitemShut {NoStop}%
\bibitem [{\citenamefont {Singh}\ and\ \citenamefont {Schwartzentruber}(2017)}]{Singh2017}%
  \BibitemOpen
  \bibfield  {author} {\bibinfo {author} {\bibfnamefont {N.}~\bibnamefont {Singh}}\ and\ \bibinfo {author} {\bibfnamefont {T.}~\bibnamefont {Schwartzentruber}},\ }\bibfield  {title} {\enquote {\bibinfo {title} {Nonequilibrium internal energy distributions during dissociation},}\ }\href {\doibase 10.1073/pnas.1713840115} {\bibfield  {journal} {\bibinfo  {journal} {Proceedings of the National Academy of Sciences}\ }\textbf {\bibinfo {volume} {115}},\ \bibinfo {pages} {47–52} (\bibinfo {year} {2017})}\BibitemShut {NoStop}%
\bibitem [{\citenamefont {Singh}\ and\ \citenamefont {Schwartzentruber}(2020{\natexlab{c}})}]{Singh2020}%
  \BibitemOpen
  \bibfield  {author} {\bibinfo {author} {\bibfnamefont {N.}~\bibnamefont {Singh}}\ and\ \bibinfo {author} {\bibfnamefont {T.}~\bibnamefont {Schwartzentruber}},\ }\bibfield  {title} {\enquote {\bibinfo {title} {Non-boltzmann vibrational energy distributions and coupling to dissociation rate},}\ }\href {\doibase 10.1063/1.5142732} {\bibfield  {journal} {\bibinfo  {journal} {The Journal of Chemical Physics}\ }\textbf {\bibinfo {volume} {152}} (\bibinfo {year} {2020}{\natexlab{c}}),\ 10.1063/1.5142732}\BibitemShut {NoStop}%
\bibitem [{\citenamefont {Boyd}(1977)}]{Boyd1977}%
  \BibitemOpen
  \bibfield  {author} {\bibinfo {author} {\bibfnamefont {R.~K.}\ \bibnamefont {Boyd}},\ }\bibfield  {title} {\enquote {\bibinfo {title} {The linear mixture formula in non-equilibrium chemical kinetics},}\ }\href {\doibase 10.1139/v77-113} {\bibfield  {journal} {\bibinfo  {journal} {Canadian Journal of Chemistry}\ }\textbf {\bibinfo {volume} {55}},\ \bibinfo {pages} {802–811} (\bibinfo {year} {1977})}\BibitemShut {NoStop}%
\bibitem [{\citenamefont {Dove}, \citenamefont {Halperin},\ and\ \citenamefont {Raynor}(1984)}]{Dove1984}%
  \BibitemOpen
  \bibfield  {author} {\bibinfo {author} {\bibfnamefont {J.~E.}\ \bibnamefont {Dove}}, \bibinfo {author} {\bibfnamefont {S.}~\bibnamefont {Halperin}}, \ and\ \bibinfo {author} {\bibfnamefont {S.}~\bibnamefont {Raynor}},\ }\bibfield  {title} {\enquote {\bibinfo {title} {Deviations from the linear mixture rule in nonequilibrium chemical kinetics},}\ }\href {\doibase 10.1063/1.447713} {\bibfield  {journal} {\bibinfo  {journal} {The Journal of Chemical Physics}\ }\textbf {\bibinfo {volume} {81}},\ \bibinfo {pages} {799–811} (\bibinfo {year} {1984})}\BibitemShut {NoStop}%
\bibitem [{\citenamefont {Carruthers}\ and\ \citenamefont {Teitelbaum}(1988)}]{Carruthers1988}%
  \BibitemOpen
  \bibfield  {author} {\bibinfo {author} {\bibfnamefont {C.}~\bibnamefont {Carruthers}}\ and\ \bibinfo {author} {\bibfnamefont {H.}~\bibnamefont {Teitelbaum}},\ }\bibfield  {title} {\enquote {\bibinfo {title} {The linear mixture rule in chemical kinetics. ii. thermal dissociation of diatomic molecules},}\ }\href {\doibase 10.1016/0301-0104(88)87133-7} {\bibfield  {journal} {\bibinfo  {journal} {Chemical Physics}\ }\textbf {\bibinfo {volume} {127}},\ \bibinfo {pages} {351–362} (\bibinfo {year} {1988})}\BibitemShut {NoStop}%
\bibitem [{\citenamefont {Mirahmadi}\ and\ \citenamefont {Pérez-Ríos}(2022)}]{Mirahmadi2022}%
  \BibitemOpen
  \bibfield  {author} {\bibinfo {author} {\bibfnamefont {M.}~\bibnamefont {Mirahmadi}}\ and\ \bibinfo {author} {\bibfnamefont {J.}~\bibnamefont {Pérez-Ríos}},\ }\bibfield  {title} {\enquote {\bibinfo {title} {Three-body recombination in physical chemistry},}\ }\href {\doibase 10.1080/0144235x.2023.2237300} {\bibfield  {journal} {\bibinfo  {journal} {International Reviews in Physical Chemistry}\ }\textbf {\bibinfo {volume} {41}},\ \bibinfo {pages} {233–267} (\bibinfo {year} {2022})}\BibitemShut {NoStop}%
\bibitem [{\citenamefont {Lindemann}\ \emph {et~al.}(1922)\citenamefont {Lindemann}, \citenamefont {Arrhenius}, \citenamefont {Langmuir}, \citenamefont {Dhar}, \citenamefont {Perrin},\ and\ \citenamefont {McC.~Lewis}}]{Lindemann1922}%
  \BibitemOpen
  \bibfield  {author} {\bibinfo {author} {\bibfnamefont {F.~A.}\ \bibnamefont {Lindemann}}, \bibinfo {author} {\bibfnamefont {S.}~\bibnamefont {Arrhenius}}, \bibinfo {author} {\bibfnamefont {I.}~\bibnamefont {Langmuir}}, \bibinfo {author} {\bibfnamefont {N.~R.}\ \bibnamefont {Dhar}}, \bibinfo {author} {\bibfnamefont {J.}~\bibnamefont {Perrin}}, \ and\ \bibinfo {author} {\bibfnamefont {W.~C.}\ \bibnamefont {McC.~Lewis}},\ }\bibfield  {title} {\enquote {\bibinfo {title} {Discussion on “the radiation theory of chemical action”},}\ }\href {\doibase 10.1039/tf9221700598} {\bibfield  {journal} {\bibinfo  {journal} {Trans. Faraday Soc.}\ }\textbf {\bibinfo {volume} {17}},\ \bibinfo {pages} {598–606} (\bibinfo {year} {1922})}\BibitemShut {NoStop}%
\bibitem [{\citenamefont {Gilbert}, \citenamefont {Luther},\ and\ \citenamefont {Troe}(1983)}]{Gilbert1983}%
  \BibitemOpen
  \bibfield  {author} {\bibinfo {author} {\bibfnamefont {R.~G.}\ \bibnamefont {Gilbert}}, \bibinfo {author} {\bibfnamefont {K.}~\bibnamefont {Luther}}, \ and\ \bibinfo {author} {\bibfnamefont {J.}~\bibnamefont {Troe}},\ }\bibfield  {title} {\enquote {\bibinfo {title} {Theory of thermal unimolecular reactions in the fall‐off range. {II.} weak collision rate constants},}\ }\href {\doibase 10.1002/bbpc.19830870218} {\bibfield  {journal} {\bibinfo  {journal} {Berichte der Bunsengesellschaft f\"{u}r physikalische Chemie}\ }\textbf {\bibinfo {volume} {87}},\ \bibinfo {pages} {169–177} (\bibinfo {year} {1983})}\BibitemShut {NoStop}%
\bibitem [{\citenamefont {Jasper}\ and\ \citenamefont {Miller}(2014)}]{Jasper2014}%
  \BibitemOpen
  \bibfield  {author} {\bibinfo {author} {\bibfnamefont {A.~W.}\ \bibnamefont {Jasper}}\ and\ \bibinfo {author} {\bibfnamefont {J.~A.}\ \bibnamefont {Miller}},\ }\bibfield  {title} {\enquote {\bibinfo {title} {{Lennard–Jones} parameters for combustion and chemical kinetics modeling from full-dimensional intermolecular potentials},}\ }\href {\doibase 10.1016/j.combustflame.2013.08.004} {\bibfield  {journal} {\bibinfo  {journal} {Combustion and Flame}\ }\textbf {\bibinfo {volume} {161}},\ \bibinfo {pages} {101–110} (\bibinfo {year} {2014})}\BibitemShut {NoStop}%
\bibitem [{\citenamefont {Pack}, \citenamefont {Snow},\ and\ \citenamefont {Smith}(1972)}]{Pack1972}%
  \BibitemOpen
  \bibfield  {author} {\bibinfo {author} {\bibfnamefont {R.~T.}\ \bibnamefont {Pack}}, \bibinfo {author} {\bibfnamefont {R.~L.}\ \bibnamefont {Snow}}, \ and\ \bibinfo {author} {\bibfnamefont {W.~D.}\ \bibnamefont {Smith}},\ }\bibfield  {title} {\enquote {\bibinfo {title} {On the mechanism of low-temperature termolecular atomic recombination},}\ }\href {\doibase 10.1063/1.1677250} {\bibfield  {journal} {\bibinfo  {journal} {The Journal of Chemical Physics}\ }\textbf {\bibinfo {volume} {56}},\ \bibinfo {pages} {926–932} (\bibinfo {year} {1972})}\BibitemShut {NoStop}%
\bibitem [{\citenamefont {Whitlock}, \citenamefont {Muckerman},\ and\ \citenamefont {Roberts}(1972)}]{Whitlock1972}%
  \BibitemOpen
  \bibfield  {author} {\bibinfo {author} {\bibfnamefont {P.}~\bibnamefont {Whitlock}}, \bibinfo {author} {\bibfnamefont {J.}~\bibnamefont {Muckerman}}, \ and\ \bibinfo {author} {\bibfnamefont {R.}~\bibnamefont {Roberts}},\ }\bibfield  {title} {\enquote {\bibinfo {title} {Classical dynamics of three-body recombination via the resonance complex mechanism},}\ }\href {\doibase 10.1016/0009-2614(72)80400-7} {\bibfield  {journal} {\bibinfo  {journal} {Chemical Physics Letters}\ }\textbf {\bibinfo {volume} {16}},\ \bibinfo {pages} {460–463} (\bibinfo {year} {1972})}\BibitemShut {NoStop}%
\bibitem [{\citenamefont {Orel}(1987)}]{Orel1987}%
  \BibitemOpen
  \bibfield  {author} {\bibinfo {author} {\bibfnamefont {A.~E.}\ \bibnamefont {Orel}},\ }\bibfield  {title} {\enquote {\bibinfo {title} {Nascent vibrational/ rotational distribution produced by hydrogen atom recombination},}\ }\href {\doibase 10.1063/1.453628} {\bibfield  {journal} {\bibinfo  {journal} {The Journal of Chemical Physics}\ }\textbf {\bibinfo {volume} {87}},\ \bibinfo {pages} {314–318} (\bibinfo {year} {1987})}\BibitemShut {NoStop}%
\bibitem [{\citenamefont {Esposito}\ and\ \citenamefont {Capitelli}(2009)}]{Esposito2009}%
  \BibitemOpen
  \bibfield  {author} {\bibinfo {author} {\bibfnamefont {F.}~\bibnamefont {Esposito}}\ and\ \bibinfo {author} {\bibfnamefont {M.}~\bibnamefont {Capitelli}},\ }\bibfield  {title} {\enquote {\bibinfo {title} {Selective vibrational pumping of molecular hydrogen via gas phase atomic recombination},}\ }\href {\doibase 10.1021/jp9061829} {\bibfield  {journal} {\bibinfo  {journal} {The Journal of Physical Chemistry A}\ }\textbf {\bibinfo {volume} {113}},\ \bibinfo {pages} {15307–15314} (\bibinfo {year} {2009})}\BibitemShut {NoStop}%
\bibitem [{\citenamefont {Forrey}(2013)}]{Forrey2013}%
  \BibitemOpen
  \bibfield  {author} {\bibinfo {author} {\bibfnamefont {R.~C.}\ \bibnamefont {Forrey}},\ }\bibfield  {title} {\enquote {\bibinfo {title} {Sturmian theory of three-body recombination: Application to the formation of {$\rm H_2$} in primordial gas},}\ }\href {\doibase 10.1103/physreva.88.052709} {\bibfield  {journal} {\bibinfo  {journal} {Physical Review A}\ }\textbf {\bibinfo {volume} {88}} (\bibinfo {year} {2013}),\ 10.1103/physreva.88.052709}\BibitemShut {NoStop}%
\bibitem [{\citenamefont {Roberts}, \citenamefont {Bernstein},\ and\ \citenamefont {Curtiss}(1969)}]{Roberts1969}%
  \BibitemOpen
  \bibfield  {author} {\bibinfo {author} {\bibfnamefont {R.~E.}\ \bibnamefont {Roberts}}, \bibinfo {author} {\bibfnamefont {R.~B.}\ \bibnamefont {Bernstein}}, \ and\ \bibinfo {author} {\bibfnamefont {C.~F.}\ \bibnamefont {Curtiss}},\ }\bibfield  {title} {\enquote {\bibinfo {title} {Resonance theory of termolecular recombination kinetics: {H}+{H}+{M}$\rightarrow${$\rm H_2$}+{M}},}\ }\href {\doibase 10.1063/1.1671032} {\bibfield  {journal} {\bibinfo  {journal} {The Journal of Chemical Physics}\ }\textbf {\bibinfo {volume} {50}},\ \bibinfo {pages} {5163–5176} (\bibinfo {year} {1969})}\BibitemShut {NoStop}%
\bibitem [{\citenamefont {Leibowitz}, \citenamefont {Menard},\ and\ \citenamefont {Stickford}(1973)}]{Leibowitz1973_2}%
  \BibitemOpen
  \bibfield  {author} {\bibinfo {author} {\bibfnamefont {L.~P.}\ \bibnamefont {Leibowitz}}, \bibinfo {author} {\bibfnamefont {W.~A.}\ \bibnamefont {Menard}}, \ and\ \bibinfo {author} {\bibfnamefont {G.~H.}\ \bibnamefont {Stickford}},\ }\bibfield  {title} {\enquote {\bibinfo {title} {Radiative relaxation behind strong shock waves in hydrogen-helium mixtures},}\ }in\ \href@noop {} {\emph {\bibinfo {booktitle} {International Shock Tube Symposium}}}\ (\bibinfo  {publisher} {Stanford University Press},\ \bibinfo {year} {1973})\ pp.\ \bibinfo {pages} {306--317}\BibitemShut {NoStop}%
\bibitem [{\citenamefont {Leibowitz}\ and\ \citenamefont {Kuo}(1976)}]{Leibowitz1976}%
  \BibitemOpen
  \bibfield  {author} {\bibinfo {author} {\bibfnamefont {L.~P.}\ \bibnamefont {Leibowitz}}\ and\ \bibinfo {author} {\bibfnamefont {T.-J.}\ \bibnamefont {Kuo}},\ }\bibfield  {title} {\enquote {\bibinfo {title} {Ionizational nonequilibrium heating during outer planetary entries},}\ }\href {\doibase 10.2514/3.61465} {\bibfield  {journal} {\bibinfo  {journal} {{AIAA} Journal}\ }\textbf {\bibinfo {volume} {14}},\ \bibinfo {pages} {1324--1329} (\bibinfo {year} {1976})}\BibitemShut {NoStop}%
\end{thebibliography}%

\end{document}